\documentclass[10pt,twocolumn,letterpaper]{article}
\PassOptionsToPackage{table}{xcolor}

\usepackage[pagenumbers]{cvpr} 
\usepackage[T1]{fontenc}
\usepackage[utf8]{inputenc}
\usepackage{microtype}
\usepackage{nicefrac}
\usepackage{adjustbox}
\usepackage{placeins}
\usepackage{xurl}
\definecolor{lightpurple}{HTML}{a56dba}

\definecolor{cvprblue}{rgb}{0.21,0.49,0.74}
\usepackage[pagebackref,breaklinks,colorlinks,linkcolor=link,citecolor=cite,breaklinks=true,allcolors=cvprblue]{hyperref}

\hypersetup{
  pdftitle={Panda Diplomacy: Sensor-level Pre-training for Particle Imaging Detectors for High Energy and Nuclear Physics},
  pdfauthor={}
}

\usepackage{booktabs}
\usepackage{subcaption}
\usepackage{xspace}
\usepackage[percent]{overpic}
\definecolor{sph}{HTML}{4970ab}     
\definecolor{lar}{HTML}{3A3A96}     
\definecolor{next}{HTML}{315e2c}     
\definecolor{wc}{HTML}{8a5733}     
\definecolor{best}{gray}{0.94}       
\usepackage{pifont}
\newcommand{\cmark}{\ding{51}}       
\newcommand{\xmark}{\ding{55}}       

\newcommand{\LArTPC}{\textcolor{lar}{LArTPC}\xspace}
\newcommand{\sPHENIX}{\textcolor{sph}{sPHENIX}\xspace}
\newcommand{\NEXT}{\textcolor{next}{NEXT}\xspace}

\newcommand{\WC}{\textcolor{wc}{Water Cherenkov}\xspace}

\newcommand{\todo}[1]{\textcolor{red}{[TODO: #1]}}

\usepackage{pgfplots}
\pgfplotsset{compat=1.18}
\usepackage{lipsum}
\usepackage{array}
\usepackage{times}
\usepackage{epsfig}
\usepackage{graphicx}
\usepackage{float}
\usepackage{wrapfig}
\usepackage{amsmath,amssymb,amsthm}
\usepackage{algorithm,algorithmicx,algpseudocode}
\usepackage{bm,xspace}
\usepackage{comment}
\usepackage{multirow}
\usepackage{balance}
\usepackage{url}
\usepackage{booktabs}
\usepackage{etoolbox,siunitx}
\usepackage{calc}
\usepackage{pifont,hologo}
\usepackage{color}
\usepackage{adjustbox}
\usepackage{amsmath}
\usepackage{enumitem}
\usepackage{bbding}  %
\usepackage[normalem]{ulem}  %
\usepackage{contour}
\usepackage[table]{xcolor}
\usepackage{soul}%
\usepackage{tocloft} %
\usepackage{dsfont}
\usepackage{etoc} %
\PassOptionsToPackage{dvipsnames}{xcolor}
\usepackage{algorithm}
\usepackage{algpseudocode}
\usepackage{listings}

\PassOptionsToPackage{square,numbers,comma,sort}{natbib}
\usepackage[switch]{lineno}

\definecolor{darkblue}{HTML}{35394B}
\definecolor{blue}{HTML}{004bb3}
\definecolor{red}{HTML}{cc1100}
\definecolor{orange}{HTML}{cc7700}
\definecolor{gray}{HTML}{efefef}
\definecolor{darkgreen}{HTML}{228B22}
\definecolor{darkgray}{HTML}{808080}
\definecolor{lightpurple}{HTML}{a56dba}
\definecolor{white}{HTML}{ffffff}
\definecolor{C0}{HTML}{1f77b4}
\definecolor{C1}{HTML}{ff7f0e}
\definecolor{C2}{HTML}{2ca02c}
\definecolor{cite}{HTML}{3270b5}
\definecolor{link}{HTML}{b53532}
\definecolor{link}{HTML}{cc1100}
\definecolor{scratch}{HTML}{001219}
\definecolor{pretrain}{HTML}{0a9396}

\contourlength{0.8pt}

\renewcommand{\eqref}[1]{Eq.~\ref{#1}}

\newcolumntype{x}[1]{>{\centering\arraybackslash}p{#1}}
\newcolumntype{y}[1]{>{\raggedright\arraybackslash}p{#1}}
\newcolumntype{z}[1]{>{\raggedleft\arraybackslash}p{#1}}

\DeclareMathSymbol{@}{\mathord}{letters}{"3B}

\makeatletter
\DeclareRobustCommand\onedot{\futurelet\@let@token\@onedot}
\def\@onedot{\ifx\@let@token.\else.\null\fi\xspace}
\def\eg{\emph{e.g}\onedot} 
\def\ie{\emph{i.e}\onedot}

\def\etal{\emph{et al}\onedot}

\newcommand*{\Rom}[1]{\expandafter\@slowromancap\romannumeral #1@}
\newcommand*{\rom}[1]{\expandafter\romannumeral #1}

\def\1{\bm{1}}

\DeclareMathAlphabet{\mathsfit}{\encodingdefault}{\sfdefault}{m}{sl}
\SetMathAlphabet{\mathsfit}{bold}{\encodingdefault}{\sfdefault}{bx}{n}

\let\originalleft\left
\let\originalright\right
\renewcommand{\left}{\mathopen{}\mathclose\bgroup\originalleft}
\renewcommand{\right}{\aftergroup\egroup\originalright}
\def\paperID{36689} 
\def\confName{CVPR}
\def\confYear{2027}

\title{Panda Diplomacy: Sensor-level Representation Learning \\
across High-Energy and Nuclear Physics}

\author{
\parbox{\linewidth}{\centering Samuel Young\textsuperscript{\rm 1, 2 $\dagger$} $\quad$ C\'esar Jes\'us-Valls\textsuperscript{\rm 3} $\quad$ Omar A. Alterkait\textsuperscript{\rm 4} $\quad$ \\ Gregor Krzmanc\textsuperscript{\rm 1, 2} $\quad$ Kazuhiro Terao\textsuperscript{\rm 2}
\\ \vspace{10pt} \small
\textsuperscript{\rm 1} Stanford University  $\quad$ \textsuperscript{\rm 2} SLAC National Accelerator Laboratory \\
\textsuperscript{\rm 3} European Organization for Nuclear Research (CERN) $\quad$ \textsuperscript{\rm 4} Tufts University \\
\textsuperscript{$\dagger$} Corresponding author \\
}
}

\begin{document}

\twocolumn[{%
\renewcommand\twocolumn[1][]{#1}%
\vspace{-17mm}
\maketitle
\vspace{-8mm}
\begin{center}
    \captionsetup{type=figure}
    \includegraphics[height=0.64\linewidth]{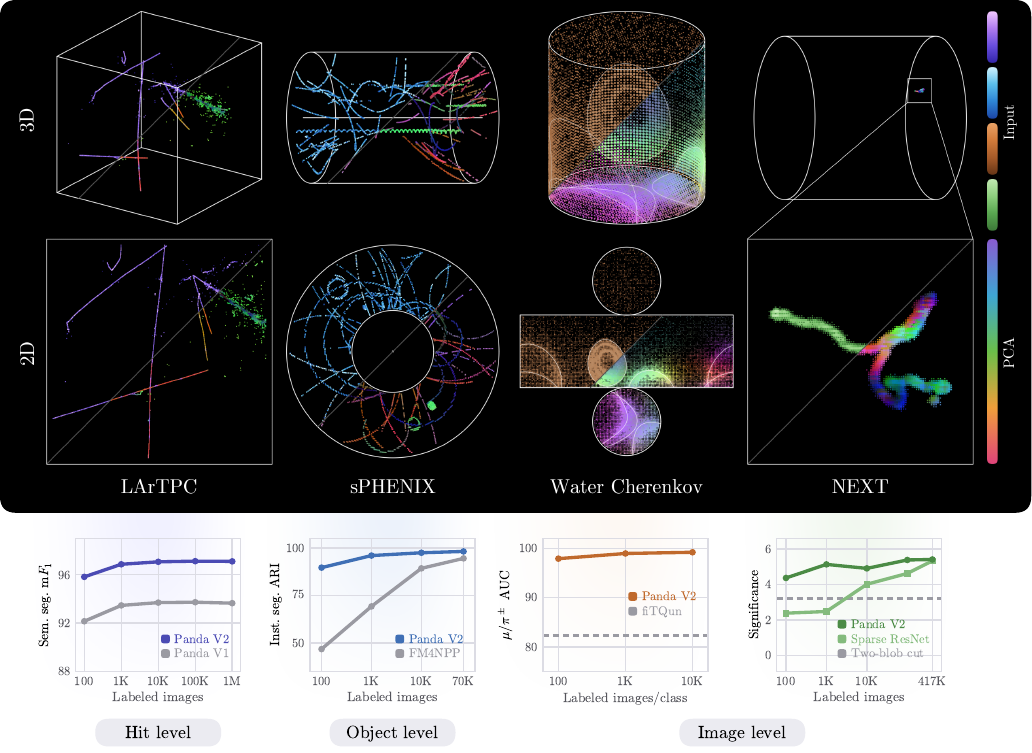}
    \captionof{figure}{
    \textbf{Learning particle and nuclear physics without labels.}
    The same model architecture and self-distillation pretraining recipe (\textit{Panda V2}) is \textbf{independently} applied to four different sensor-level detectors in high energy and nuclear physics: a liquid argon time projection chamber (\LArTPC), a collider TPC (\sPHENIX), a water Cherenkov detector (\textcolor{wc}{WC}), and \NEXT. \textbf{Top:} each point is a detector measurement, colored by its measured signal in the upper left of each view and, in the lower right, by the features that the Panda V2 Encoder produces, projected to RGB via PCA. \textbf{Bottom:} with the encoder frozen, lightweight heads trained on few labeled images match or outperform prior methods at the three separate levels of analysis: hit, object, and image.
    }\label{fig:1}
\end{center}%
}]

\addtocontents{toc}{\protect\setcounter{tocdepth}{-1}} 
\begin{abstract}

Foundation models trained on raw sensor data are increasingly being developed for particle and nuclear physics, but existing approaches have largely been designed and evaluated for individual detector technologies, and whether a common self-supervised learning framework can generalize across different detectors remains an open question.
We introduce Panda V2, a simplified encoder, a refined self-distillation objective, and a new fine-tuning recipe for learning dense, sensor-level representations. We apply it independently, with limited modifications, to four qualitatively different detectors: a liquid argon time projection chamber (LArTPC), the sPHENIX collider TPC, a water Cherenkov detector, and the NEXT high-pressure xenon TPC.
With frozen encoders and lightweight heads, the models achieve state-of-the-art or competitive performance in semantic segmentation, instance segmentation, event classification, and kinematics regression, often with orders of magnitude fewer labeled events than existing methods.
On sPHENIX, Panda V2 exceeds a recent foundation model's particle identification performance with 700$\times$ fewer labeled events; on the LArTPC, a linear probe trained with 100 labeled events outperforms the corresponding Panda V1 probe trained with any number of labels.
We further perform a cross-detector transfer learning study, finding that an encoder pretrained on one detector recovers 83--98\% of the in-domain performance on three of the four detectors. Finally, we show that the learned representations encode physically meaningful quantities (\eg, energy loss rate) not provided during pretraining. These results establish a common pretraining and adaptation framework for developing sensor-level foundation models across diverse particle detector technologies. Project site: \url{https://youngsm.com/panda-v2}.

\end{abstract}
    
\section{Introduction}
\label{sec:introduction}

Experiments in high-energy and nuclear physics (HENP) aim to answer fundamental questions about the Universe: why it contains more matter than antimatter, why and how neutrinos oscillate, what dark matter is, how supernovae explode, and how quarks and gluons behaved in the first microseconds of the Big Bang~\cite{Dine2003,GonzalezGarcia2003,Bertone2018,Burrows2013, Shuryak2017}. Many experiments image physics events with \textit{particle detectors} that record charge or light as 3D point clouds. Within this common format, the signature of a particle varies drastically across detectors (Fig.~\ref{fig:1}): whereas a muon leaves a straight track in a liquid argon time projection chamber (\LArTPC), it casts a sharp ring of light on the walls of a water Cherenkov (\textcolor{wc}{WC}) detector. Each experiment, and often each analysis, therefore builds its own reconstruction pipeline from hand-crafted pattern recognition~\cite{Acciarri2018}, likelihood fits~\cite{fitqunUB, fitqunT2K, fitqunSK}, or cascades of task-specific supervised networks~\cite{domine2020scalable}, trained on simulation that itself must be calibrated against data.

Recent progress in self-supervised learning (SSL) has produced strong foundation models (FMs)~\cite{bommasani2021opportunities} for natural images~\cite{caron2021emerging, oquab2024dinov2learningrobustvisual}, language~\cite{brown2020language}, and the natural sciences~\cite{Rives2021, Bodnar2025, Zeni2025} that, once pretrained on a large unlabeled corpus, transfer to downstream tasks with little or no fine-tuning. Within a HENP experiment, one such encoder would replace much of this siloed development: a new analysis would need only a small task-specific model on the encoder's features, reducing its development time and the labeled simulation it requires. The encoder could also be pretrained directly on real detector data, lessening the reliance on simulation, and would provide smaller collaborations, which often lack the resources to optimize a modern pipeline of their own, a strong starting point.

In this work, we focus on FMs that operate at the \textit{sensor-level}, \ie those that learn from raw detector hits rather than from reconstructed physics objects like particles, and are \textit{dense}, \ie those which provide a feature for every hit. Such a model can serve reconstruction at any granularity: hits, particles, groups of particles, and whole images. Dense sensor-level models have produced useful features~\cite{polarmae,panda,park2025fm4npp}, but each relies on a detector-specific architecture or pretraining objective. A recipe that each experiment could apply to its own data with the same architecture and objective, adapting only a few settings tied to its detector's geometry, would make these benefits available across the field at the cost of one pretraining run per experiment. It is also a prerequisite for the longer-term and more speculative goal of pretraining one model across detectors~\cite{doi:10.1142/S0217751X26500867, aarrestad2026building}.

In this work, we take a step toward a general sensor-level foundation model (FM) for HENP and show that a single encoder architecture, pretraining recipe, and fine-tuning recipe, which we name \textbf{Panda V2}, can be applied \textit{independently} to four types of detector data: a \LArTPC, the \sPHENIX collider TPC, a \textcolor{wc}{water Cherenkov} detector, and the \NEXT high-pressure xenon TPC. These detectors span argon, xenon, and water media; charge- and light-based readouts; and record particles whose energies span over four orders of magnitude. Starting from Panda V1~\cite{panda}, we revise the encoder, the pretraining recipe, and the fine-tuning recipe (\S\ref{sec:method}), reducing the backbone size by 41\% and simplifying the recipe. Across detectors, the architecture, objective, optimizer, and schedule are shared, and, apart from input preprocessing, only a handful of dataset-dependent hyperparameters differ. We vary both the pretraining set and the labeled set from 100 to 1M images, and fine-tune using a small task-specific head that operates on frozen encoder outputs.

On the \LArTPC, a linear layer on frozen features trained with only 100 labeled images surpasses the same probe on Panda V1 trained with 1M (95.8 vs.\ 93.7 m$F_1$). On \sPHENIX, a small MLP trained with 100 labeled images matches the particle identification of FM4NPP~\cite{park2025fm4npp} trained with 70K (82.1 vs.\ 81.8 m$F_1$), a \textbf{700$\times$ improvement} with a 3.5$\times$ smaller backbone. In \NEXT, a linear layer trained on mean-pooled features with 1K labeled images separates signal from background with a figure of merit ($\varepsilon_s/\sqrt{\varepsilon_b}$) of 5.13, against 3.21 for the standard hand-designed selection~\cite{ferrario2016topological,ferrario2019nextwhite}, and is comparable to a supervised network~\cite{kekic2021next} trained with all 417K (5.33). In the \textcolor{wc}{WC} detector, a linear layer trained with 10K labeled images per class separates muons from charged pions (99.2 vs.\ 82.3 AUC) and electrons from neutral pions (92.4 vs.\ 86.6 AUC) better than the classical reconstruction, which itself performs better than Panda V2 for particle kinematics prediction.

In sum, this paper makes the following contributions:
\begin{itemize}
\item (\S\ref{sec:method}) \textbf{Panda~V2}, a refined sensor-level FM pretraining and fine-tuning recipe for point clouds with structure at the scale of individual points.

\item (\S\ref{sec:results}) \textbf{In-depth evaluations of pre-trained models} at three levels of granularity using the same pretrained encoder: hits (semantic segmentation), particles (instance segmentation), and events (image classification and particle kinematics regression).

\item (\S\ref{sec:ood}) \textbf{A cross-detector transfer study}, in which a frozen encoder from another detector reaches 83--98\% of the in-domain encoder's score on the \LArTPC, \sPHENIX, and \textcolor{wc}{WC} detectors, but less than half of it on \NEXT, and never matches the in-domain encoder.

\item (\S\ref{sec:interp}) \textbf{A study of learned representation} via an analysis of the manifold structure within the learned embedding space of each model.
\end{itemize}

We release all model weights and training code.\footnote{\url{https://github.com/DeepLearnPhysics/Panda}}   
\section{Related work}
\label{sec:related}

\textbf{Self-supervised learning on point clouds.} Advances in self-supervised learning for 3D point clouds have provided strong, reusable generic features that speed up applications prototyping and often beat state-of-the-art solutions~\cite{zhang2026utonia,concerto}. There are three categories of training objectives which dominate large scale 3D pre-training: \textit{masked autoencoding} (\eg, PointMAE~\cite{pang2022maskedautoencoderspointcloud}, PointBERT~\cite{yu2022pointbertpretraining3dpoint}), \textit{autoregressive prediction} (\eg, PointGPT~\cite{chen2023pointgptautoregressivelygenerativepretraining}, PointMamba~\cite{liang2024pointmambasimplestatespace}), and \textit{contrastive learning} (\eg PointContrast~\cite{xie2020pointcontrastunsupervisedpretraining3d}, Utonia~\cite{zhang2026utonia}, and recently VINO~\cite{kruse2026vinorevisiting3drepresentation}). These methods rely on either UResNet~\cite{he2015deepresiduallearningimage, ronneberger2015unetconvolutionalnetworksbiomedical}- or ViT~\cite{dosovitskiy2020image}-style point cloud encoders (\eg, Sparse UResNet~\cite{choy20194dspatiotemporalconvnetsminkowski}, Point Transformer V3~\cite{wu2024pointtransformerv3simpler}, Volume Transformer~\cite{yilmaz2026volumetransformerrevisitingvanilla}). These methods were optimized for \textit{natural scene understanding}, and not for sensor-level data, where structure exists at the scale of individual points. We engineer Panda V2 with this in mind, finding that changes to the encoder architecture and pre-training scheme are necessary to achieve full performance.

\textbf{Foundation models in HENP.} 
FMs across HENP differ in what level of supervision is provided, what they take as input, and in whether they have been shown to transfer to other detectors. Generally, the level of supervision is chosen based on how well physicists trust their simulations of their detector. For particle colliders, often the simulation is extremely well-calibrated, so collider FMs are typically trained with supervised, generative, or simulator-derived objectives~\cite{Mikuni_2025, mikuni2024solving, bhimji2025omnilearnedfoundationmodelframework, hsu2026evenet, mokhtar2026mlpffm, kakati2026prompting, r3sl, vigl2026scalinglaws, vigl2026scalerecipe}.
Those in astrophysics have multiple observations of the same object by different instruments, naturally allowing for learning the alignment between views contrastively or via cross-modal masked modeling~\cite{lanusse2023astroclip, parker2025aion}. Neutrino physics has neither advantage: models of neutrino interactions with nuclei are poorly constrained, and a new detector can take more than a year to model. This motivates self-supervision on unlabeled detector data~\cite{wilkinson2025contrastivelearningrobustrepresentations, polarmae, panda, magel2026curtainfoundationmodelneutrino}, the setting we adopt, though we note some neutrino FMs still pretrain partly on simulation labels~\cite{AlonsoMonsalve2026}.

\textbf{Sensor-level and dense foundation models.} Most collider FMs take reconstructed objects as input, such as the particles of a jet~\cite{golling2024maskedparticlemodelingsets, birk2024omnijet}; as such, a point cloud format alone does not make a model sensor-level (\S\ref{sec:introduction}). Among sensor-level FMs, most produce one representation per image and are evaluated only on image-level tasks~\cite{wilkinson2025contrastivelearningrobustrepresentations, magel2026curtainfoundationmodelneutrino, AlonsoMonsalve2026}. Dense sensor-level FMs, which keep a feature for every hit, have so far been demonstrated only on single detectors~\cite{polarmae, panda, park2025fm4npp}, and all but Panda~\cite{panda} are pretrained via masked or autoregressive token prediction. Transfer across detectors has been demonstrated for object-level models~\cite{krzmanc2026cross, elsharkawy2026omnimol}, between geometries of a single detector technology~\cite{magel2026curtainfoundationmodelneutrino}, and with full fine-tuning on event-level tasks~\cite{AlonsoMonsalve2026}. In Panda V2, we apply one dense self-distillation recipe, with the same architecture and objective, to four sensor-level modalities, evaluate the same frozen features at the hit, particle, and event levels, and measure a full transfer matrix between encoders of different detectors. 
\section{Datasets}
\label{sec:data}

\begin{table}[t]\centering\scriptsize\setlength{\tabcolsep}{2.5pt}
\begin{tabular}{llrlrr}\toprule
Detector & Dataset & Events & A point is... & pts/img & obj/img \\\midrule
\LArTPC & PILArNet-M~\cite{polarmae} & 1.2M & $\vec{x}$, energy & 0.03-11.5K & 1--32 \\
\sPHENIX & TPCpp-10M~\cite{tpcpp10mZenodo} & 10.1M & $\vec{x}$, energy & 0.05--3.2K & 1--67 \\
\textcolor{wc}{WC} & WAND (this work) & 1.25M & static $\vec{x}$, photons, $t$ & 0.04-11.0K & 1-5 \\
\NEXT & NEXT~\cite{nextaissZenodo} & 1.1M &  $\vec{x}$, energy & 0.001-7.0K & 1--2 \\\bottomrule
\end{tabular}
\caption{\textbf{Datasets overview.} All data are simulated. Ranges are minimum and maximum possible values.}
\label{tab:interface}
\end{table}

\begin{figure*}
    \centering
    \includegraphics[width=\linewidth]{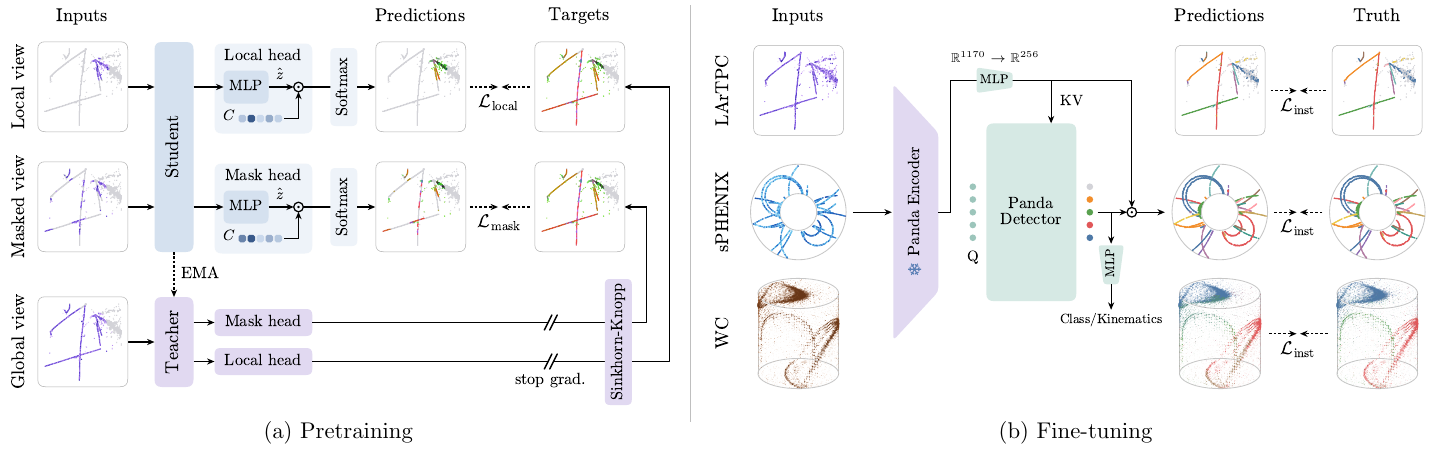}
    \caption{\textbf{Panda V2 overview.}}
    \label{fig:pretraining}
\end{figure*}

Here we briefly introduce each type of data and downstream physics tasks we evaluate our models on. Examples of each dataset's point cloud structure can be found in Fig.~\ref{fig:1}.
We refer the reader to the appendix for an in-depth description of these data and our full evaluation configuration.

\subsection{\LArTPC}

\textbf{Definition.} Liquid argon time projection chambers (LArTPCs)~\cite{rubbia1977liquid}, the detector of choice in US-based neutrino experiments and DUNE, the United States' flagship particle physics project, are high-resolution 3D imaging detectors that record millimeter-scale charge depositions over meter-scale volumes.

\textbf{Downstream tasks.} Following Panda V1~\cite{panda}, we choose evaluate our pre-trained models on the \textbf{semantic segmentation} of geometric motifs\footnote{Track, shower, delta ray, Michel electron, and low energy deposits (LED).} (reporting macro $F_1$), and \textbf{instance segmentation} of individual particles (reporting ARI).

\subsection{\sPHENIX}

\textbf{Definition.} The sPHENIX experiment~\cite{klest2022compacttpc} is a particle collider experiment at the Relativistic Heavy Ion Collider (RHIC)~\cite{busza2026rhic}, where heavy ions and polarized protons are collided together at energies of 200 GeV, allowing studies of the quark-gluon plasma and the structure of protons and nuclei. We use the TPCpp-10M~\cite{tpcpp10mZenodo} dataset of 10M images.

\textbf{Downstream tasks.} We follow FM4NPP's~\cite{park2025fm4npp} full set of evaluations, performing instance segmentation\footnote{Also known as track finding in the collider context.} (reporting adjusted rand score, \ie ARI) and \textbf{semantic segmentation} for particle identification (PID)\footnote{Proton, electron, kaon, pion, and other.} (reporting macro $F_1$\footnote{Note that FM4NPP reports macro-averaged precision and recall; we take their harmonic mean, and compute ours the same way. Note that this is not the normal definition we use in \LArTPC and \textcolor{wc}{WC}.}).

\subsection{\WC (WAND)}

\textbf{Definition.} In a water Cherenkov detector, neutrinos interacting in water produce charged particles that, when above the Cherenkov threshold, emit light in a cone around their direction of travel. This light is recorded by photon detectors (PMTs) lining the cylindrical detector walls, forming characteristic ring-like patterns (Fig.~\ref{fig:1}). We use WAND, a new dataset of 1.25M simulated events in a Super-Kamiokande-like~\cite{fukuda2003superkamiokande} detector, each with one to five particles ($e^\pm$, $\mu^\pm$, $\pi^\pm$, $\pi^0$, $\gamma$) of up to 2~GeV kinetic energy.

\textbf{Downstream tasks.} We focus on evaluating models' performance in the (i) discrimination between pairs of single particle images (muon/electron, electron/neutral pion, and muon/charged pion) (reporting AUC), (ii) prediction of the initial position, direction, and momentum of muons and electrons (reporting $68^\text{th}$ percentile absolute error $\sigma_{68}$), and (iii) multi-particle instance segmentation (reporting particle $F_1$\footnote{Correct matches are defined as predicted particles whose rings are correctly matched to another particle ($>0.5$ IoU over charge-weighted points) and whose type is correctly classified according to ground truth.}).

\subsection{\NEXT}

\textbf{Definition.} The NEXT experiment~\cite{nextcollaboration2021sensitivitytonnescaledetectorneutrinoless, lvarez2012} is a high-pressure xenon gaseous time projection chamber (TPC) searching for neutrinoless double beta decay ($0\nu\beta\beta$) with the $^{136}$Xe isotope. A major background in the NEXT experiment comes from the decay of $^{214}$Bi, which can emit a gamma ray that can produce electron tracks of similar energy. We utilize the publicly released NEXT dataset~\cite{nextaissZenodo}, which contains both signal and background events.

\textbf{Downstream tasks.} We perform image classification between signal ($0\nu\beta\beta$, two electrons) and background ($^{214}$Bi, one electron). We report community-standard maximum significance\footnote{Defined as the maximum value of $\varepsilon_s/\sqrt{\varepsilon_b}$, where $\varepsilon_s$ is signal selection efficiency, and $\varepsilon_b$ is background acceptance. The intuition is that one wants to increase signal ($\varepsilon_s$) while reducing the amount of background events ($\varepsilon_b$) that pass the classifier cut.} achieved. We compare with two baselines from prior work on the NEXT detector: a cut on the energies of the two blobs~\cite{ferrario2016topological,ferrario2019nextwhite}, whose voxel size and blob radius we choose on validation data, and a supervised sparse ResNet~\cite{kekic2021next}, which we retrain on this dataset with its published architecture and training recipe.

\subsubsection{Splits.} On every detector, pretraining uses only training split images, and checkpoints are selected on validation images and scored once on the test split. Two datasets limit $K$: TPCpp-10M labels only 70K of its training events, and in \NEXT, classification is defined on the 417K training events that pass the experiment's fiducial single-track selection. Split sizes are listed in the Supplementary Material.

\section{Panda V2}
\label{sec:method}

\begin{figure}
    \centering
    \input{figs/ladder_combined.pgf}
    \vspace{-4pt}
    \caption{\textbf{Improving self-supervised pretraining.}}
    \label{fig:pt_ladder}
\end{figure}

\begin{figure}

\vspace{0pt}
\centering
\includegraphics[width=0.9\linewidth]{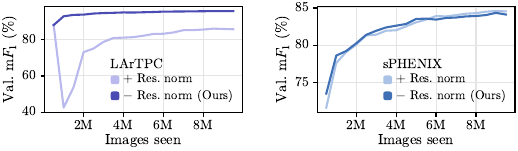}
\captionof{figure}{\textbf{Removing residual pre-LayerNorm.} With residual pre-LN, the \LArTPC probe collapses early in training, whereas \sPHENIX is unaffected.}
\label{fig:norm}
\end{figure}

The signature that a particle leaves in a HENP detector is, to good approximation, independent of where in the detector the particle travels, such that two cropped and strongly augmented views of one image still appear, to a human, as the same type of particle. Additionally, detector images are spatially \textit{redundant}: what a small mask removes from an image can usually be inferred from the visible surroundings. Together, these two properties can be exploited to provide a training signal for learning meaningful representations that requires no labels.

From each unlabeled image in a dataset, we sample two global, two masked, and six local views, and apply hard augmentations\footnote{We follow \cite{wu2025sonataselfsupervisedlearningreliable, zhang2026utonia} in using per-point sub-voxel jittering, random 3D rotations, flips, and coordinate scaling.} to each view. 
A student network encodes the local and mask views, while a teacher network encodes the global views. The student is trained to match the teacher's assignment of each point to a learnable bank of prototypes~\cite{swav}, via a cross-entropy loss. As SwAV, the teacher's prototype assignments are balanced  across the entire batch via Sinkhorn-Knopp~\cite{Sinkhorn1967ConcerningNM}. The teacher does not receive any gradients, and instead is updating as an exponential moving average of the student network. This process is illustrated in Fig.~\ref{fig:pretraining}a.

\subsection{Robust pretraining with the Panda Encoder}
\label{sec:pretraining}

Starting from Panda V1~\cite{panda} with the Point Transformer V3 (PTv3)~\cite{wu2024pointtransformerv3simpler} backbone, we modify the encoder and the self-distillation objective, one change at a time, measuring each on both \LArTPC and \sPHENIX with a linear probe on frozen features. We make changes that both simplify the recipe and increase performance. We report the cumulative m$F_1$ after each change, as well as the score lost when that change alone is removed from the final recipe ($\Delta$) in Fig.~\ref{fig:pt_ladder}. All configurations are pretrained over 10 epochs on 1M images at a batch size of 48 unless stated otherwise.

\newcommand{\rung}[1]{\hyperref[fig:pt_ladder]{Fig.~\ref*{fig:pt_ladder}.#1}}

Starting from Panda V1~\cite{panda}, which uses a Point Transformer V3 (PTv3)~\cite{wu2024pointtransformerv3simpler} backbone with attention windows of 256 points (\rung{1}), we change one component at a time. We add random isotropic scaling of the coordinates, by a factor drawn from $[0.9, 1.2]$ for the full image and from $[0.9, 1.1]$ for each view (\rung{2})~\cite{zhang2026utonia}. We replace PTv3 with LitePT-B~\cite{litept} without its non-residual pre-LayerNorm~\cite{ba2016layernormalization} in the attention blocks, which destabilized pretraining (\rung{3}), and reduce the encoder from five resolutions to three, with strides of 2, 1 and 4 between its four stages (widths 54/108/432/576, blocks 3/6/12/6), moving compute to the finer stages where representations are more important (\rung{4}); we call the result the \textbf{Panda V2 Encoder}. Additionally, we hold the weight decay, masking ratio and mask size constant at 0.05, 60\% and 0.04 instead of increasing them over the course of training (\rung{5}), double the number of prototypes to 8{,}192 (\rung{6}), and widen the attention windows to 1{,}024 points (\rung{7}). Following recent work~\cite{dos, yang2026foundationmodels3dscene}, we remove mask tokens from the encoder input, so that the loss is computed only on points visible to the student (\rung{8}), and we apply the loss to features upcast through every stage to the finest resolution (1{,}170 channels), a choice that, unlike in prior work~\cite{zhang2026utonia}, affects performance (\rung{9}). Finally, we increase the batch size to 96 and train for 20 epochs (\rung{10}). Between detectors, we change only the voxel size,\footnote{We define the voxel size in each detector to be 0.9$\times$ the minimum distance between two 3D points.} per-event minimum point cuts, mask size, whether scale augmentation is applied, and the level at which the loss is applied (Table~\ref{tab:recipe}). We provide reasons for each in the Supplementary Material.

\begin{table}[t]\centering\scriptsize\setlength{\tabcolsep}{3pt}
\begin{tabular}{lcccc}\toprule
Hyperparameter Setting & \LArTPC & \sPHENIX & \textcolor{wc}{WC} & \NEXT \\\midrule
Mask cell size (voxels) & 40 & 13 & 17 & 15 \\
Min.\ points per event & 1{,}024 & 1{,}024 & 1{,}024 & 512 \\
Global crop fraction & 0.4--1.0 & 0.85--1.0 & 0.4--1.0 & 0.4--1.0 \\
Local crop fraction & 0.1--0.4 & 0.3--0.6 & 0.1--0.4 & 0.1--0.4 \\
Scale augmentation (\rung{2}) & \checkmark & \checkmark & \xmark & \checkmark \\
Upcast level (\rung{9}) & 3 (finest) & 3 (finest) & 2 (second finest) & 3 (finest) \\\bottomrule
\end{tabular}
\caption{\textbf{Self-distillation hyperparameters that differ between detectors.}
}
\label{tab:recipe}
\end{table}

\subsection{Enabling data-efficient task transfer}
\label{sec:transfer}

\begin{figure}
    \centering
    \includegraphics[width=\linewidth]{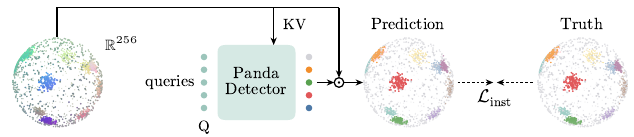}
    \caption{\textbf{Synthetic pretraining for object detection}}
    \label{fig:synthpd}
\end{figure}

\begin{table}[t]
\centering
\footnotesize
\definecolor{gain}{HTML}{2E7D32}
\newcommand{\up}[1]{{\scriptsize\textcolor{gain}{+#1}}}
\begin{tabular}{l c@{\hspace{4pt}}l c@{\hspace{4pt}}l}
\toprule
Instance segmentation (ARI) & \multicolumn{2}{c}{\LArTPC} & \multicolumn{2}{c}{\sPHENIX} \\
\midrule
Panda V1 decoding~\cite{panda} & 76.9 & & 63.8 & \\
+ optimized postprocessing & 78.8 & \up{1.9} & 70.0 & \up{6.2} \\
+ synthetic pretraining & 84.1 & \up{5.3} & 84.9 & \up{14.9} \\
+ image mixing & \cellcolor{best}\textbf{86.6} & \up{2.5} & \cellcolor{best}\textbf{89.6} & \up{4.7} \\
\bottomrule
\end{tabular}
\vspace{-0.5em}
\caption{\textbf{Panda Detector optimizations.}}
\vspace{-5pt}
\label{tab:ft_ladder}
\end{table}
\begin{table}[t]
\centering
\scriptsize
\setlength{\tabcolsep}{2.5pt}
\begin{tabular*}{\linewidth}[t]{l@{\extracolsep{\fill}}r@{\extracolsep{0pt}\hspace{2\tabcolsep}}cccc@{\hspace{0.5\tabcolsep}}c@{\hspace{0.5\tabcolsep}}cccc}
\toprule
\multicolumn{2}{l}{\setbox0\hbox{Sem.\ Seg.\ (mF$_1$)}\rlap{\smash{\raisebox{-3.6pt}{\rule{\wd0}{\cmidrulewidth}}}}\box0} & \multicolumn{4}{c}{PT$_K$ + FT$_K$} &  & \multicolumn{4}{c}{PT$_\text{1M}$ + FT$_K$} \\
\cmidrule(lr){3-6} \cmidrule(lr){8-11}
Method / $K$ & Param. & 100 & 1K & 10K & 1M &  & 100 & 1K & 10K & 1M \\
\midrule
\multicolumn{11}{l}{\textit{Supervised}} \\
PTv3~\cite{wu2024pointtransformerv3simpler} & 107.1M & 63.9 & 77.9 & 94.0 & \cellcolor[HTML]{E3E3FF}98.1 & \phantom{\raisebox{0.12ex}{\tiny=}} & \llap{\raisebox{0.12ex}{\tiny=}\hspace{0.5\tabcolsep}}63.9 & 77.9 & 94.0 & \cellcolor[HTML]{E3E3FF}98.1 \\
Panda V2 & 53.7M & \cellcolor[HTML]{D6D6FF}\textbf{90.4} & \cellcolor[HTML]{D6D6FF}\textbf{95.5} & \cellcolor[HTML]{D6D6FF}\textbf{97.4} & \cellcolor[HTML]{D6D6FF}\textbf{98.2} & \phantom{\raisebox{0.12ex}{\tiny=}} & \llap{\raisebox{0.12ex}{\tiny=}\hspace{0.5\tabcolsep}}90.4 & \cellcolor[HTML]{F4F4FF}95.5 & \cellcolor[HTML]{E3E3FF}97.4 & \cellcolor[HTML]{D6D6FF}\textbf{98.2} \\
\midrule
\multicolumn{11}{l}{\textit{Self-supervised}} \\
PoLAr-MAE (dec.)~\cite{polarmae} & 4.4M  & 71.2 & 80.6 & 84.6 & 83.4 &  & -- & -- & -- & 83.4 \\
Panda V1 (lin.) & \llap{\raisebox{0.12ex}{\tiny$<$}\kern0.5pt}0.01M & 78.0 & 89.9 & 92.9 & 93.7 &  & \cellcolor[HTML]{F4F4FF}92.2 & 93.5 & 93.7 & 93.7 \\
Panda V2 (lin.) & \llap{\raisebox{0.12ex}{\tiny$<$}\kern0.5pt}0.01M & \cellcolor[HTML]{F4F4FF}82.3 & \cellcolor[HTML]{F4F4FF}94.1 & \cellcolor[HTML]{F4F4FF}96.7 & 97.1 &  & \cellcolor[HTML]{E3E3FF}95.8 & \cellcolor[HTML]{E3E3FF}96.9 & \cellcolor[HTML]{F4F4FF}97.1 & 97.1 \\
Panda V2 (dec.) & 0.3M & \cellcolor[HTML]{E3E3FF}83.0 & \cellcolor[HTML]{E3E3FF}94.6 & \cellcolor[HTML]{E3E3FF}97.2 & \cellcolor[HTML]{F4F4FF}97.8 &  & \cellcolor[HTML]{D6D6FF}\textbf{95.9} & \cellcolor[HTML]{D6D6FF}\textbf{97.2} & \cellcolor[HTML]{D6D6FF}\textbf{97.6} & \cellcolor[HTML]{F4F4FF}97.8 \\
\color{black!45}PoLAr-MAE (full)~\cite{polarmae} & 24.4M \color{black!45} & \color{black!45}73.5 & \color{black!45}84.0 & \color{black!45}86.5 & \color{black!45}85.8 & \color{black!45} & \color{black!45}-- & \color{black!45}-- & \color{black!45}-- & \color{black!45}85.8 \\
\color{black!45}Panda V1 (full) & \color{black!45}107.1M & \color{black!45}84.3 & \color{black!45}93.4 & \color{black!45}96.2 & \color{black!45}98.2 & \color{black!45} & \color{black!45}92.6 & \color{black!45}95.1 & \color{black!45}96.4 & \color{black!45}98.2 \\
\bottomrule
\end{tabular*}
\vspace{-6pt}
\caption{\label{tab:semseg_lar}\textbf{Semantic segmentation on LArTPC.}}
\end{table}

\begin{table}[t]
\centering
\scriptsize
\setlength{\tabcolsep}{2.5pt}
\begin{tabular*}{\linewidth}[t]{l@{\extracolsep{\fill}}r@{\extracolsep{0pt}\hspace{2\tabcolsep}}cccc@{\hspace{0.5\tabcolsep}}c@{\hspace{0.5\tabcolsep}}cccc}
\toprule
\multicolumn{2}{l}{\setbox0\hbox{Sem.\ Seg.\ (mF$_1$)}\rlap{\smash{\raisebox{-3.6pt}{\rule{\wd0}{\cmidrulewidth}}}}\box0} & \multicolumn{4}{c}{PT$_K$ + FT$_K$} &  & \multicolumn{4}{c}{PT$_\text{1M}$ + FT$_K$} \\
\cmidrule(lr){3-6} \cmidrule(lr){8-11}
Method / $K$ & Param. & 100 & 1K & 10K & 70K &  & 100 & 1K & 10K & 70K \\
\midrule
\multicolumn{11}{l}{\textit{Supervised}} \\
OneFormer~\cite{kolodiazhnyi2023oneformer3dtransformerunifiedpoint,park2025fm4npp} & 45.0M & -- & -- & -- & 53.0 & \raisebox{0.12ex}{\tiny=} & -- & -- & -- & 53.0 \\
SAGEConv~\cite{park2025fm4npp} & 0.9M & -- & -- & -- & 53.6 & \raisebox{0.12ex}{\tiny=} & -- & -- & -- & 53.6 \\
Panda V2 & 54.0M & \cellcolor[HTML]{C4D7F4}\textbf{56.2} & \cellcolor[HTML]{C4D7F4}\textbf{72.9} & \cellcolor[HTML]{C4D7F4}\textbf{83.3} & \cellcolor[HTML]{D8E5F8}83.7 & \phantom{\raisebox{0.12ex}{\tiny=}} & \cellcolor[HTML]{EEF3FC}\llap{\raisebox{0.12ex}{\tiny=}\hspace{0.5\tabcolsep}}56.2 & \cellcolor[HTML]{EEF3FC}72.9 & \cellcolor[HTML]{EEF3FC}83.3 & \cellcolor[HTML]{D8E5F8}83.7 \\
\midrule
\multicolumn{11}{l}{\textit{Self-supervised}} \\
FM4NPP (dec.){$^\dagger$}~\cite{park2025fm4npp} & 0.7M & -- & -- & -- & -- &  & -- & -- & -- & \cellcolor[HTML]{EEF3FC}81.8 \\
Panda V2 (lin.) & \llap{\raisebox{0.12ex}{\tiny$<$}\kern0.5pt}0.01M & \cellcolor[HTML]{EEF3FC}51.3 & \cellcolor[HTML]{EEF3FC}69.0 & \cellcolor[HTML]{EEF3FC}78.1 & \cellcolor[HTML]{EEF3FC}83.1 &  & \cellcolor[HTML]{D8E5F8}79.9 & \cellcolor[HTML]{D8E5F8}82.9 & \cellcolor[HTML]{D8E5F8}84.3 & \cellcolor[HTML]{D8E5F8}83.7 \\
Panda V2 (mlp) & 0.3M & \cellcolor[HTML]{D8E5F8}52.8 & \cellcolor[HTML]{D8E5F8}71.0 & \cellcolor[HTML]{D8E5F8}80.7 & \cellcolor[HTML]{C4D7F4}\textbf{85.1} &  & \cellcolor[HTML]{C4D7F4}\textbf{82.1} & \cellcolor[HTML]{C4D7F4}\textbf{84.6} & \cellcolor[HTML]{C4D7F4}\textbf{85.1} & \cellcolor[HTML]{C4D7F4}\textbf{85.3} \\
\bottomrule
\multicolumn{11}{@{}l}{$^\dagger$Pretrained on 10M.} \\
\end{tabular*}
\vspace{-6pt}
\caption{\label{tab:semseg_sph}\textbf{Semantic segmentation on sPHENIX.}}
\end{table}

\begin{table}[t]
\centering
\scriptsize
\setlength{\tabcolsep}{2.5pt}
\begin{tabular*}{\linewidth}[t]{l@{\extracolsep{\fill}}r@{\extracolsep{0pt}\hspace{2\tabcolsep}}cccc@{\hspace{0.5\tabcolsep}}c@{\hspace{0.5\tabcolsep}}cccc}
\toprule
\multicolumn{2}{l}{\setbox0\hbox{Instance Seg.\ (ARI)}\rlap{\smash{\raisebox{-3.6pt}{\rule{\wd0}{\cmidrulewidth}}}}\box0} & \multicolumn{4}{c}{PT$_K$ + FT$_K$} &  & \multicolumn{4}{c}{PT$_\text{1M}$ + FT$_K$} \\
\cmidrule(lr){3-6} \cmidrule(lr){8-11}
Method / $K$ & Param. & 100 & 1K & 10K & 1M &  & 100 & 1K & 10K & 1M \\
\midrule
\multicolumn{11}{l}{\textit{Supervised}} \\
Panda V2 & 58.2M & \cellcolor[HTML]{D6D6FF}\textbf{93.5} & \cellcolor[HTML]{D6D6FF}\textbf{96.8} & \cellcolor[HTML]{D6D6FF}\textbf{97.4} & \cellcolor[HTML]{E3E3FF}97.9 & \phantom{\raisebox{0.12ex}{\tiny=}} & \cellcolor[HTML]{D6D6FF}\llap{\raisebox{0.12ex}{\tiny=}\hspace{0.5\tabcolsep}}\textbf{93.5} & \cellcolor[HTML]{D6D6FF}\textbf{96.8} & \cellcolor[HTML]{D6D6FF}\textbf{97.4} & \cellcolor[HTML]{E3E3FF}97.9 \\
\midrule
\multicolumn{11}{l}{\textit{Self-supervised}} \\
Panda V1 (dec.) & 4.5M & -- & -- & -- & -- &  & 73.2 & 87.3 & \cellcolor[HTML]{E3E3FF}95.0 & 97.0 \\
Panda V2 (dec.) & 4.7M & \cellcolor[HTML]{E3E3FF}67.2 & \cellcolor[HTML]{E3E3FF}92.0 & \cellcolor[HTML]{E3E3FF}97.2 & \cellcolor[HTML]{D6D6FF}\textbf{98.0} &  & \cellcolor[HTML]{E3E3FF}86.6 & \cellcolor[HTML]{E3E3FF}95.7 & \cellcolor[HTML]{D6D6FF}\textbf{97.4} & \cellcolor[HTML]{D6D6FF}\textbf{98.0} \\
\color{black!45}Panda V1 (full) & \color{black!45}95.2M & \color{black!45}-- & \color{black!45}-- & \color{black!45}-- & \color{black!45}97.9 & \color{black!45} & \color{black!45}-- & \color{black!45}-- & \color{black!45}-- & \color{black!45}97.9 \\
\bottomrule
\end{tabular*}
\vspace{-4pt}
\caption{\label{tab:instseg_lar}\textbf{Instance segmentation on LArTPC.}}
\end{table}

\begin{table}[t]
\centering
\scriptsize
\setlength{\tabcolsep}{3.4pt}
\begin{tabular*}{\linewidth}[t]{l@{\extracolsep{\fill}}r@{\extracolsep{0pt}\hspace{2\tabcolsep}}cccc@{\hspace{0.5\tabcolsep}}c@{\hspace{0.5\tabcolsep}}cccc}
\toprule
\multicolumn{2}{l}{\setbox0\hbox{Instance Seg.\ (ARI)}\rlap{\smash{\raisebox{-3.6pt}{\rule{\wd0}{\cmidrulewidth}}}}\box0} & \multicolumn{4}{c}{PT$_K$ + FT$_K$} &  & \multicolumn{4}{c}{PT$_\text{1M}$ + FT$_K$} \\
\cmidrule(lr){3-6} \cmidrule(lr){8-11}
Method / $K$ & Param. & 100 & 1K & 10K & 70K &  & 100 & 1K & 10K & 70K \\
\midrule
\multicolumn{11}{l}{\textit{Supervised}} \\
EggNet~\cite{calafiura2024eggnetevolvinggraphbasedgraph, park2025fm4npp} & 0.2M & -- & -- & -- & 72.6 & \raisebox{0.12ex}{\tiny=} & -- & -- & -- & 72.6 \\
Exa.TrkX~\cite{exatrk, park2025fm4npp} & 3.9M & -- & -- & -- & \cellcolor[HTML]{EEF3FC}87.7 & \raisebox{0.12ex}{\tiny=} & -- & -- & -- & 87.7 \\
Panda V2 & 57.6M & \cellcolor[HTML]{C4D7F4}\textbf{86.3} & \cellcolor[HTML]{C4D7F4}\textbf{94.3} & \cellcolor[HTML]{C4D7F4}\textbf{96.7} & \cellcolor[HTML]{D8E5F8}97.6 & \phantom{\raisebox{0.12ex}{\tiny=}} & \cellcolor[HTML]{D8E5F8}\llap{\raisebox{0.12ex}{\tiny=}\hspace{0.5\tabcolsep}}86.3 & \cellcolor[HTML]{D8E5F8}94.3 & \cellcolor[HTML]{D8E5F8}96.7 & \cellcolor[HTML]{D8E5F8}97.6 \\
\midrule
\multicolumn{11}{l}{\textit{Self-supervised}} \\
FM4NPP (dec.){$^\dagger$}~\cite{park2025fm4npp} & 2.4M & -- & -- & -- & -- &  & \cellcolor[HTML]{EEF3FC}46.8 & \cellcolor[HTML]{EEF3FC}69.3 & \cellcolor[HTML]{EEF3FC}89.3 & \cellcolor[HTML]{EEF3FC}94.5 \\
Panda V2 (dec.) & 4.2M & \cellcolor[HTML]{D8E5F8}73.5 & \cellcolor[HTML]{D8E5F8}91.5 & \cellcolor[HTML]{D8E5F8}96.2 & \cellcolor[HTML]{C4D7F4}\textbf{98.0} &  & \cellcolor[HTML]{C4D7F4}\textbf{89.6} & \cellcolor[HTML]{C4D7F4}\textbf{96.0} & \cellcolor[HTML]{C4D7F4}\textbf{97.4} & \cellcolor[HTML]{C4D7F4}\textbf{98.1} \\
\bottomrule
\multicolumn{11}{@{}l}{$^\dagger$Pretrained on 10M.} \\
\end{tabular*}
\vspace{-6pt}
\caption{\label{tab:instseg_sph}\textbf{Instance segmentation on sPHENIX.}}
\end{table}

\begin{table}[t]
\centering
\scriptsize
\setlength{\tabcolsep}{2.9pt}
\begin{tabular}{l cccc}
\toprule
PT$_\text{1M}$ $\downarrow$ FT$_\text{1K}$ $\rightarrow$ & LArTPC (ARI) & sPHENIX (ARI) & WC ($F_1$) & NEXT (f.o.m.) \\
\midrule
LArTPC & \cellcolor[HTML]{D6D6FF}\textbf{96.0}\rlap{\,\tiny$\checkmark$} & \cellcolor[HTML]{D8E5F8}88.0\rlap{\,\tiny$\checkmark$} & 35.5\rlap{\,\tiny$\checkmark$} & \cellcolor[HTML]{DFEEDB}2.53\rlap{\,\tiny$\times$} \\
sPHENIX & \cellcolor[HTML]{E3E3FF}93.7\rlap{\,\tiny$\checkmark$} & \cellcolor[HTML]{C4D7F4}\textbf{96.2}\rlap{\,\tiny$\checkmark$} & \cellcolor[HTML]{FDF4EC}35.8\rlap{\,\tiny$\checkmark$} & \cellcolor[HTML]{F1F8EF}2.34\rlap{\,\tiny$\times$} \\
WC & 89.5\rlap{\,\tiny$\checkmark$} & 82.2\rlap{\,\tiny$\checkmark$} & \cellcolor[HTML]{F7D9BD}\textbf{42.5}\rlap{\,\tiny$\checkmark$} & 2.31\rlap{\,\tiny$\times$} \\
NEXT & \cellcolor[HTML]{F4F4FF}91.2\rlap{\,\tiny$\checkmark$} & \cellcolor[HTML]{EEF3FC}84.0\rlap{\,\tiny$\checkmark$} & \cellcolor[HTML]{FAE6D3}37.2\rlap{\,\tiny$\checkmark$} & \cellcolor[HTML]{CFE6C8}\textbf{5.53}\rlap{\,\tiny$\checkmark$} \\
\midrule
Baseline & 87.3\rlap{$^{a}$} & 69.3\rlap{$^{b}$} & 33.9\rlap{$^{c}$} & 3.21\rlap{$^{d}$} \\
\bottomrule
\multicolumn{5}{@{}l}{$^a$Panda V1~\cite{panda}. $^b$FM4NPP~\cite{park2025fm4npp}. $^c$fiTQun~\cite{fitqunSK}. $^d$Two-blob cut~\cite{ferrario2019nextwhite,ferrario2016topological}.} \\
\end{tabular}
\vspace{-4pt}
\caption{\label{tab:transfer}\textbf{Out-of-domain transfer with a frozen encoder.} $\checkmark$/$\times$: above/below the baseline.}
\end{table}


Pretraining a strong encoder is half the battle; we also need to make its features usable with task-specific heads that are small, simple to optimize, and label efficient.

Many reconstruction tasks in HENP can be abstracted as a single mapping $f$ from an image, a set of $N$ points with
coordinates $\mathbf{x}_i \in \mathbb{R}^3$ and measured features $\mathbf{e}_i$ to a set of $G$ objects,
\begin{equation}
  f:\; \{(\mathbf{x}_i, \mathbf{e}_i)\}_{i=1}^{N} \;\longmapsto\; \{(y_g,\, c_g,\, \mathbf{r}_g)\}_{g=1}^{G},
  \label{eq:setpred}
\end{equation}
in which each object consists of a binary mask $y_g \in \{0,1\}^N$ that selects the points belonging to the object, a
class $c_g$, and, optionally, regressed properties $\mathbf{r}_g$.
This style of problem has recently been increasingly addressed by DETR-style decoders~\cite{detr,m2f,kolodiazhnyi2023oneformer3dtransformerunifiedpoint}, and, as in Panda V1~\cite{panda}, we use a module of this type----which we call Panda Detector---for every such instance segmentation task (illustrated in Fig.~\ref{fig:pretraining}b). Such decoders are, however, difficult to train with few labels~\cite{m2f,zhu2021deformabledetrdeformabletransformers, li2022dndetracceleratedetrtraining, zhang2022dinodetrimproveddenoising}, and we therefore focus on improving Panda Detector at 100 labeled images using a frozen backbone, ablating with one change at a time (Table~\ref{tab:ft_ladder}).

\textbf{Optimized postprocessing.}
We find that the decoding rule inherited from Panda V1 discards low-confidence true objects and leaves their points unassigned. We instead keep every query whose most likely real class has probability at least 0.3, even if the no-object class is more likely. Each point not covered by a kept mask is then assigned to the kept query with the highest mask probability at that point, provided this probability is at least 0.1. This increases ARI by (\textcolor{lar}{+1.9}/\textcolor{sph}{+6.2}) for \LArTPC/\sPHENIX.

\textbf{Synthetic pretraining.} In the few-label regime, instance segmentation often lags far behind semantic segmentation. We attribute this gap to the two skills that a transformer-based~\cite{vaswani2017attention} set prediction network must learn: (i) a clustering procedure that partitions the encoder's features, with each query choosing one cluster in latent space, and (ii) a classification of each cluster. We argue that only the latter is much more detector-specific and easier to learn than the former. 
Thus, following prior work in amortized clustering~\cite{lee2019deepamortizedclustering,lee2019set} and prior-data fitted networks (PFNs)~\cite{muller2021transformers}, we fit Panda Detector V2 to synthetic point sets drawn from a Gaussian mixture model (GMM), so that a single forward pass of the decoder approximates Bayesian inference of the cluster assignments under a Gaussian prior. Each point cloud is sampled from a GMM in $\mathbb{R}^{256}$ whose component centroids lie in a random 32-dimensional subspace, and after adding uniform background noise, every point is projected onto the unit hypersphere (illustrated in Fig.~\ref{fig:synthpd}).
We find this prior beneficial (\textcolor{lar}{+5.3}/\textcolor{sph}{+14.9} ARI), and thus use this pretrained decoder as an initialization for downstream tasks. We refer the reader to the Supplementary Material for a full description and evaluation of synthetic pretraining.

\textbf{Image mixing.} We introduce image mixing~\cite{nekrasov2021mix3doutofcontextdataaugmentation}, \ie the introduction of an additional labeled image within each training image with probability 0.5, further increasing ARI by (\textcolor{lar}{+2.5}/\textcolor{sph}{+4.7}).

Taken together, we improve Panda Detector's performance for \LArTPC and \sPHENIX by 9.7 and 25.8 ARI relative to Panda Detector V1.

\begin{table*}[t]
\scriptsize
\begin{minipage}[t]{0.648\linewidth}
\centering
\setlength{\tabcolsep}{2.0pt}
\begin{tabular}[t]{lrccc@{\hspace{0.8em}}ccc@{\hspace{0.8em}}ccc}
\toprule
\multicolumn{2}{l}{\setbox0\hbox{Single Particle Reco.}\rlap{\smash{\raisebox{-3.6pt}{\rule{\wd0}{\cmidrulewidth}}}}\box0} & \multicolumn{3}{c}{Classification (AUC)} & \multicolumn{3}{c}{Muon Kinematics ($\sigma_{68}$ $\downarrow$)} & \multicolumn{3}{c}{Electron Kinematics ($\sigma_{68}$ $\downarrow$)} \\
\cmidrule(lr){3-5} \cmidrule(lr){6-8} \cmidrule(lr){9-11}
Method & Param. & $\mu/e^\dagger$ & $\mu/\pi^\pm$ & $e/\pi^0$ & Vtx.\ (cm) & Dir.\ ($^\circ$) & Mom.\ (\%) & Vtx.\ (cm) & Dir.\ ($^\circ$) & Mom.\ (\%) \\
\midrule
\multicolumn{11}{l}{\textit{Classical}} \\
fiTQun & -- & \multicolumn{1}{c}{\cellcolor[HTML]{FDF4EC}17} & \multicolumn{1}{c}{\cellcolor[HTML]{FDF4EC}82.3} & \multicolumn{1}{c}{\cellcolor[HTML]{FDF4EC}86.6} & \multicolumn{1}{c}{\cellcolor[HTML]{F7D9BD}\textbf{19.0}} & \multicolumn{1}{c}{\cellcolor[HTML]{F7D9BD}\textbf{1.13}} & \multicolumn{1}{c}{\cellcolor[HTML]{F7D9BD}\textbf{1.9}} & \multicolumn{1}{c}{\cellcolor[HTML]{F7D9BD}\textbf{20.9}} & \multicolumn{1}{c}{\cellcolor[HTML]{F7D9BD}\textbf{1.79}} & \multicolumn{1}{c}{\cellcolor[HTML]{F7D9BD}\textbf{2.3}} \\
\midrule
\multicolumn{11}{l}{\textit{Self-supervised}} \\
Panda V2 (lin.) & \llap{\raisebox{0.12ex}{\tiny$<$}\kern0.5pt}0.01M & \cellcolor[HTML]{FAE6D3}0.14 & \cellcolor[HTML]{F7D9BD}\textbf{99.2} & \cellcolor[HTML]{FAE6D3}92.4 & \cellcolor[HTML]{FDF4EC}37.0 & \cellcolor[HTML]{FDF4EC}2.29 & \cellcolor[HTML]{FDF4EC}2.2 & \cellcolor[HTML]{FDF4EC}41.3 & \cellcolor[HTML]{FDF4EC}3.22 & \cellcolor[HTML]{FAE6D3}3.3 \\
Panda V2 (mlp) & 0.3M & \cellcolor[HTML]{F7D9BD}\textbf{0.03} & \cellcolor[HTML]{FAE6D3}99.0 & \cellcolor[HTML]{F7D9BD}\textbf{93.0} & \cellcolor[HTML]{FAE6D3}27.0 & \cellcolor[HTML]{FAE6D3}1.42 & \cellcolor[HTML]{FAE6D3}2.0 & \cellcolor[HTML]{FAE6D3}30.3 & \cellcolor[HTML]{FAE6D3}2.51 & \cellcolor[HTML]{FDF4EC}3.4 \\
\bottomrule
\multicolumn{11}{@{}l}{$^\dagger$$1-\mathrm{AUC}$, in units of $10^{-5}$ (lower is better).} \\
\end{tabular}
\caption{\label{tab:wc}\textbf{Single-particle reconstruction in a water Cherenkov detector.}}
\end{minipage}\hfill
\begin{minipage}[t]{0.347\linewidth}
\centering
\setlength{\tabcolsep}{2.8pt}
\begin{tabular}[t]{lrccccc}
\toprule
\multicolumn{2}{l}{\setbox0\hbox{Class.\ (max.\ f.o.m.)}\rlap{\smash{\raisebox{-3.6pt}{\rule{\wd0}{\cmidrulewidth}}}}\box0} & \multicolumn{5}{c}{PT$_\text{1M}$ + FT$_K$} \\
\cmidrule(lr){3-7}
Method / $K$ & Param. & 100 & 1K & 10K & 100K & 417K \\
\midrule
\multicolumn{7}{l}{\textit{Supervised / classical}} \\
Two-blob cut & -- & \multicolumn{5}{c}{3.21} \\
Sparse ResNet & 33.1M & \cellcolor[HTML]{F1F8EF}2.39 & \cellcolor[HTML]{F1F8EF}2.48 & \cellcolor[HTML]{DFEEDB}4.02 & \cellcolor[HTML]{DFEEDB}4.63 & \cellcolor[HTML]{DFEEDB}5.33 \\
\midrule
\multicolumn{7}{l}{\textit{Self-supervised}} \\
Panda V2 (lin.) & \llap{\raisebox{0.12ex}{\tiny$<$}\kern0.5pt}0.01M & \cellcolor[HTML]{CFE6C8}\textbf{4.37} & \cellcolor[HTML]{CFE6C8}\textbf{5.13} & \cellcolor[HTML]{CFE6C8}\textbf{4.91} & \cellcolor[HTML]{CFE6C8}\textbf{5.38} & \cellcolor[HTML]{CFE6C8}\textbf{5.41} \\
\bottomrule
\end{tabular}
\vspace{\baselineskip}
\caption{\label{tab:next}\textbf{Signal classification on NEXT.}}
\end{minipage}
\end{table*}
  
\section{Results}
\label{sec:results}

We pretrain one encoder for each detector on 1M unlabeled images for 20 epochs, and keep the checkpoint whose linear probe performs best on a subset of held-out validation data. To measure what the encoder has learned, we train on $K$ labeled images with a linear layer (lin.) or a small decoder (dec.) on top of the frozen encoder. We write PT$_\text{1M}$ when the encoder is pretrained on all 1M images and PT$_K$ when it is pretrained on $K$ images only; similarly, FT$_K$ refers to fine-tuning using $K$ events. All results are on held-out test sets. We provide qualitative examples and the full evaluation methods with additional results in the Supplementary Material.

\subsection{In-domain transfer}
\label{sec:indomain}

\textbf{\LArTPC.}
Panda V2 is a step change over Panda V1 in both downstream tasks and at every label count (Tables~\ref{tab:semseg_lar},~\ref{tab:instseg_lar}). Using the frozen PT$_{1M}$ encoder, a linear layer trained on 100 labeled images reaches 95.8 m$F_1$, which Panda V1 does not reach with 1M, making our features 10{,}000$\times$ more label-efficient; for \textbf{instance segmentation}, the frozen encoder gains 13.4 ARI over Panda V1 decoding at 100 labels. With all labels, a frozen encoder comes within 0.4 m$F_1$ of full supervision.

\textbf{\sPHENIX.}
We compare against FM4NPP and its reported baselines in Tables~\ref{tab:semseg_sph},~\ref{tab:instseg_sph}. For \textbf{semantic segmentation}, a decoder trained on 100 labeled images matches FM4NPP trained on 70K, at 82.1 against 81.8 m$F_1$. For \textbf{instance segmentation}, our frozen encoder with 1K labeled images reaches 96.0 ARI and exceeds FM4NPP with 70K (94.5), while at equal label counts the gap is as large as 43 ARI. Performance improves further with more labels, up to 85.3 m$F_1$ and 98.1 ARI at 70K.

\textbf{\WC.}
With 10K labeled images per class, a linear layer on frozen features outperforms fiTQun at particle identification by a wide margin (Table~\ref{tab:wc}): it separates muons from charged pions at 99.2 AUC vs. 82.3, and electrons from neutral pions at 92.4 vs. 86.6. Kinematic quantities are less accessible, with fiTQun surpassing Panda V2 across label counts.

\textbf{\NEXT.}
At every label count, a linear layer on frozen features is better than both the classical and neural network baselines (Table~\ref{tab:next}), and with 1K labeled images it comparable to the Sparse ResNet trained on all 417K. Because this dataset is new and the baseline were developed on different data, we regard these results as indicative rather than authoritative.

\subsection{Out-of-domain transfer}
\label{sec:ood}

We test out-of-domain transfer by applying each pretrained encoder to the downstream tasks of the other three detectors, training only a task head on 1,000 labeled images while the encoder stays frozen. To enable transfer fairly, we leave (1) the input linear feature projection and (2) all batch normalized~\cite{ioffe2015batchnormalizationacceleratingdeep} layers in Panda Encoder trainable, so as to allow the backbone to adapt to both the different statistics of the inputs and batch.
We find that the amount of transfer varies by detector: on the \LArTPC, \sPHENIX, and \textcolor{wc}{WC} detectors, an out-of-domain encoder reaches 83--98\% of the in-domain score, whereas on \NEXT it reaches less than half (Table~\ref{tab:transfer}).

\subsection{Geometry of the Learned Representation}
\label{sec:interp}

\begin{figure*}[t]
    \centering
    \includegraphics[width=0.9\textwidth]{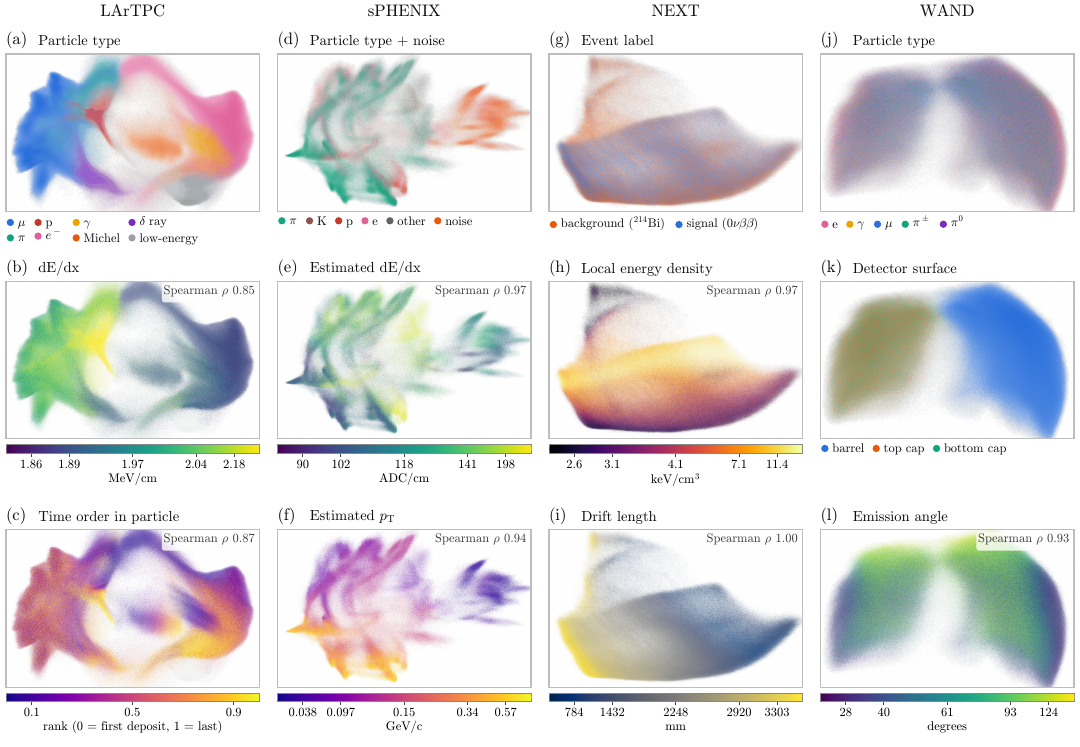}
    \vspace{-4pt}
    \caption{\textbf{Manifold maps of the learned representation.} 
    We provide one map per detector (columns). Points are colored by simulation truth or estimated quantities not provided in the dataset: particle type or event label (a,d,g,j); deposited energy per unit length $dE/dx$ (b,e; estimated in \sPHENIX), time ordering within individual particle's lifetime (c); estimated transverse momentum $p_\mathrm{T}$ (f), local energy density (h), drift length (i), and, in \WC, detector surface (k) and photon emission angle (l). For each continuous quantity, we report the Spearman correlation $\rho$ between the truth and the prediction of one linear probe on the full point features of 1,000 images, evaluated on held-out images.
    }
    \label{fig:atlas}
\end{figure*}

\textbf{Examining the low dimensional manifold of learned representations.} Prior work has shown that ``concepts" found in foundation models often correspond to internal geometries or \textit{sub-manifolds} of the latent space of model outputs, rather than simply directions~\cite{sun2026dense, fel2026rabbithulltaskrelevantconcepts}. Linear probing methods ignore this fact, and thus qualitative visualizations like PCA ignore important learned structure. We find this to be true with our pre-trained models as well.\footnote{Despite residing in a 1170-dimensional space and requiring 85-254 PCA directions to describe 90\% of each model's learned latent space, each model's TwoNN~\cite{Facco_2017} intrinsic dimensions vary between 16.2--16.8.}. Inspired by recent work attempting to find and visualize the internal geometry of FMs~\cite{fel2026rabbithulltaskrelevantconcepts}, we construct the underlying manifold by (i) splitting a set of encoded features from 1{,}000 events into 4{,}096 small neighborhoods via k-means, (ii) approximating each as a 24-dimensional linear subspace via probabilistic PCA (pPCA)~\cite{Tipping1999-hy}, and (iii) arranging the structure of the neighborhoods~\cite{Brands_NIPS2002_8929c70f} by measuring the co-activation statistics\footnote{We associate each point with the two clusters whose pPCA models assign it the highest likelihood.} of individual points within each subspace and embedding them in 2D with a mass-weighted~\cite{LafonLee2006} variant of PHATE~\cite{Moon2019-lf}, in which each neighborhood is weighted by the number of points it contains. We perform this process for each detector, yielding the ``manifold maps" visualized in Fig.~\ref{fig:atlas}; additional details on this method and further evaluations of representations can be found in the Supplementary Material.

Fig.~\ref{fig:atlas} shows that physics and detector geometry are the main organizing axes of the representation. In addition to organizing by particle type with varying amounts of success, each model organizes by dense properties it was never given during pretraining, such as particle energy deposition rate ($dE/dx$) and ordering within individual particle clusters over time (\ie, thereby representing a notion of causality). For \textcolor{wc}{WC}, which contain static 3D point clouds, we find that representations are chiefly organized by where each point is in the detector geometry.

We additionally test the linear decodability of these properties by training linear probes to predict each continuous property based on single per-point features, providing Spearman correlation values computed on held-out events in Fig.~\ref{fig:atlas}. 

\section{Conclusion}
\label{sec:conclusion}

We have shown that a single encoder architecture, self-supervised pretraining recipe, and fine-tuning toolkit can be applied to create four sensor-level FMs that successfully transfer to downstream tasks using a fully frozen backbone. Despite this success, there are limitations to our approach: for example, (i) pre-training on scenes with static geometries (\ie, WC) appears to lag behind in performance compared to those with arbitrary geometries, and (ii) we did not measure the scalability of this recipe with respect to model size.

We hope this work inspires others in the field of HENP to work on cross-detector applications and improving these methods; we expand on additional future directions in the Supplementary Material.



\section*{Acknowledgments}
We thank the authors of TPCpp-10M and the sPHENIX collaboration for open sourcing their dataset and providing a common baseline to compare our method to, and the NEXT collaboration for open sourcing the NEXT dataset.
We also thank Junjie Xia, Taritree Wongjirad, and Vinicius Da Silva for helpful discussions.
This work is supported by the U.S. Department of Energy, Office of Science, and Office of High
Energy Physics under Contract No. DE-AC02-76SF00515 as a part of the Genesis Mission National Science and Technology Challenges and the U.S.-Japan Science and Technology Cooperation Program under the award DE-SCL0000225.

\section*{LLM Usage Statement}

The authors have utilized generative AI tools (Anthropic's Claude Opus 5.5 and OpenAI's GPT Sol 6.1) to aid in model code implementation, evaluation, and orchestration of running experiments across several computing clusters; refining, correcting, and formatting the manuscript to improve the clarity of language; and refining or formatting of data reported. Responsibility for the manuscript text, including scientific ideas communicated, figures, and the presentation and interpretations of results lie entirely with the authors.

\FloatBarrier 
{\small
  \bibliographystyle{ieeenat_fullname}
  \bibliography{main}
}
\clearpage
\setcounter{page}{1}
\maketitlesupplementary
\pdfbookmark[0]{Supplementary Material}{supplement-start}
\addtocontents{toc}{\protect\setcounter{tocdepth}{2}} 
{\small\renewcommand{\numberline}[1]{#1\hskip 1em}
\tableofcontents}
\section{Extended discussion and future work}

In this work, we have taken a step toward a general pretraining and fine-tuning recipe for sensor-level foundation models in HENP. We showed that encoders pretrained with this recipe transfer to a wide range of tasks with few labels while staying frozen, and that an encoder pretrained on one detector can even be used on another, reaching 83--98\% of the in-domain encoder's score on three of the four detectors, though never matching it (\S\ref{sec:ood}). Despite these advances, a number of important questions remain.

\paragraph{Scaling studies for sensor-level SSL.} Self-distillation yields representations that can be used for tasks at the level of sensors, objects, and whole images with very few labels. However, this study tests a single recipe with a fixed model size, varying only the amount of pretraining data. Neural scaling laws in language and vision show that the best performance comes from increasing model and data size together~\cite{kaplan2020scaling,hoffmann2022training}, and similar trends have begun to appear in HENP, both for supervised jet tagging~\cite{vigl2026scalinglaws} and for self-supervised pretraining on TPC data~\cite{park2025fm4npp}. Joint-embedding objectives like ours (\eg, SimCLR~\cite{chen2020simple}, Barlow Twins~\cite{zbontar2021barlowtwinsselfsupervisedlearning}, VICReg~\cite{bardes2021vicreg}, and DINOv2~\cite{oquab2024dinov2learningrobustvisual}), however, are difficult to scale predictably, since their training loss says little about the quality of the learned representation; LeJEPA~\cite{balestriero2025lejepaprovablescalableselfsupervised} and TESSERA v2~\cite{feng2026tesserav2scalingpixelwise} have recently begun to address this. We hope that future work shows whether, and how, sensor-level pretraining scales.

\paragraph{Multi-modal pretraining.} Detectors in HENP more often than not include more than one sensor technology; for example, many LArTPCs contain photomultiplier tubes that collect the scintillation light generated as particles move through the detector. Though recent work has begun to pretrain on several subdetectors at once~\cite{AlonsoMonsalve2026}, and astronomy has made progress on aligning observations of different modalities~\cite{lanusse2023astroclip,parker2025aion}, the optimal way to perform self-supervised learning on multiple sensor modalities has yet to be fully explored.

\paragraph{Quantifying cross-domain transfer.} We have looked at transferring model weights pretrained on one domain and fine-tuned on another, with varying amounts of fine-tuning, and have shown that pretrained models can indeed transfer (\S\ref{sec:ood}), a result common to a number of recent cross-domain transfer studies~\cite{krzmanc2026cross,elsharkawy2026omnimol,Mikuni_2026,wu2026lightweightfoundationmodelcollider}. But what exactly is being transferred? Is it an understanding of how particles interact with matter, or simply the inductive biases of a trained network, \eg, looking at both local and global context to generate predictions? A rigorous study quantifying what transfers, and what doesn't, is needed.

\paragraph{Application to real data.} There is strikingly little literature on applying SSL to real HENP data, in part because collaborations keep their data closed. We expect this to change as these methods show more promise and as collaborations become more open.

\paragraph{Utilizing high-fidelity physics simulators.} The fidelity of simulation, and the trust placed in it, varies widely across HENP experiments. At the LHC, for example, detector simulation~\cite{geant4} is trusted enough to allow SAM-style~\cite{kirillov2023segment} pretraining on vast amounts of labeled simulation, in the hope that what the model learns transfers easily to other tasks; we note that this is most often done at the object level~\cite{mikuni2024solving,mokhtar2026mlpffm,hsu2026evenet}, and not at the sensor level. In neutrino physics, on the other hand, simulation is harder to trust and expensive to produce, which places the emphasis on SSL. Differentiable simulators~\cite{Gasiorowski_2024,lucid, krupa2026exploringboundariesdifferentiableradiation} have also become an important part of HENP: with them, simulation parameters, if modeled correctly, can be calibrated to data by gradient descent in a fully data-driven way. How such simulators can be paired with foundation models, during pre-training and/or during inference, to improve the scientific reach of HENP experiments is still unresolved.

\paragraph{Fusing symbolic physics knowledge with sensory encoders.} Recent studies have used LLMs to perform physics analyses~\cite{eckner2025automating,badea2026agents}, read rasterized images of sensor-level data~\cite{sagar2025finetuningvisionlanguagemodelsneutrino}, and act as a knowledge base for individual experiments~\cite{chatlas2026}. However, passing the representations of a sensor-level encoder directly into an LLM's latent space, as vision-language models do with image encoders~\cite{liu2023llava}, has yet to be demonstrated.

\paragraph{}We hope that this work, with the open sourcing of the model training and inference code, weights, and dataset promote further research from the community in these areas. 

\section{Dataset description}
\label{app:data}

\begin{figure*}[t]
    \centering
    \includegraphics[width=\linewidth]{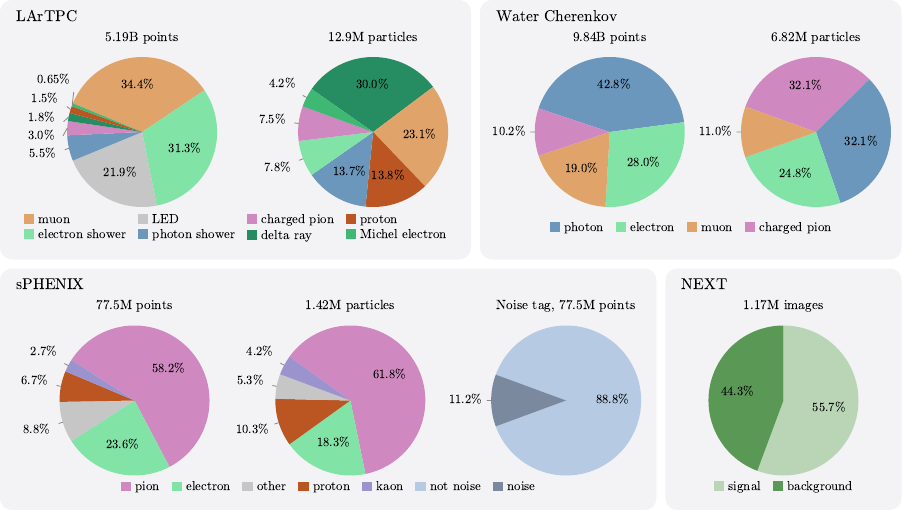}
    \caption{\textbf{Class composition of each dataset}: the share of the points, particles or images that each class holds, over all labeled images of the dataset (for the WC detector, the particle-bomb training sample). For the \LArTPC, a class is a pair of a motif and a particle type, and points are counted as stored, before the 0.13\,MeV threshold; for the WC detector, a point is a hit PMT of a particle, and particles that light fewer than three PMTs are not counted.}
    \label{fig:app-classes}
\end{figure*}

This section extends \S\ref{sec:data}. For each detector, we provide a description of each detector's detection principles, a description of the dataset, and the metrics performed for validation.
In Table~\ref{tab:app-splits} we provide the splits of the four datasets and in Table~\ref{tab:app-eval} the evaluation protocol of each downstream task; the detection principle of each detector and an example image are shown in Figs.~\ref{fig:app-lartpc}--\ref{fig:app-next}, and the class composition of each task in Fig.~\ref{fig:app-classes}. For every detector, the label sets we use are nested, \ie, the $K$ labeled images are the first $K$ images of one fixed order of the labeled training images, and the out-of-domain runs of \S\ref{sec:ood} use the same label sets, validation images and test protocol as the in-domain runs at $K=1{,}000$.

\begin{table*}[t] 
\centering
\footnotesize
\setlength{\tabcolsep}{6pt}
\begin{tabular}{@{}l r r r r r r r@{}}
\toprule
 & \multicolumn{3}{c}{Pretraining} & \multicolumn{2}{c}{Labeled} & & \\
\cmidrule(lr){2-4}\cmidrule(lr){5-6}
Detector & total & post-cut$^\dagger$ & used$^\ddagger$ & total & used$^\ddagger$ & Validation & Test \\
\midrule
\LArTPC & 1{,}082{,}400 & 1{,}045{,}215 & 1{,}000{,}000 & 1{,}082{,}400 & 1{,}000{,}000 & 66{,}800 & 50{,}000 \\
\sPHENIX & 10{,}000{,}000 & 3{,}075{,}482 & 1{,}000{,}000 & 70{,}000 & 70{,}000 & 13{,}000 & 7{,}000 \\
\textcolor{wc}{WC} & 981{,}599 & 961{,}202 & 961{,}202 & 981{,}599 & 981{,}599 & 32{,}960 & 374{,}927 \\
\NEXT & 1{,}155{,}489 & 1{,}151{,}389 & 1{,}000{,}000 & 1{,}000{,}000 & 416{,}993\rlap{$^{*}$} & 10{,}000 & 151{,}389 \\
\bottomrule
\end{tabular}
\caption{\textbf{Splits}, in images. $^\dagger$The images with at least 1{,}024 points for \LArTPC, \sPHENIX, and \textcolor{wc}{WC} and at least 512 for \NEXT. $^\ddagger$The images of the largest pretraining pool and of the largest label set. $^{*}$The labeled images that pass the event selection of \S\ref{app:data_next}.}
\label{tab:app-splits}
\end{table*}

\begin{table*}[t]
\centering
\footnotesize
\setlength{\tabcolsep}{2.5pt}
\begin{tabular}{@{}l l r l l l@{}}
\toprule
Detector & Task & Test & Scored on & Reported metric & Compared methods \\
\midrule
\LArTPC & Sem.\ seg. & 50{,}000 & points, pooled & macro $F_1$ (5 motifs) & Panda V1$^{a}$, PTv3$^{a}$, PoLAr-MAE \\
 & Inst.\ seg. & 50{,}000 & points, per image & ARI & Panda V1$^{a}$ \\
\sPHENIX & Sem.\ seg. & 6{,}943 & points, pooled & $F_1$ (\S\ref{sec:data}) & FM4NPP, its baselines \\
 & Inst.\ seg. & 6{,}943 & points, per image & ARI & FM4NPP, its baselines \\
\textcolor{wc}{WC} & Classification & 16{,}464 & images & pairwise AUC & fiTQun \\
 & Kinematics & 8{,}807 & images & $\sigma_{68}$ of the error & fiTQun \\
 & Inst.\ seg. & 72{,}255 & rings, pooled & particle $F_1$ & fiTQun \\
\NEXT & Classification & 63{,}190 & images & $\varepsilon_s/\sqrt{\varepsilon_b}$ & two-blob cut, sparse ResNet \\
\bottomrule
\end{tabular}
\caption{\textbf{Evaluation protocol.} Test column: the number of test images. $^{a}$Panda V1's own code, checkpoints and postprocessing, run on our test images. Metrics are defined in the Evaluation paragraph of each detector.}
\label{tab:app-eval}
\end{table*}

\subsection{\LArTPC}
\label{app:data_lar}

\begin{figure}[t]
    \centering
    \includegraphics[width=\linewidth]{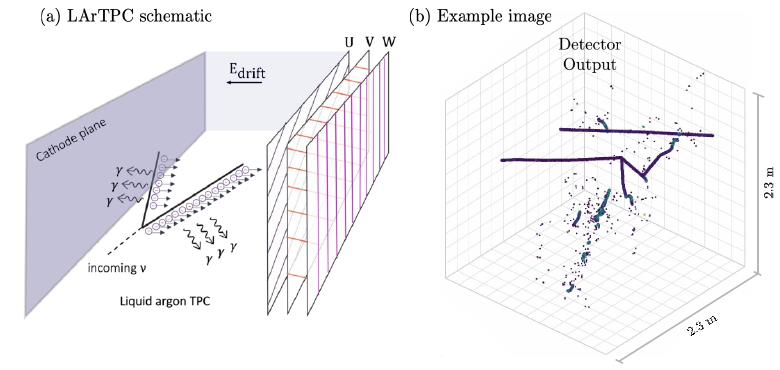}
    \caption{\textbf{LArTPC detection principle and example image.} (a) A charged particle ionizes the argon along its path, and the electric field $E_\text{drift}$ carries the freed electrons to the readout planes, which record where and when the charge arrives. (b) The resulting 3D image, in which every point is one recorded energy deposition. (a) is adapted from~\cite{Abud_2025} and (b) from~\cite{panda}, CC-BY 4.0.}
    \label{fig:app-lartpc}
\end{figure}


\paragraph{Detection mechanism.} A LArTPC is a volume of liquid argon between a cathode and an anode, across which a uniform electric field is applied, with readout planes at the anode (Fig.~\ref{fig:app-lartpc}). A charged particle that crosses the argon ionizes the argon along its path, and the freed electrons drift toward the anode at a constant velocity. The position at which the charge arrives on the readout planes gives two coordinates of every energy deposition, and the time that the charge took to arrive, measured from the time of the interaction, gives the third. In the resulting 3D image, muons, pions and protons appear as tracks, whereas electrons and photons appear as branching showers. The charge collected along a trajectory also gives the energy that the particle lost per unit length ($dE/dx$), which is used to tell apart particles whose trajectories look alike.

\paragraph{Dataset.} We use PILArNet-M~\cite{polarmae}, a dataset of 1.2M simulated images of a $(2.3\,\text{m})^3$ LArTPC volume, voxelized at a 3\,mm pitch ($768^3$ voxels). Primary particles\footnote{A \emph{primary} particle is generated by the source of an event, \eg a neutrino interaction or a cosmic ray, whereas \emph{secondary} particles are the descendants of primaries, which the simulation produces as it propagates the primaries through the detector.} are generated with the community-standard multi-particle vertex (MPV) and multi-particle rain (MPR) generators and are propagated through the liquid argon. MPV produces several particles from a common vertex, as a neutrino interaction would, and MPR produces particles whose starting points are distributed uniformly over the volume. Neither follows a neutrino event generator: the physics of neutrino--argon interactions is poorly constrained, so primaries drawn from such a generator would be biased with respect to reality and would make sim-to-real transfer harder. Neutrino experiments therefore cover as much of the phase space as possible with uniform assortments of primaries of random energies. The response of the detector, including charge diffusion and attenuation, is not simulated, which makes PILArNet-M roughly equivalent to the deconvolved response of a pixel-based LArTPC, \ie the 3D charge image that remains once the response of the readout has been removed. Each occupied voxel stores the deposited energy and four labels: the particle that deposited the energy, the \emph{interaction} that this particle belongs to (the group of particles that share a vertex), the type of the particle (photon, electron, muon, pion or proton), and a geometric motif.
The five motifs are tracks, showers, Michel electrons (the electron into which a stopped muon decays), delta rays (electrons knocked out of the argon by a passing particle), and low energy deposits (LED), amorphous charge depositions that are treated as background in instance segmentation. Particles of different motifs differ in size by orders of magnitude, even within one particle type: electrons alone form delta rays of about ten points, Michel electrons of several tens of points and showers of hundreds to thousands of points, so delta rays make up 30\% of all particles but only 2\% of all points (Fig.~\ref{fig:app-classes}).

\paragraph{Splits.} We pretrain the PT$_\text{1M}$ encoder on the first 1M images of a fixed random order of the training images with at least 1{,}024 stored points, and each PT$_K$ encoder on the first $K$ images of the same order (Table~\ref{tab:app-splits}). We apply no cut on the number of points during fine-tuning and evaluation. As in previous work~\cite{polarmae,panda}, we keep only points with more than 0.13\,MeV of deposited energy.

\paragraph{Evaluation.} \textit{Semantic segmentation.} We pool all test points into a single confusion matrix and score each motif by its $F_1$ score, the harmonic mean of its precision $P=\textrm{TP}/(\textrm{TP}+\textrm{FP})$, the share of the points predicted as the motif that truly belong to it, and its recall $R=\textrm{TP}/(\textrm{TP}+\textrm{FN})$, the share of the motif's points that the model finds, where TP, FP and FN count true positives, false positives and false negatives. Because precision can be traded for recall and vice versa, $F_1$ is high only when both are. We report the per-class $F_1$ and the \textbf{macro $F_1$}, the mean of the per-class $F_1$ scores over the five motifs.

\textit{Instance segmentation.} The adjusted Rand index (ARI)~\cite{Hubert1985} measures the agreement between the predicted and the true assignment of an image's points to particles, corrected for chance, such that a perfect prediction scores 1 and a random prediction scores 0 on average. We compute the ARI of an image over all of the image's points, with all unassigned and background points forming one predicted cluster and all LED points forming one true cluster, and report the mean over test images. This ARI departs in two respects from the ARI with which Panda V1 and PTv3 were originally reported~\cite{panda}. First, Panda V1 computed the ARI of an image over its non-LED points only, factoring LED segmentation out of the score. Second, Panda V1 and PTv3 were scored on 10{,}000 test images with at least 1{,}024 points, whereas we score every method on all 50{,}000 test images. We therefore re-score Panda V1 and PTv3 under our definition, and their numbers differ from the numbers published in~\cite{panda}.

\subsection{\sPHENIX}
\label{app:data_sph}

\begin{figure}[t]
    \centering
    \includegraphics[width=\linewidth]{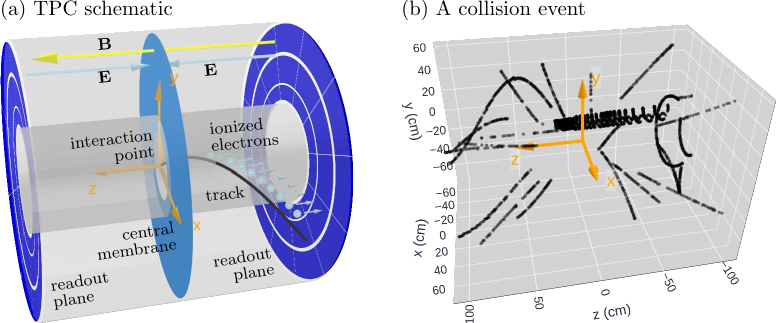}
    \caption{\textbf{sPHENIX detection principle and example image.} (a) A charged particle that leaves the interaction point ionizes the gas along its track, the electric field $\mathbf{E}$ carries the freed electrons to the readout planes at the two ends of the cylinder, and the magnetic field $\mathbf{B}$ bends the track. (b) A proton--proton collision at $\sqrt{s}=200$~GeV, in which every point is one reconstructed cluster of charge. Reproduced from~\cite{tpcpp10m}, CC-BY 4.0.}
    \label{fig:app-sphenix}
\end{figure}

\paragraph{Detection mechanism.} The sPHENIX TPC~\cite{klest2022compacttpc} is a cylindrical, gas-filled TPC whose principal axis lies along a proton-proton beam. It records energy deposits in an annulus around the beamline, between radii of 32 and 78\, cm, and has a length of 211\,cm (Fig.~\ref{fig:app-sphenix}). It is placed in a 1.4\,T magnetic field parallel to the beam. When a charged particle is produced in a proton-proton collision, it ionizes the gas along its path, and the freed electrons drift along the cylinder axis to readout planes at the two ends, where the position in the plane transverse to the beams is determined by the readout electrode (pad) that collects the charge and the position along the axis by the arrival time of the charge. The sPHENIX experiment's reconstruction software is then used to cluster these signals into points (spacepoints in the terminology of sPHENIX), with a resolution of about 150\,\textmu m in the azimuthal direction and 750\,\textmu m along the axis. The magnetic field bends each charged particle onto a helix, which the TPC records as a few tens of points and from whose curvature the momentum of the particle transverse to the beams (so-called $p_\textrm{T}$) is measured.

\paragraph{Dataset.} We utilize the TPCpp-10M~\cite{tpcpp10mZenodo} dataset, which includes proton-proton collisions at a center-of-mass energy of $\sqrt{s}=200$ GeV. The dataset consists of 10M unlabeled images and 90{,}000 labeled images, and each labeled point carries the identity of the particle that produced the point (pion, electron, proton, kaon, and ``other")\footnote{Charge-conjugates of particles, \eg $\pi^+$ and $\pi^-$, are folded into a single class definition (\eg $\pi^\pm$).} and a noise tag, which marks the points of particles with a transverse momentum below 60\,MeV/$c$.

\paragraph{Splits.} We pretrain on unlabeled images only, with the pools chosen as for the \LArTPC (Table~\ref{tab:app-splits}). We fine-tune on the 70{,}000 labeled training images and evaluate on the validation and test partitions of the release.

\paragraph{Evaluation.} We follow the evaluation protocol of FM4NPP~\cite{park2025fm4npp} as closely as possible. Predictions are scored on point predictions and points tagged as noise are kept in every metric. As FM4NPP does, we score only the test images with at most 3{,}200 points, which keeps 6{,}943 of the 7{,}000 test images (Tables~\ref{tab:app-splits} and~\ref{tab:app-eval}).

\textit{Semantic segmentation.} For particle identification, all test points are pooled, and we compute the macro precision and the macro recall as the means of the per-class precision and recall over the five particle types. Previous work reports these two numbers, and not macro $F_1$. Therefore, we use the less common definition of $F_1$~\cite{opitz2021macrof1macrof1}: the harmonic mean between macro precision and recall. Note that this $F_1$ differs from the macro $F_1$ for \LArTPC. Beside these macro averages, we report the accuracy, \ie the fraction of points assigned their true class, and the $F_1$ of each class.

In this Supplementary Material, we also report scores on an additional noise tagging task, for which we report accuracy and $F_1$.

\textit{Instance segmentation.} The ARI is defined as for the \LArTPC (\S\ref{app:data_lar}): the ARI of an image is computed over all of the image's points, with unassigned points forming one predicted cluster, and the reported ARI is the mean over test images.\footnote{\citet{park2025fm4npp} reports using pooled events to construct ARI. However, the official code repository explains that they use this per-image mean. Thus, we use this per-image mean as a basis for comparison instead of the pooled ARI.} We also report the track finding metrics of~\cite{park2025fm4npp}, pooled over the test images. A predicted and a true track are matched if their shared points make up more than half of each track; additionally, track efficiency and purity defined as the fractions of true and of predicted tracks that are matched. Spacepoint efficiency is defined the fraction of the points of true tracks that lie in the predicted track holding most of each true track's points, such that points assigned to no predicted track count against the spacepoint efficiency, and spacepoint purity is the fraction of the points of predicted tracks that belong to the true track holding most of each predicted track's points.

\subsection{\WC}
\label{app:data_wc}

\begin{figure}[t]
    \centering
    \includegraphics[width=\linewidth]{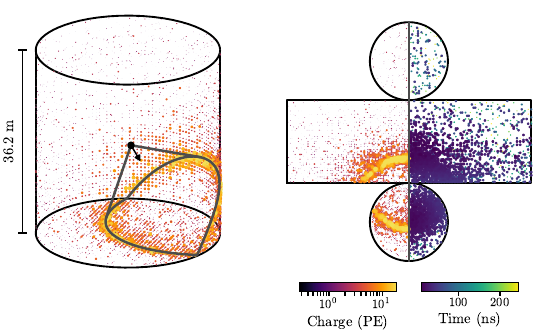}
    \caption{\textbf{Water Cherenkov detection principle and example image.} \textbf{Left:} A particle (black dot and arrow) emits a cone of Cherenkov light, and the PMTs on the detector walls that this light reaches record a ring; every point is one hit PMT. \textbf{Right:} The same single-particle event with the detector walls unrolled, colored by the charge (in photoelectrons, left half) and by the arrival time (in nanoseconds, right half) of every hit PMT; larger markers indicate a higher charge or an earlier time.}
    \label{fig:app-wc}
\end{figure}

\begin{figure*}[t]
    \centering
    \includegraphics[width=0.78\textwidth]{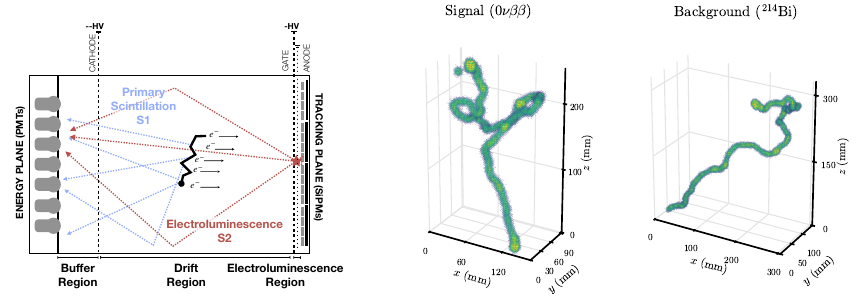}
    \caption{\textbf{NEXT detection principle and example images.} \textbf{Left:} A fast electron ionizes the xenon along its path, and the freed electrons drift to the anode, where they are converted into light (electroluminescence, S2) that the tracking plane images; the prompt scintillation light (S1) marks the start of the drift. Reproduced from~\cite{NEXT2026}, CC-BY 4.0. \textbf{Right:} A signal ($0\nu\beta\beta$) image, with a dense energy deposition (blob) at both ends of the track, and a background ($^{214}$Bi) image, with a blob at one end.}
    \label{fig:app-next}
\end{figure*}

\paragraph{Detection mechanism.} A water Cherenkov detector consists of a large tank of water whose walls are lined with photodetectors (photomultiplier tubes, or PMTs) (Fig.~\ref{fig:app-wc}). A charged particle that travels through the water faster than the speed that light travels through water emits Cherenkov light in a cone around its direction of travel, and the PMTs that this light reaches record the amount of light (charge) and the arrival time of the light, such that every particle above the Cherenkov threshold appears as a ring of hit PMTs on the detector walls. In the resulting image, a muon appears as a sharp ring and an electron, which showers and scatters repeatedly, as a diffuse ring, whereas a charged pion that undergoes a hadronic interaction and a neutral pion, which decays into two photons, each produce several rings; a muon that stops in the water often decays into an electron, whose light arrives later than the light of the muon. The momentum, the starting point and the direction of a particle determine the thickness and the shape of the particle's ring, and a reconstruction algorithm infers the type and these three quantities of every particle from the charges and times of the hit PMTs.

\paragraph{Dataset.} WAND (\textbf{W}ater-cherenkov \textbf{A}nnotated \textbf{N}eutrino \textbf{D}ataset) is a newly released dataset which contains simulated images of a Super-Kamiokande~\cite{fukuda2003superkamiokande}-like cylindrical detector of radius 16.96\,m and height 36.38\,m with 11{,}096 20-inch PMTs on its walls. Each image starts from one or more primary particles that share a single vertex, which is drawn uniformly within 90\% of the detector radius and half-height, and each primary particle has an isotropic direction and a kinetic energy drawn uniformly in 1--2000\,MeV for electron/positrons $e^\pm$, 1--2130\,MeV for photons $\gamma$ (up to the highest photon energy from a 2\,GeV neutral pion $\pi^0$) and 200--2000\,MeV for muons/antimuons $\mu^\pm$, charged pions $\pi^\pm$ and neutral pions $\pi^0$, where the lower limits lie somewhat above the Cherenkov thresholds such that every primary particle produces light. The primary particles and their secondary particles are propagated with 
a Geant4~\cite{geant4} application that also generates the Cherenkov photons, and the photons are transported through the detector with LUCiD~\cite{lucid}
under Super-Kamiokande's parametrization of photon absorption, scattering and reflection~\cite{Abe:2013gga}. The response of the electronics is emulated as in the \texttt{SKI} implementation of WCSim, the WC community's mature detector simulation software.\footnote{\url{https://github.com/WCSim/WCSim}} Specifically, the photons that arrive at a PMT are converted into hits, each with a jittered\footnote{Also called smearing in HENP.} charge and time of first detection; a photon that arrives more than 200\,ns after the start of a hit creates a new hit in the same PMT, a hit-multiplicity trigger then groups the hits into readout windows, and dark noise is added to every window at a rate of 4.2\,kHz. Each point in an image corresponds to one hit. Apart from dark noise, more than 99\% of the light beyond the first readout window comes from decay electrons: on average, a particle-bomb image has about 33k photoelectrons in the first window, whereas a decay electron contributes about 0.4k photoelectrons and dark noise about 35 photoelectrons to a later window.

Labels are stored for the primary particles and for three kinds of secondary particles: decay electrons above 1\,MeV, the two photons of each $\pi^0$, and charged pions above 195\,MeV/$c$ that start a new trajectory, \ie pions that are produced in a hadronic interaction or deflected by more than $5^\circ$. Every other simulated particle, such as a $\delta$-ray or a shower particle, is attributed to its nearest labeled ancestor, such that the charge of every hit is divided among the labeled particles. For each labeled particle, WAND stores these charge fractions, which are the segmentation targets, and the type, momentum, direction, starting point and containment (whether the particle stops inside the detector), which are the classification and regression targets.

WAND consists of three principal groups of samples. The training sample contains $10^6$ \emph{particle bomb} images, in which one to five primary particles are drawn from $\{e^\pm,\mu^\pm,\pi^\pm,\pi^0,\gamma\}$. The validation and test sets include a second particle-bomb sample of $5\times10^4$ images, as well as the following configurations: $e^\pm$, $\gamma$, $\mu^\pm$, $\pi^\pm$ and $\pi^0$ ($5\times10^4$ images each), muon pairs ($5\times10^4$), and the pairs $\mu^-\pi^+$, $e^-\pi^+$, $e^-\pi^0$ and $\mu^-\pi^0$ ($10^4$ each, with electrons $\geq100$\,MeV), which resemble charged-current neutrino interactions.

\paragraph{Splits.} We pretrain on the images of the particle bomb training sample in which more than 1{,}024 PMTs recorded light, and we fine-tune on all images; for the single particle tasks, the labeled images are the single particle images of this sample whose vertex lies at least 2\,m from every wall and whose particle stops inside the detector, and 8{,}000 further single particle images of this sample are held out for validation. The second particle-bomb sample is divided into a validation half and a test half, and the single-particle and two-particle samples are used for testing only (Table~\ref{tab:app-splits}).

\paragraph{Evaluation.} We score our models and fiTQun on identical images, and we apply the event selection under which fiTQun~\cite{fitqunSK,fitqunUB,fitqunT2K}, our baseline, is used in Super-Kamiokande analyses~\cite{fitqunSK}, such that fiTQun is only evaluated in the regime for which fiTQun was designed and tuned (\S\ref{app:fitqun}). We apply four requirements:

\begin{enumerate}
    \item The true vertex of an image lies at least 2\,m from every wall.
    \item The particles are \emph{contained}, \ie the primary particle and every particle that the primary particle produces stop inside the detector. For instance segmentation, a true ring whose particle is not contained is left out of the metric calculation, together with any predicted ring that is matched to this true ring. For single-ring events, the image is dropped entirely if the primary is uncontained.
    \item fiTQun fits only the first readout window of an image, such that a labeled particle counts as a true ring only if the particle lights at least three PMTs in the first readout window; a decay electron that lights only later windows is therefore a true ring for neither method.
    \item The cut constants of fiTQun's decision tree were tuned on the even-numbered files of the single-particle test samples of electrons, muons, charged pions and neutral pions. These four samples are therefore evaluated on their odd-numbered files only; the photon sample and the two-particle samples, which were not used for tuning, are used in full.
\end{enumerate}

The second particle-bomb sample is not used in the comparison with fiTQun, because the particle-bomb images on which fiTQun was run include images of our validation half.

\textit{Image Classification.} To separate two particle types, each method assigns a score to the images of these two types. We report area under the receiver operating characteristic curve (AUROC, or AUC) of this score, \ie the probability that a randomly chosen image of the first type receives a higher score than a randomly chosen image of the second type.

\textit{Kinematics.} The error of an image is the distance, in centimeters, between the predicted and the true starting point of the particle, the angle, in degrees, between the predicted and the true direction, or the absolute difference between the predicted and the true momentum as a fraction of the true momentum, \ie  $\Delta p/p :=|p_\text{pred}-p_\text{true}|/p_\text{true}$. We report the 68th percentile of this error $\sigma_{68}$ over the test images.

\textit{Instance segmentation.} Predicted and true rings are matched as defined in \S\ref{sec:data}, and a match counts as a true positive only if the predicted type equals the true type. With TP the number of true positives, $N_\text{pred}$ the number of predicted rings and $N_\text{true}$ the number of true rings, each summed over the test samples, the particle $F_1$ score is $2\,\mathrm{TP}/(N_\text{pred}+N_\text{true})$. With $M$ the number of matches regardless of type, the ring finding $F_1$ score is $2M/(N_\text{pred}+N_\text{true})$, and the ring classification accuracy is $\mathrm{TP}/M$, the fraction of matched rings whose predicted type equals the true type.

\subsection{\NEXT}
\label{app:data_next}

\paragraph{Detection mechanism.} NEXT searches for $0\nu\beta\beta$ decays of $^{136}$Xe, in which two electrons share the decay energy, against a background of single electrons of nearly the same energy, chiefly from the gamma rays of $^{214}$Bi decays (\S\ref{sec:data}). In the NEXT TPC (Fig.~\ref{fig:app-next}, left), an electron that travels through the xenon gas ionizes the xenon along its path. The freed electrons drift to the anode at one end of the chamber, where a region of high electric field converts them into a proportional amount of light (electroluminescence), which an array of silicon photomultipliers images. The position along the drift direction is given by the delay between this light and the prompt scintillation light that the xenon emits when the electron passes, and photomultiplier tubes measure the total energy of the event with a resolution better than 1\% full width at half maximum (FWHM) near the energy released by the decay, $Q_{\beta\beta}=2.458$\,MeV. The total energy therefore cannot separate signal from background; however, the topology of the track can. An electron loses energy most densely just before it stops, so the two electrons of a $0\nu\beta\beta$ decay leave a single track with a dense energy deposition (``blob'') at both ends, whereas the single electron of a $^{214}$Bi background event leaves a track with only one blob (Fig.~\ref{fig:app-next}, right).

\paragraph{Dataset.} We use the public NEXT dataset~\cite{nextaissZenodo}, which contains simulated events in a cylindrical xenon chamber 3.6\,m in diameter and 3.6\,m long at a pressure of 5\,bar: 648{,}793 $0\nu\beta\beta$ decays in the gas and 516{,}696 $^{214}$Bi decays in the surrounding copper shielding. Energy depositions are displaced according to the diffusion of the drifting electrons, which is set below the diffusion of pure xenon to mimic the experiment's conventional deconvolution algorithm~\cite{JHEP2021}, and are rebinned to 3\,mm voxels. Only events whose total energy, smeared by a 1\% FWHM resolution, falls in a 50\,keV window around $Q_{\beta\beta}$ are kept, such that the two classes have nearly the same total energy. Each point carries a position and a deposited energy, and the only label is the class of the image.

\paragraph{Splits.} We hold out 5{,}000 images of each class for validation and put the remaining images with at least 512 points in one fixed random order. The first 1M images in this order form the pool of the PT$_\text{1M}$ encoder, each PT$_K$ pool is the first $K$ of them, and the remaining 151{,}389 images, on which no encoder was pretrained, form the test set.

\paragraph{Evaluation.} \textit{Event selection.} Following the topological analyses of NEXT~\cite{ferrario2019nextwhite, JHEP2021}, we define classification on the images that pass a fiducial single-track selection: every point must lie at least 20\,mm inside the instrumented volume, and the points must form a single connected component when grouped into 15\,mm cubes. Without this selection, energy depositions away from the main track make the two classes nearly trivially separable, so we apply it to the labeled, validation and test images of every method. The selection acts on the inputs, not on the outputs of a model: an image that fails the selection is removed before any model is fine-tuned or evaluated, rather than classified and then discarded. We do not apply the selection to the pretraining pools. The labeled images are thus the selected images among the first 1M, the validation images are the 3{,}849 selected images of the holdout, and the test images are the selected images of the test set (Tables~\ref{tab:app-splits} and~\ref{tab:app-eval}). At $K\leq10$K, the labeled images are balanced between the two classes, and at larger $K$ they keep the class ratio of the selected images, about one background image for every ten signal images.

\textit{Classification.} A classifier assigns a score to every image, and we keep the images whose score exceeds a threshold, which accepts a fraction $\varepsilon_s$ of the signal images and a fraction $\varepsilon_b$ of the background images. We report the figure of merit (f.o.m.) $\varepsilon_s/\sqrt{\varepsilon_b}$ on the test images, at the threshold that maximizes the f.o.m. on the validation images.

\section{Additional analyses}
\label{app:analyses}

In this section, we report analyses that support the choices made in the main text or extend its results. Each subsection describes its setup and states whether it is evaluated on validation or test images.

\subsection{Pretraining ablations}
\label{app:pt_ablate}

In Table~\ref{tab:ablations}, we change one pretraining setting for \LArTPC and \sPHENIX at a time, starting from the final recipe at batch size 48 (\rung{9} of the main text), including the settings that differ between detectors (Table~\ref{tab:recipe}): the mask size, scale augmentation, the upcast level, and the minimum number of points per pretraining image.

\begin{table*}[t]
\centering
\scriptsize
\setlength{\tabcolsep}{3pt}
\captionsetup[subtable]{font=scriptsize, justification=centering}
\begin{subtable}[t]{0.24\linewidth}
\centering
\begin{tabular}[t]{ccc}
\toprule
points & \textcolor{sph}{sPHENIX} & \textcolor{lar}{LArTPC} \\
\midrule
256 & 83.7 & 96.2 \\
512 & 85.1 & \cellcolor{best}\textbf{96.6} \\
1024 & 86.0 & \cellcolor{best}\textbf{96.6} \\
2048 & \cellcolor{best}\textbf{86.4} & \cellcolor{best}\textbf{96.6} \\
\bottomrule
\end{tabular}
\caption{\textbf{Attention patch size.}
}
\end{subtable}
\hfill
\begin{subtable}[t]{0.24\linewidth}
\centering
\begin{tabular}[t]{ccc}
\toprule
voxels & \textcolor{sph}{sPHENIX} & \textcolor{lar}{LArTPC} \\
\midrule
10 & 85.6 & 96.6 \\
20 & \cellcolor{best}\textbf{86.2} & 96.7 \\
30 & \cellcolor{best}\textbf{86.2} & \cellcolor{best}\textbf{96.8} \\
40 & 85.9 & 96.6 \\
50 & 85.9 & 96.5 \\
\bottomrule
\end{tabular}
\caption{\textbf{Mask size.}
}
\end{subtable}
\hfill
\begin{subtable}[t]{0.24\linewidth}
\centering
\begin{tabular}[t]{ccc}
\toprule
count & \textcolor{sph}{sPHENIX} & \textcolor{lar}{LArTPC} \\
\midrule
2048 & \cellcolor{best}\textbf{86.2} & \cellcolor{best}\textbf{96.6} \\
4096 & 85.9 & 96.5 \\
8192 & 86.0 & \cellcolor{best}\textbf{96.6} \\
16384 & 85.9 & \cellcolor{best}\textbf{96.6} \\
\bottomrule
\end{tabular}
\caption{\textbf{Prototypes.}
}
\end{subtable}
\hfill
\begin{subtable}[t]{0.24\linewidth}
\centering
\begin{tabular}[t]{ccc}
\toprule
level (stride) & \textcolor{sph}{sPHENIX} & \textcolor{lar}{LArTPC} \\
\midrule
0 (8) & 83.1 & 96.1 \\
1 (2) & 84.0 & 96.4 \\
2 (2) & 84.0 & 96.3 \\
3 (1) & \cellcolor{best}\textbf{86.0} & \cellcolor{best}\textbf{96.6} \\
\bottomrule
\end{tabular}
\caption{\textbf{Upcast level.} 0=coarsest.
}
\end{subtable}
\par\medskip
\begin{subtable}[t]{0.24\linewidth}
\centering
\begin{tabular}[t]{ccc}
\toprule
physical? & \textcolor{sph}{sPHENIX} & \textcolor{lar}{LArTPC} \\
\midrule
\xmark & \cellcolor{best}\textbf{86.0} & \cellcolor{best}\textbf{96.6} \\
\cmark & 82.4 & 96.1 \\
\bottomrule
\end{tabular}
\caption{\textbf{Physical augmentations.}
}
\end{subtable}
\hfill
\begin{subtable}[t]{0.24\linewidth}
\centering
\begin{tabular}[t]{ccc}
\toprule
mask tokens & \textcolor{sph}{sPHENIX} & \textcolor{lar}{LArTPC} \\
\midrule
\xmark & \cellcolor{best}\textbf{86.0} & 96.6 \\
\cmark & 85.5 & \cellcolor{best}\textbf{96.7} \\
\bottomrule
\end{tabular}
\caption{\textbf{Masking.}
}
\end{subtable}
\hfill
\begin{subtable}[t]{0.24\linewidth}
\centering
\begin{tabular}[t]{ccc}
\toprule
scaling? & \textcolor{sph}{sPHENIX} & \textcolor{lar}{LArTPC} \\
\midrule
\cmark & \cellcolor{best}\textbf{86.0} & \cellcolor{best}\textbf{96.6} \\
\xmark & 85.5 & 96.5 \\
\bottomrule
\end{tabular}
\caption{\textbf{Scale augmentation.}}
\end{subtable}
\hfill
\begin{subtable}[t]{0.24\linewidth}
\centering
\begin{tabular}[t]{ccc}
\toprule
min.\ points & \textcolor{sph}{sPHENIX} & \textcolor{lar}{LArTPC} \\
\midrule
32 & 81.8 & 96.5 \\
1024 & \cellcolor{best}\textbf{86.0} & \cellcolor{best}\textbf{96.6} \\
\bottomrule
\end{tabular}
\caption{\textbf{Minimum points.}}
\end{subtable}
\caption{\textbf{Ablating Panda V2 pre-training.} We evaluate learned features using a linear probe for semantic segmentation on \LArTPC and \sPHENIX. We provide macro $F_1$ scores for PID (\textcolor{sph}{sPHENIX}) and motif (\textcolor{lar}{LArTPC}) as defined in (\S\ref{sec:data}). The minimum number of points applies to the images of the pretraining pool. Each column's best value is \textbf{highlighted}.}
\label{tab:ablations}
\end{table*}

\subsubsection{Detector-specific settings}

Aiming for a universal recipe for sensor-level detectors, we investigate the necessity of detector-specific configurations by modulating mask size, a minimum point cut, detector-specific physical augmentations, coordinate scaling, local and global crop sizing, and distillation loss granularity.

\paragraph{Mask sizes (Table~\ref{tab:ablations}b).} Linear probes are generally insensitive to mask sizes: between 10 and 50 voxels, sPHENIX m$F_1$ stays between 85.6 and 86.2, and LArTPC m$F_1$ between 96.5 and 96.8.

\paragraph{Unphysical augmentations (Table~\ref{tab:ablations}e).} Panda V2's augmentations involve rotating each view about all three axes, flipping along all three axes and randomly scaling the entire image; this means that for example, for sPHENIX, we rotate the entire beamline indiscriminately, enforcing consistency in the learned representations across these views.
These augmentations are necessary for the best results: keeping only the rotations and flips that respect the symmetries of the detector and dropping the scaling (``physical'' augmentations\footnote{Rotations about the $y$ axis and flips of $x$ and $z$ for LArTPC, rotations about the beam ($z$) axis and flips of $z$ for sPHENIX}) lowers the sPHENIX m$F_1$ by 3.6 and the LArTPC m$F_1$ by 0.5. Additionally, one may assume that by allowing for unphysical augmentations, information that should depend on absolute position or direction in the detector are lost; however, in \S\ref{app:invariance} we show this is not necessarily true due to the position-dependent features being encoded in the inputs via local geometric or charge patterns.

\paragraph{Minimum point cut.} A cut on the minimum number of points per pretraining image is necessary: lowering the minimum from 1{,}024 points to 32 lowers the sPHENIX m$F_1$ by 4.2 and the LArTPC m$F_1$ by 0.1. We therefore keep a minimum for every detector as a precaution, but lower it to 512 points for \NEXT, because a minimum of 1{,}024 points removed too many \NEXT images.

\paragraph{Coordinate scaling.}
In a WC image, the vertex is encoded almost entirely in the positions of the hit PMTs together with the times of the hits, and scaling the positions without the times would break the relation between hit positions and hit times. Thus, we do not perform random coordinate scaling in WC.

\paragraph{Local and global crop settings.} sPHENIX images are the smallest of the four detectors, and with the crop fractions of the other detectors (0.4--1.0 of the image for global views and 0.1--0.4 for local views), the crops were so small that the self-distillation task became too difficult and pretraining was unstable; we note that this is not unrelated to the minimum point cut above. We therefore raise the crop fractions for sPHENIX's global and local views to 0.85--1.0 and 0.3--0.6.

\paragraph{Distillation loss granularity.} We apply the loss at the finest level for every detector except WC, because pretraining on WC at the finest level suffered from a geometric collapse of the features that we could only partially resolve, and we apply the loss one level coarser instead (that is, we apply the loss after upcasting and concatenating features up to the penultimate stage, starting from the coarsest stage).

\subsection{Invariance of the learned features}
\label{app:invariance}

\begin{figure*}[t]
    \centering
    \includegraphics[width=\linewidth]{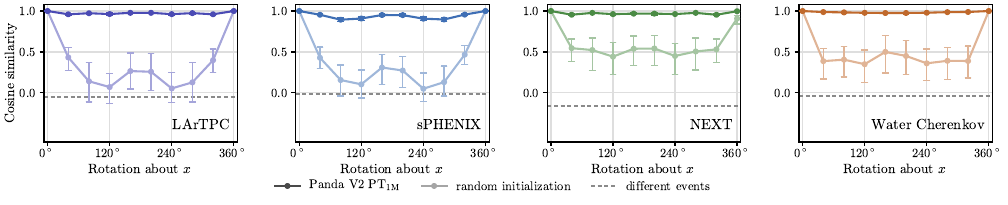}
    \caption{\textbf{Rotation invariance of the frozen features.} Cosine similarity between the mean feature of a validation image and that of the same image rotated about the $x$ axis, after subtracting the mean feature over 500 images; median and interquartile range. Dashed line: the median similarity between different images.}
    \label{fig:app-rotation}
\end{figure*}

During pretraining, each view is centered and rotated along all three axes by random angles. The self-distillation objective rewards the student for creating predictions which are invariant to these views; therefore, the encoder, which is never given absolute position of sensor data, is trained to learn weak invariances to both translations and rotations. Figure~\ref{fig:app-rotation} shows that this indeed is the case empirically: rotating a validation image about the $x$ axis, which is not an axis of symmetry of any of the four detectors, leaves its centered mean feature almost unchanged, with a median cosine similarity of 0.89 to 0.98 in the worst case. For reference, a randomly initialized encoder of the same architecture gives 0.05 to 0.44.

This invariance may cause discomfort. While at a high level particles do often look similar in difference parts of a detector, this invariance isn't strictly true. For example, while charge drifts in a TPC towards a collection plane, the cloud of charge diffuses, leaving particle trajectories that were created further along the drift axis to appear thicker than those which were created closer to it. And in WC, the light that a photomultiplier tube (PMT) records clearly depends on its distance and angle to the particle.


Invariance to the coordinates of an image, however, does not necessarily remove information that the image itself carries. One needs no coordinate system to see that a strongly diffused track was far from the collection plane, and nothing in the objective makes images at different drift distances look alike. We test this with NEXT, which includes a diffusion effect on drifted charge. On rotated and centered NEXT validation images, one linear map from the frozen per-hit features, averaged over the hits of an image, recovers its absolute drift position to 4.9\,cm RMSE (over a 3.6\,m distance), compared to 19.9\,cm for a randomly initialized encoder (which we note is already somewhat impressive) and 104\,cm for the same map on the input point cloud. As discussed in the main text, drift length also plays a large part in the organization of the manifold of NEXT features (Fig.~\ref{fig:atlas}i). 

But how might one predict a particle's vertex $\mathbf{p}$ or direction $\hat{\mathbf{d}}$ from frozen, approximately invariant features? Let an image have $N$ points with features $\mathbf{z}_i \in \mathbb{R}^D$ at positions $\mathbf{x}_i \in \mathbb{R}^3$. A linear probe on the mean feature $\bar{\mathbf{z}} = \frac{1}{N}\sum_i \mathbf{z}_i$ has no equivariant mechanism: its output does not change when the image is rotated by $R$ or translated by $\mathbf{t}$, whereas the direction becomes $R\hat{\mathbf{d}}$ and the vertex $R\mathbf{p} + \mathbf{t}$. We therefore need a probe that combines the approximately invariant features with the positions at which they occur. With the mean position $\bar{\mathbf{x}} = \frac{1}{N}\sum_i \mathbf{x}_i$, we define the centered coordinate--feature moment
\begin{equation}
M = \frac{1}{N}\sum_{i=1}^{N} (\mathbf{x}_i - \bar{\mathbf{x}})(\mathbf{z}_i - \bar{\mathbf{z}})^\top \in \mathbb{R}^{3\times D}.
\label{eq:h1}
\end{equation}
Each column of $M$ is a three-dimensional vector that describes how one feature channel varies across the image. To predict a vector, we combine the $D$ columns with learned weights $\mathbf{a} \in \mathbb{R}^D$,
\begin{equation}
M\mathbf{a} = \frac{1}{N}\sum_{i=1}^{N} (\mathbf{x}_i - \bar{\mathbf{x}})\,\mathbf{a}^\top(\mathbf{z}_i - \bar{\mathbf{z}}) \in \mathbb{R}^{3},
\label{eq:naive}
\end{equation}
where $s_i = \mathbf{a}^\top(\mathbf{z}_i - \bar{\mathbf{z}})$ is a scalar corresponding to the activation of one learned combination of features at point $i$; $M\mathbf{a}$ thus becomes an average of all positions in the image relative to $\bar{\mathbf{x}}$, weighted by $s_i$. If the features are invariant, then under $\mathbf{x}_i \mapsto R\mathbf{x}_i + \mathbf{t}$, for any rotation or reflection $R$ (pretraining also flips the views), we have $\bar{\mathbf{z}} \mapsto \bar{\mathbf{z}}$ and $M \mapsto RM$, so $M\mathbf{a}$ is E(3) equivariant; because the encoder's features are only approximately invariant (Fig.~\ref{fig:app-rotation}), $M\mathbf{a}$ is only approximately equivariant.\footnote{We note that some molecular models predict dipole moments from atomic partial charges in the same way~\cite{unke2019physnet,schutt2021painn}, and equivariant networks build vector features similarly~\cite{deng2021vectorneurons,satorras2021egnn}.}

However, because $M\mathbf{a}$ is a weighted sum of the centered positions, it depends on how the points of an image are spread. To correct for this, we compute $M\mathbf{a}$ from whitened positions $\Sigma^{-1/2}(\mathbf{x}_i - \bar{\mathbf{x}})$---which by definition are spread equally in every direction---and map the result back with $\Sigma^{-1/2}$, where $\Sigma = \frac{1}{N}\sum_i (\mathbf{x}_i - \bar{\mathbf{x}})(\mathbf{x}_i - \bar{\mathbf{x}})^\top$ is the covariance of the positions, \ie, apply $\Sigma^{-1}$ to $M\mathbf{a}$. $\Sigma$ is invertible unless all positions of an image lie in one plane, and we add $10^{-9}\operatorname{tr}(\Sigma)\,I$ to it for numerical stability.

For the vertex, we can ask what invariant quantity found within each per-point feature would let a probe recover it exactly. A natural candidate is the distance $d_i$ from point $i$ to the vertex: it does not change when the image is moved or rotated, and the detector response depends on it, as in WC, where the time at which a photomultiplier records light depends on its distance to the particle. If the activations encoded the squared distances, the vertex would be a linear function of them, and thus a linear probe would be able to read it out. Writing $\mathbf{u} = \mathbf{p} - \bar{\mathbf{x}}$ for the vertex and $\Delta\mathbf{x}_i = \mathbf{x}_i - \bar{\mathbf{x}}$ for point $i$, both relative to $\bar{\mathbf{x}}$, we have $d_i^2 = \lVert \Delta\mathbf{x}_i - \mathbf{u}\rVert^2 = \lVert \Delta\mathbf{x}_i\rVert^2 - 2\Delta\mathbf{x}_i^\top \mathbf{u} + \lVert \mathbf{u}\rVert^2$, which is linear in $\mathbf{u}$ apart from the constant $\lVert \mathbf{u}\rVert^2$. Multiplying both sides by $\Delta\mathbf{x}_i$ and averaging over the points removes that constant (because $\sum_i \Delta\mathbf{x}_i = 0$) and leaves $\frac{1}{N}\sum_i \Delta\mathbf{x}_i d_i^2 = \frac{1}{N}\sum_i \Delta\mathbf{x}_i\lVert \Delta\mathbf{x}_i\rVert^2 - 2\Sigma \mathbf{u}$. Solving for $\mathbf{u}$ gives
\begin{equation}
\mathbf{u} = \frac{1}{2}\,\Sigma^{-1}\,\frac{1}{N}\sum_{i=1}^{N} \Delta\mathbf{x}_i\big(\lVert \Delta\mathbf{x}_i\rVert^2 - d_i^2\big),
\label{eq:trilateration}
\end{equation}
the least-squares location of a point from its distances to known points, known as trilateration~\cite{beck2008source}. An activation $s_i$ can therefore, in principle, equal $-d_i^2/2$ up to a constant, which drops out because the $\Delta\mathbf{x}_i$ average to zero; $\Sigma^{-1}M\mathbf{a}$ then supplies the term of \eqref{eq:trilateration} that contains $d_i$. The other term depends only on the positions, so we compute it exactly and add it with a learned weight $b$ rather than asking the activations to represent it,
\begin{equation}
\mathbf{y} = \Sigma^{-1}\Big[M\mathbf{a} + b\,\frac{1}{N}\sum_{i=1}^{N} \Delta\mathbf{x}_i\lVert \Delta\mathbf{x}_i\rVert^2\Big],
\label{eq:moment}
\end{equation}
which equals \eqref{eq:trilateration} when $s_i = -d_i^2/2 + c$, for any constant $c$, and $b = 1/2$; thus, if the learned  features do encode the squared distances, then the vertex is able to be exactly solved. 

Note that the position-only term above contains no information about the vertex: on its own, it locates the vertex within 12.7\,m, and with randomly initialized features the full probe reaches only 3.7\,m and 5.9\,m for electrons and muons. It only allows a linear probe to express the exact solution \eqref{eq:trilateration} once the features supply the distances. Without this term, the probe still locates the vertex within 0.48\,m and 0.47\,m with 10K labeled images. We define the predicted vertex as $\bar{\mathbf{x}} + \mathbf{y}$. For direction prediction, $\Sigma^{-1}$ removes the same dependence on the shape of the image, but \eqref{eq:trilateration} has no counterpart; we nonetheless use the same form \eqref{eq:moment} with a separate $\mathbf{a}$ and $b$ to predict $\mathbf{y}/\lVert\mathbf{y}\rVert$. Each probe therefore has just $D + 1$ parameters, one weight per feature channel shared by the three axes and $b$, whereas a linear probe on the mean feature has $3(D + 1)$ and, because its input does not rotate, cannot represent the rotation at all. The equivariant probe consequently needs very few labels: with 1K instead of 10K labeled images, its vertex $\sigma_{68}$ grows only to 0.44\,m and 0.47\,m for electrons and muons, whereas that of the linear probe on the mean feature grows to 3.4\,m and 3.6\,m. The vertex and direction of Table~\ref{tab:wc} use this probe, with $s_i$ a linear function or an MLP of $\mathbf{z}_i$ (\S\ref{app:exp-finetune}); because a constant added to every $s_i$ drops out, the output of the MLP needs no centering.

In this section, we instead fit $\mathbf{a}$ and $b$ by ridge regression on the frozen features of the first 10K training images of each particle and report the 68th percentile of the error, $\sigma_{68}$, on validation images. The probe reaches 0.42\,m and 0.37\,m on the vertex and 3.2$^\circ$ and 2.3$^\circ$ on the direction for electrons and muons, against 2.0\,m, 2.2\,m, 8.1$^\circ$ and 9.3$^\circ$ for a linear probe on the mean feature, and 13$^\circ$ and 19$^\circ$ on the direction for the same probe on a randomly initialized encoder.

\subsection{Synthetic pretraining enables task transfer in fewer steps}

We find that while synthetic pre-training yields gains in downstream performance, it also reaches near peak performance (99\% of peak) at a certain label count in a training run drastically faster for LArTPC and sPHENIX than without it. With the pretrained decoder, Panda V2 reaches 99\% of peak on 1K labeled images after seeing just 520K images on LArTPC (\ie 520 epochs), whereas the randomly initialized decoder requires 1.24M images (\ie 1,240 epochs). On sPHENIX, the pretrained decoder reaches 99\% of peak after seeing just 80K images, whereas the randomly initialized decoder requires 1.42M images--18$\times$ more images. The synthetic decoder does not appear to help much with WC, where we expect that features are not well-specified with a Gaussian mixture prior.

\begin{figure*}[t]
    \centering
    \includegraphics[width=0.85\linewidth]{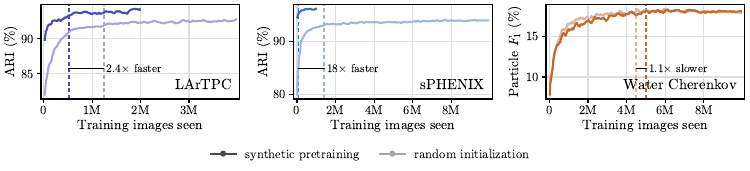}
    \caption{\textbf{Synthetic pretraining of the decoder.} Validation score against training images seen (16 per step) for the decoder with synthetic pretraining and with random initialization, on the frozen PT$_\text{1M}$ encoder with 1K labeled images: ARI on \LArTPC and \sPHENIX, particle $F_1$ on the WC validation images. Curves are running means over the last eight evaluations (160K images). Dashed lines mark where each curve first reaches 99\% of its best, and the factor is the ratio of the two positions.}
    \label{fig:app-synth}
\end{figure*}

In Figure~\ref{fig:app-synth}, we show the validation score over training for the decoder with and without synthetic pretraining.

\subsection{Transferring fine-tuning change to another backbone}

It is well-known that the self-distillation pretraining objective clusters representations on a hypersphere~\cite{caron2021emerging,swav,govindarajan2024dinovonmisesfishermixture}. Thus, it may be natural that a Gaussian mixture prior projected onto a hypersphere provides a good initialization point for fine-tuning. But does it apply elsewhere? More generally, do the optimizations presented in \S\ref{sec:transfer} transfer to other HENP FMs? We test this by performing the same ablation ladder of Table~\ref{tab:ft_ladder} with FM4NPP's released weights.

\begin{figure}[!ht]
    \centering
    \includegraphics[width=0.9\linewidth]{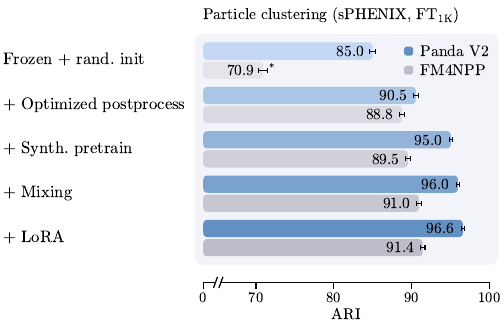}
    \caption{\textbf{Do fine-tuning changes transfer to other pretrained sensor-level FMs?} We find that our fine-tuning recipe transfers to FM4NPP, whose backbone is detector-specific, based on Mamba, and is pretrained with an autoregressive prediction task. $^*$Our reproduction using Panda Detector and FM4NPP's backbone. For reference, FM4NPP reports 69.3 with its own decoder, whereas we report 70.9. Error bars are 95\% CI.}
    \label{fig:synth_ladder}
\end{figure}

Figure~\ref{fig:synth_ladder} shows the ladder on \sPHENIX with 1K labeled images, scored by test ARI, for Panda V2 and for FM4NPP. We first fine-tune randomly initialized Panda Detector on the frozen FM4NPP encoder and decode as FM4NPP does, assigning every hit to its highest-scoring query. This reproduction reaches 70.9 ARI, close to the 69.3 ARI that FM4NPP reports with its own decoder, so Panda Detector alone does not account for the gains that follow. Each later step of the ladder also improves FM4NPP, and the full recipe raises its ARI from 70.9 to 91.4 (+20.5).

The steps do not help the two encoders equally. Optimized postprocessing gives FM4NPP its largest gain (17.9 ARI, against 5.5 ARI for Panda V2), whereas synthetic pretraining adds 0.7 ARI to FM4NPP against 4.5 ARI to Panda V2. Image mixing (1.5 ARI for FM4NPP and 1.0 for Panda V2) and LoRA (0.4 and 0.6 ARI) add similar amounts to the two encoders. The small gain from synthetic pretraining on FM4NPP is consistent with the hypothesis above: FM4NPP is pretrained with an autoregressive objective that does not organize its features into clusters on a hypersphere, so a decoder pretrained on such clusters gives a weaker starting point. Optimized postprocessing, image mixing, and LoRA do transfer to FM4NPP, so most of the recipe does not depend on the encoder's pretraining objective.

\begin{figure*}[t]
    \centering
    \includegraphics[width=\linewidth]{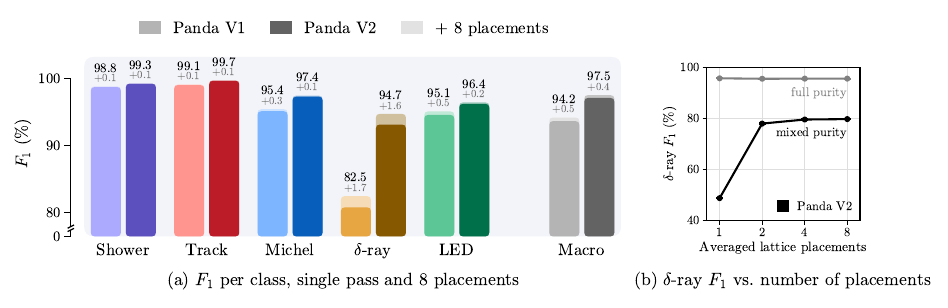}
    \caption{\textbf{Test-time placement averaging on \LArTPC.} (a) Motif $F_1$ of the linear heads of Table~\ref{tab:semseg_lar} (frozen PT$_\text{1M}$ encoder, 1M labeled images) on the test set, for Panda V1 and V2, with one placement (bar) and with eight placements averaged (cap). (b) Panda V2's $\delta$-ray $F_1$ on hits whose single-pass $2{\times}2{\times}2$ cell holds one class (full purity) or several (mixed purity). Error bars are 95\% CI.}
    \label{fig:app-placements}
\end{figure*}

\subsection{Test-time placement averaging}
\label{app:ttt}



We find that most of the linearly decodable semantic meaning in learned representations appears only after the first attention blocks have been applied, \ie, at the second finest resolution. As such, the frozen encoder's outputs are usable only at that resolution, where one cell is equivalent to 2x2x2 voxels. This bounds the highest possible $F_1$ a model can score, and makes the output depend on which origin is chosen for the voxel lattice.

Consider a point from a $\delta$-ray that lies directly next to a muon. Depending on the origin of the lattice, the coarse voxel that contains that $\delta$-ray point after the first pooling contains only $\delta$-ray points or points of the muon as well. In the latter case, the highest possible $\delta$-ray $F_1$ score is less than 100\%.\footnote{The muon point that shares the cell is affected the same way, but a muon track has far more points than a $\delta$-ray, so its $F_1$ moves less.} This is a significant issue for cases where particles touch: with a single forward pass, the $\delta$-ray $F_1$ of a linear probe on frozen Panda V2 outputs, fit and evaluated on validation images, is 49.4\% on the voxels whose coarse cell also holds another particle, compared with 95.2\% on all other voxels (Figure~\ref{fig:app-placements}b). 

Assuming that a point usually lies next to points of its own particle, we propose \textbf{test-time placement averaging}, that is, averaging out these unlucky placements by shifting the input grid by one voxel along a different subset of the three axes (\ie the eight corners of a 2x2x2 cube), running the encoder on all eight inputs, and averaging the per-point features. This technique, known as ``cycle spinning"~\cite{Coifman1995}, is commonly used to reduce shift-variance artifacts in the discrete wavelet transform (DWT), which like the Panda Encoder, performs successive downsamples and is thus is affected by shift variance~\cite{1525499}.

We find this issue also occurs in Panda V1; Panda V1 notably contains attention blocks in the earlier stages, which suggests that the lack of early attention blocks in V2 is not the issue.

Figure~\ref{fig:app-placements}a shows the gain from averaging the eight placements for each motif on validation images. On the test images, averaging the eight placements raises the motif macro $F_1$ of Panda V2's linear probe from 97.1\% to 97.5\% and its $\delta$-ray $F_1$ from 93.1\% to 94.7\%, and most of the gain is on hits in mixed cells, where the $\delta$-ray $F_1$ rises from 48.8\% to 79.7\%. In Fig.~\ref{fig:app-placements}b, we find that the performance gains plateau after the second placement.

We do not apply test-time placement averaging elsewhere in this work, though we expect it to be beneficial for LArTPC and sPHENIX. Note that this is necessarily an issue about the interaction between the pre-training objective and backbone architecture; in both Panda V1 and V2, it appears that the pre-training objective encourages the first stage of the encoder to focus on organizing information for later stages to perform computations on. We hope further work expands on this issue.

\subsection{Qualitative results}
\label{app:qualitative}

\begin{figure*}[p]
    \centering
    \includegraphics[width=0.8\linewidth]{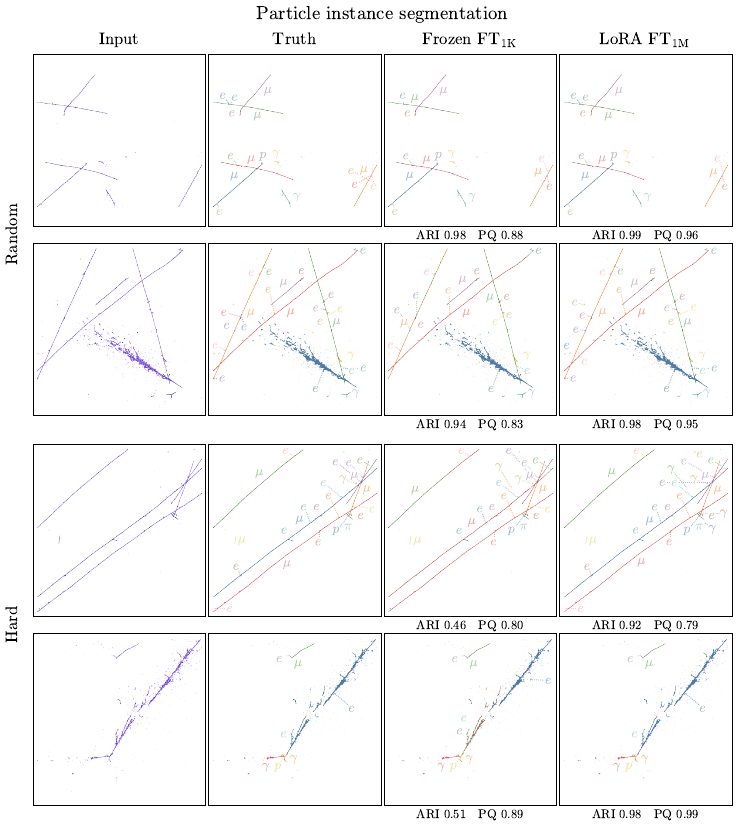}
    \caption{\textbf{Qualitative LArTPC particle clustering examples.} We provide the input to the PT$_\text{1M}$ model, ground truth, as well as reconstructions from the model in the label constrained (frozen FT$_\text{1K}$) and label unconstrained (LoRA on FT$_\text{1M}$) cases. Here, ``hard" images are picked by randomly sampling from the bottom quartile of scores on the data-constrained fine-tune, while ``random" images are random samplings of the test set. Particles are matched to truth and receive the same coloring to promote visual ease. Particle type truth and predictions are placed next to each particle. We provide both ARI and PQ scores for each reconstruction. Low energy deposits, which are uncountable in PILArNet-M, are colored in grey.}
    \label{fig:app-lar-qual}
\end{figure*}

\paragraph{\LArTPC.} In Figures~\ref{fig:app-lar-qual}, we show the same four test images segmented into particles by two models: the frozen PT$_\text{1M}$ encoder with a decoder trained on 1K labeled images, and the PT$_\text{1M}$ encoder fine-tuned with LoRA on 1M labeled images. We draw two of the images at random, and two from the quartile of images on which the frozen model scores worst. 

\begin{figure*}[p]
    \centering
    \includegraphics[width=0.8\linewidth]{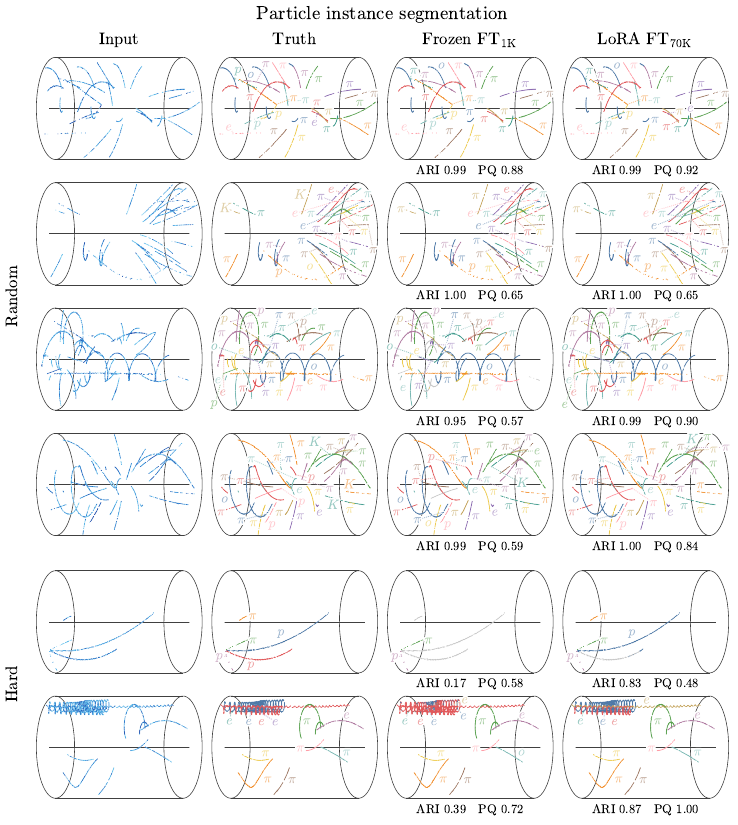}
    \caption{\textbf{Qualitative sPHENIX particle clustering examples.} We provide the input to the PT$_\text{1M}$ model, ground truth, as well as reconstructions from the model in the label constrained (frozen FT$_\text{1K}$) and label unconstrained (LoRA on FT$_\text{70K}$) cases. Here, ``hard" images are picked by randomly sampling from the bottom quartile of scores on the data-constrained fine-tune, while ``random" images are random samplings of the test set. Particles are matched to truth and receive the same coloring to promote visual ease. Particle type truth and predictions are placed next to each particle. We provide both ARI and PQ scores for each reconstruction.}
    \label{fig:app-sph-qual}
\end{figure*}

\paragraph{\sPHENIX.} In Figure~\ref{fig:app-sph-qual}, we provide six test images segmented into particles by two models: the frozen PT$_\text{1M}$ encoder with a decoder trained on 1K labeled images, and the PT$_\text{1M}$ encoder fine-tuned with LoRA on 70K labeled images. We draw four of the images at random, and two from the quartile of images on which the frozen model scores worst.

\begin{figure*}[p]
    \centering
    \includegraphics[width=0.8\linewidth]{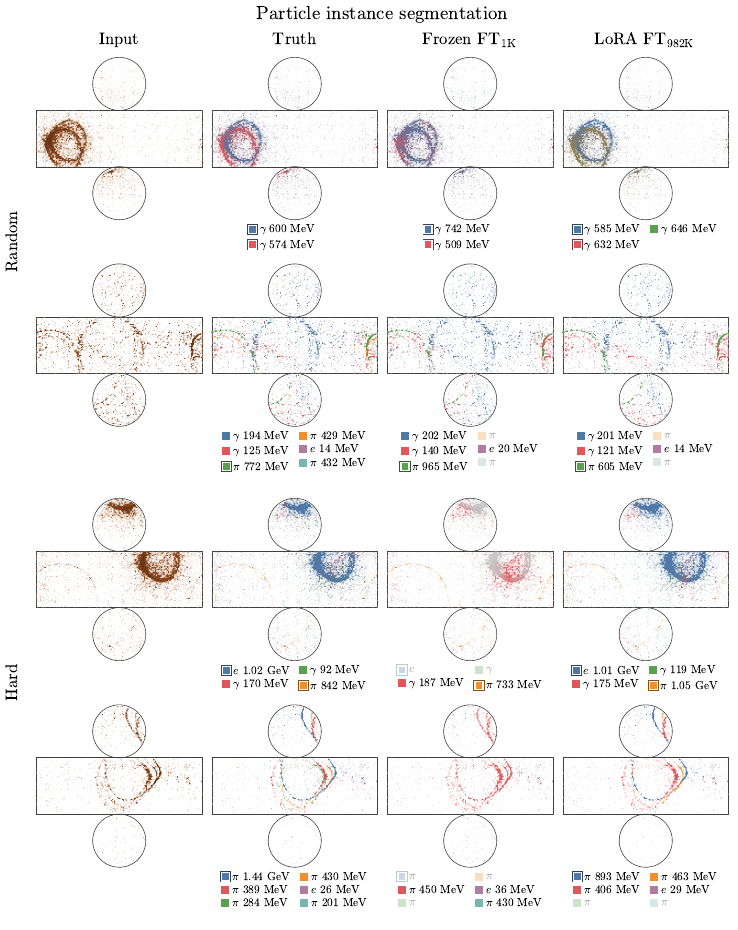}
    \caption{\textbf{Qualitative WC image reconstructions.} We provide the input to the PT$_\text{1M}$ model, ground truth, as well as reconstructions from the model in the label constrained (frozen FT$_\text{1K}$) and label unconstrained (LoRA on FT$_\text{982K}$) cases. Here, ``hard" images are picked by randomly sampling from the bottom quartile of scores on the data-constrained fine-tune, while ``random" images are random samplings of the test set. Particles are matched to truth and receive the same coloring to promote visual ease. In the legend, we provide true and predicted particle momenta, and highlight primary particles by giving their legend marker a dark outline. Particles which Panda V2 misses are faded in the legend. We note that the model receives and segments the 3D point cloud, not the unrolled 2D view.}
    \label{fig:app-wc-qual}
\end{figure*}

\paragraph{\WC.} In Figure~\ref{fig:app-wc-qual}, we show four test images segmented into particles by two models: the frozen PT$_\text{1M}$ encoder with a decoder trained on 1K labeled images, and the PT$_\text{1M}$ encoder fine-tuned with LoRA on 982K labeled images. We draw two of the images at random, and two from the quartile of images on which the frozen model scores worst.


\section{Experimental setup}
\label{app:exp}

In this section, we provide details required to reproduce each model in this work: the inputs to the encoder (\S\ref{app:exp-inputs}), its architecture (\S\ref{app:exp-encoder}), pretraining (\S\ref{app:exp-pretrain}), fine-tuning for each task (\S\ref{app:exp-finetune}), the baselines that we train or reproduce (\S\ref{app:exp-baselines}), and the test protocol (\S\ref{app:exp-test}).

\subsection{Inputs}
\label{app:exp-inputs}


(1) Pretraining images with fewer points than the minimum are removed; for \LArTPC, points below 0.13\,MeV are removed first. (2) Coordinates $\mathbf{x}$ are normalized to $(\mathbf{x} - \mathbf{c})/s$. (3) Points are voxelized on a regular grid, and the values $E$ of the points in a voxel are summed. (4) $E$ is mapped to
\begin{equation}
E' = 2\,\frac{\log_{10}(E + e_\mathrm{lo}) - \log_{10} e_\mathrm{lo}}{\log_{10}(e_\mathrm{hi} + e_\mathrm{lo}) - \log_{10} e_\mathrm{lo}} - 1,
\end{equation}
which maps $[0, e_\mathrm{hi}]$ to $[-1, 1]$. The encoder receives the normalized coordinates and $E'$ of each voxel. Training augmentations are applied between steps (2) and (3) (\S\ref{app:exp-pretrain} and \S\ref{app:exp-finetune}).

In WC, points correspond to the static positions of photomultipliers that recorded a signal. Each position contains both the summed charge (in photoelectrons, or PE) and the earliest hit time $t_\mathrm{hit}$ at that position. We measure each time relative to a reference time $t_0$ of the image, $t = t_\mathrm{hit} - t_0$, and the encodings below receive $t$.\footnote{Specifically, $t_0$ is the earliest hit time $t'$ of the image for which at least eight of the image's hit times, $t'$ included, lie in $[t', t' + 10\,\mathrm{ns}]$, or the earliest hit time of the image if no time meets this condition. The earliest hit time itself is usually a dark-noise hit hundreds of nanoseconds before the light from the particle, so this condition keeps $t_0$ from falling on noise.} The encoding of $t$ must resolve differences of nanoseconds, from which the arrival times of the Cherenkov light give the vertex and direction, over a range of microseconds, which includes delayed light such as that of decay electrons. To do so, we represent each time $t$ with two encodings. The first is a Fourier encoding,
\begin{equation}
\boldsymbol{\phi}(t) = \big(\sin(2\pi t/T_k),\, \cos(2\pi t/T_k)\big)_{k=0}^{10},
\label{eq:wc-fourier}
\end{equation}
with periods $T_k = 8\cdot 4^k$\,ns from 8\,ns to $T_{10} \approx 8.4$\,ms. The shortest periods resolve differences of nanoseconds and the longest exceed every time in the WC dataset, so no two times may share all phases. the range of times in an image, however, the phases of the longest periods change by small amounts (a delay of 10\,$\mu$s moves the phase of $T_{10}$ by less than 0.01\,rad), so $\boldsymbol{\phi}$ does not give the elapsed time on a usable scale. We therefore include a notion of absolute time, that is,

\begin{equation}
\ell(t) = \operatorname{sign}(t)\,\frac{\log(1 + |t|/T_0)}{\log(1 + T_{10}/T_0)},
\label{eq:wc-logtime}
\end{equation}
which orders all times on one monotonic scale, from the prompt Cherenkov light tens of nanoseconds after $t_0$ to delayed light microseconds later. The logarithmic time is 0 at $t_0$ and $\pm1$ at $\pm T_{10}$, and its sign separates the hits before $t_0$ from those after it. Each position therefore has 27 input channels: 3 coordinates, charge (PE), and 23 time channels.

\begin{figure*}[t]
\centering
\includegraphics[width=\linewidth]{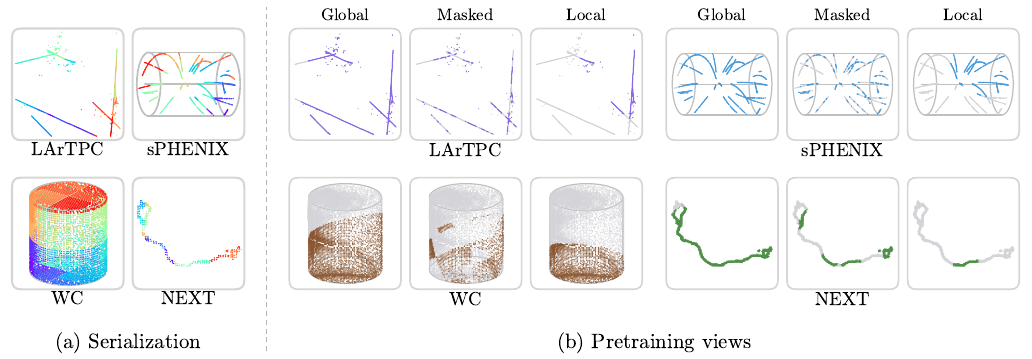}
\caption{\textbf{Serialization and pretraining views.} (a) The serialization used for attention: the voxels of one image per detector, colored by their position along a Hilbert curve, from the first (purple) to the last (red). The voxels are those of the third stage, twice the size of the input voxels, where attention is first applied. (b) A global view, its masked copy, and a local view of the same images. The points in each view are colored, and the other points of the image, including the points removed by the mask, are grey.}
\label{fig:app-serial-views}
\end{figure*}

\paragraph{Serialization.} To linearize attention later on, we perform block-wise attention within patches of 1{,}024 points. We serialize entire images by ordering all points on a number of space-filling curves (Hilbert, Morton, trans-Hilbert, trans-Morton), as in PTv3~\cite{wu2024pointtransformerv3simpler}; this is a form of locality-sensitive hashing~\cite{Indyk1998}, that is, this process approximates applying attention within a local $k$-NN. We provide examples of serialized events across the four detectors in Fig.~\ref{fig:app-serial-views}a.

\subsection{Panda V2 Encoder}
\label{app:exp-encoder}

The Panda V2 Encoder is UNet~\cite{ronneberger2015unetconvolutionalnetworksbiomedical}-style encoder composed of four separate stages (channels 54, 108, 432, 576) over three different resolutions (1, 2, and 8 voxels), with the middle two stages staying at the same resolution. We derive it from LitePT-B~\cite{litept} (\S\ref{sec:pretraining}), which has five stages (channels 54, 108, 216, 432, 576 and blocks 3, 3, 3, 12, 3), each at half the resolution of the previous one, with sparse convolutions in the first three stages and attention in the last two. We reduce these to the four stages above, with sparse convolutions in the first two and attention in the last two, and we change LitePT in two further ways, described below: we remove one LayerNorm from its attention blocks, and we replace its decoder by feature up-casting. The architecture is identical for all four detectors with the exception of the number of input channels (\S\ref{app:exp-inputs}). A $5^3$ sparse convolution~\cite{graham20183d} followed by layer normalization~\cite{ba2016layernormalization} and a GELU activation~\cite{hendrycks2023gaussianerrorlinearunits} first maps the inputs of each voxel to 54 channels. The first two stages consist of sparse convolution blocks (kernel of 3$^3$), a linear layer and layer normalization inside a residual connection. The third and fourth stages consist of standard pre-norm transformer~\cite{vaswani2017attention} blocks with an MLP ratio of 4, in which self-attention among voxels is computed within windows of 1{,}024 contiguous points along a space-filling curve. Each of these blocks applies a LayerNorm at the start of its attention branch and of its MLP branch. LitePT-B additionally applies a LayerNorm to the residual stream itself at the start of each attention block, outside both branches; we remove this LayerNorm, because with it the pretraining of the \LArTPC encoder became unstable (Fig.~\ref{fig:norm}). We set the number of channels per head to 18 across all attention stages. Each successive block cycles through four curves~\cite{wu2024pointtransformerv3simpler}: the Hilbert and z-order curves and their transposes, whose assignment to blocks is shuffled at every forward pass. Queries and keys are rotated by 3D rotary position encoding (PointROPE)~\cite{litept} with a base of 100, which rotates each third of a head's channels by one integer voxel coordinate. Between stages, the points within a voxel $s$ times larger are merged by a linear layer, a max-pool, batch normalization~\cite{ioffe2015batchnormalizationacceleratingdeep} and a GELU, and their coordinates are averaged; with $s = 1$, the pooling only changes the width. We do not use a decoder. Instead, as in the feature up-casting of Sonata~\cite{wu2025sonataselfsupervisedlearningreliable}, the features of each stage are copied back to the input voxels by inverting the pooling maps and are concatenated together, which gives $54 + 108 + 432 + 576 = 1{,}170$ channels per voxel. The four stages have 3, 6, 12 and 6 blocks, and the encoder has 53.4M parameters (53.6M for WC, whose stem takes more input channels).

\subsection{Pretraining}
\label{app:exp-pretrain}

\begin{table}[t]
\centering
\footnotesize
\setlength{\tabcolsep}{4pt}
\begin{tabular}{@{}l>{\raggedright\arraybackslash}p{0.6\columnwidth}@{}}
\toprule
Batch size & 96 images \\
Steps & 208{,}334 (20M images) \\
Optimizer & AdamW~\cite{loshchilov2019decoupledweightdecayregularization}, $\beta_2 = 0.996$ \\
Peak learning rate & $5.94\times10^{-3}$; conv and attention blocks $5.94\times10^{-4}$ \\
Schedule & 10\% linear warmup from $1/10$ of peak, cosine decay to $10^{-4}$ of peak; $\beta_1$ follows the inverse~\cite{smith2018superconvergencefasttrainingneural}, $0.95 \to 0.85 \to 0.95$ \\
Weight decay & 0.05 (biases and norms excluded) \\
Gradient clipping & 2.4 \\
Teacher momentum & cosine, 0.996 to 1 \\
Teacher temperature & 0.04 to 0.07 over the first 10\% of steps \\
Student temperature & 0.1 \\
Drop path~\cite{huang2016deepnetworksstochasticdepth} (teacher, student) & 0.0, 0.3 \\
LayerScale~\cite{touvron2021goingdeeperimagetransformers} & no \\
Precision & bf16 \\
\bottomrule
\end{tabular}
\caption{\textbf{Pretraining hyperparameters}, shared by all detectors and all pool sizes $K$.}
\label{tab:app-pretrain}
\end{table}

\paragraph{Views.} From each image of $N$ points, we sample two global views and six local views (Fig.~\ref{fig:app-serial-views}b). Each view is the set of the $k$ points nearest to a center, with $k$ a fraction of $\min(N, 30{,}000)$ drawn from $U(0.4, 1.0)$ for global views and $U(0.1, 0.4)$ for local views ($U(0.85, 1.0)$ and $U(0.3, 0.6)$ for \sPHENIX; see Table~\ref{tab:recipe}). Before cropping, we scale the coordinates of the image by a factor drawn from $U(0.9, 1.2)$, and, with probability 0.8, multiply the value $E$ of each point in the global views by $1 + \epsilon$, with $\epsilon \sim \mathcal{N}(0, 0.05^2)$ clipped to $\pm 0.1$. We then center each view, scale it by a factor drawn from $U(0.9, 1.1)$, rotate it about the $x$, $y$ and $z$ axes, each with probability 0.8 and by an angle drawn from $U(-\pi, \pi)$, flip it along each axis with probability 0.5, and add Gaussian noise to its coordinates with a standard deviation of $1/8$ (global) or $1/4$ (local) of a voxel, clipped to one voxel. In WC, we apply neither scaling (Table~\ref{tab:recipe}), and we shift all times $t$ of a view by one global offset drawn from $U(-10, 10)$\,ns; we compute $t_0$ once per image, before this shift, so that the offset moves the relative times of the view. The two masked views are copies of the global views from which the points in 60\% of the occupied cells of a cubic grid are removed; the cell size of each detector is given in Table~\ref{tab:recipe}. Example local, global, and masked views for the four detectors are shown in Fig.~\ref{fig:app-serial-views}b.

\paragraph{Objective.} The objective is illustrated in Fig.~\ref{fig:pretraining}a. The teacher encodes the global views, and the student encodes the masked and local views. Two heads score the features of each point against $C=8{,}192$ prototypes. Each head maps the features through a two-layer MLP ($1{,}170 \to 4{,}096 \to 256$, with GELU activations) and computes the cosine similarity between its output and each prototype. The mask head scores the points of the global and masked views, and the local head scores the points of the global and local views. Each student point is paired with the nearest teacher point in the coordinates of the image before augmentation, within a radius of 2 voxels; student points with no teacher point within this radius are excluded from the loss. We convert the teacher's scores into targets with three Sinkhorn-Knopp~\cite{Sinkhorn1967ConcerningNM} iterations over groups of 24 images, and the student's scores into predictions with a softmax. Thus, for each point, we have a probability distribution over all $C$ prototypes. The loss applied is the cross-entropy between the matched points between the teacher and student, averaged over the points of each view and then over views. There are three terms: the masked views against their own global view (weight 0.25), the masked views against the other global view (0.25), and the local views against the first global view (0.5). The teacher is an exponential moving average of the student; as such, no gradients flow through to the teacher. Settings are listed in Table~\ref{tab:app-pretrain}.

\paragraph{Per-detector settings.} The settings in Table~\ref{tab:recipe} are the only differences between detectors, and we give the reason for each in \S\ref{app:pt_ablate}. For WC, where the loss is applied one level coarser, the heads read the features of stages 2--4 at the stage-2 voxels (1{,}116 channels), and the matching radius is 2 stage-2 voxels.

\paragraph{Dataset pools and checkpoints.} The pool of $K$ images for PT$_K$ is the first $K$ images of one fixed random permutation of the training split of each detector. Every run is scheduled for 208{,}334 steps, or 20M image presentations, regardless of $K$. Unfortunately, the self-distillation loss does not have much meaning~\cite{feng2026tesserav2scalingpixelwise}, as the student constantly predicts the outputs of a moving target. Thus, we select checkpoints by taking periodic linear probes throughout training (every 500K images for $K\geq10$K, every 250K for $K=1$K, and every 100K for $K=100$). We save the teacher and evaluate a probe on its frozen features of validation images---looking at semantic segmentation performance for \LArTPC and \sPHENIX, particle kinematics regression for \textcolor{wc}{WC}, and image classification for \NEXT. We set our final checkpoints for each $K$ as the checkpoint whose evaluation achieved the peak score.

\subsection{Fine-tuning}
\label{app:exp-finetune}

\begin{table*}[t]
\centering
\footnotesize
\setlength{\tabcolsep}{4pt}
\begin{tabular}{@{}l>{\raggedright\arraybackslash}p{0.23\linewidth}>{\raggedright\arraybackslash}p{0.28\linewidth}>{\raggedright\arraybackslash}p{0.26\linewidth}@{}}
\toprule
 & Semantic segmentation & Instance segmentation & Image-level tasks \\
\midrule
Head & linear, or MLP (256 hidden) & Panda Detector V2 (3 blocks, width 256, 16 heads), from synthetic pretraining & \NEXT: logistic regression (frozen), linear (LoRA); WC particle type: linear or MLP; WC kinematics: equivariant probe of \eqref{eq:moment} \\
Loss & CE + 0.05 Lov\'asz~\cite{berman2018lovaszsoftmaxlosstractablesurrogate} & Hungarian matching; focal~\cite{lin2018focallossdenseobject} 2, dice~\cite{milletari2016v} 5, class 2, no object 0.5; WC adds per-query momentum, vertex and direction regression & CE; WC kinematics: squared error \\
Learning rate & frozen: head $3\times10^{-4}$; LoRA: head and LoRA $10^{-4}$ & decoder $3\times10^{-4}$, LoRA $10^{-4}$ & frozen: head $3\times10^{-4}$; LoRA: LoRA $10^{-4}$ \\
Selection criteria & macro $F_1$ & harmonic mean of ARI and panoptic quality~\cite{kirillov2019panopticsegmentation}; WC: muon direction $\sigma_{68}$ & validation AUROC; WC kinematics: each head's own score \\
\midrule
Optimization & \multicolumn{3}{>{\raggedright\arraybackslash}p{\dimexpr0.77\linewidth+16pt\relax}}{AdamW~\cite{loshchilov2019decoupledweightdecayregularization} (weight decay 0.01), batch size 16, OneCycle~\cite{smith2018superconvergencefasttrainingneural} cosine schedule over 625{,}000 steps (10M images) with 625 warmup steps} \\
\bottomrule
\end{tabular}
\caption{\textbf{Fine-tuning settings.}}
\label{tab:app-ft}
\end{table*}

\paragraph{Protocol.} We fine-tune with the settings of Table~\ref{tab:app-ft}. Every run is scheduled for 10M images, regardless of the number of labeled images $K$. The label set of size $K$ is the first $K$ images of the training split. With a frozen encoder, only the head or decoder is trained; with LoRA~\cite{hu2021loralowrankadaptationlarge}, we add adapters of rank 16 to every linear layer of the encoder; the supervised baselines train every weight of the same architecture from random initialization.

\paragraph{WC kinematics.} The vertex and direction heads are the equivariant probe of \eqref{eq:moment}, with $s_i$ either a linear function of the point features or an MLP of them with one hidden layer of 256 units, ReLU and dropout 0.3. The momentum head is a linear map or an MLP of the mean point feature that predicts the logarithm of the momentum. We train the three heads with AdamW on the squared error, either on the frozen encoder or jointly with LoRA adapters on the encoder, with one set of adapters shared by the three heads of each particle type.

\paragraph{Synthetic decoder pretraining.} We pretrain the three blocks of Panda Detector V2 on 720{,}000 synthetic scenes, each a set of 2{,}048 unit vectors in $\mathbb{R}^{256}$. A scene has $G$ groups, with $G$ drawn log-uniformly from 2 to 40, whose centers are random unit vectors in a random 32-dimensional subspace. Each point of a group with center $\boldsymbol{\mu}_g$ is drawn from an isotropic Gaussian with mean $\cos\theta\,\boldsymbol{\mu}_g$ and covariance $(\sin^2\theta/255)\,I$ in the 255-dimensional subspace orthogonal to $\mathbf{1}$, and is then projected onto the unit sphere, so that each scene is a Gaussian mixture projected onto the sphere (\S\ref{sec:transfer}). Because a 255-dimensional Gaussian concentrates near its typical radius, the points of a group lie close to the angle $\theta$ from its center. We draw $\theta$ once per scene between $2^\circ$ and $80^\circ$, and all groups of a scene have the same number of points. In 65\% of the scenes, a fraction of the points drawn from $U(0, 0.5)$ belongs to no group and is uniform on the sphere. The decoder receives the points directly as keys and values and is trained to output one mask per group, with the losses of Table~\ref{tab:app-ft}, until convergence (90{,}000 steps at a batch size of 8) using AdamW and a learning rate of $3\times10^{-4}$. As in prior-data fitted networks~\cite{muller2021transformers}, training on scenes drawn from this prior makes the decoder's output an approximation, learned rather than exact, to the posterior over the assignment of points to groups under the prior. For fine-tuning, we copy its three blocks and its final layer normalization and initialize all other parameters of the decoder anew.

\subsection{Baselines}
\label{app:exp-baselines}

For \LArTPC and \sPHENIX, we compare against published models. For \NEXT and \textcolor{wc}{WC}, however, no baseline has been published. Thus, we build our own baselines from the standard tools of each collaboration and tune them on validation images, so that each baseline is as strong as possible.

\subsubsection{\NEXT}

\paragraph{Event selection.} Every method, ours included, is evaluated on the events that pass the topology selection of~\cite{JHEP2021}: every hit must lie at least 20\,mm inside the drift volume, and the hits must form one connected track when binned into cubes of 15\,mm.

\paragraph{Two-blob cut.} A $0\nu\beta\beta$ decay emits two electrons, and each electron ends its track in a dense deposit of energy, a ``blob," whereas the background electron from a $^{214}$Bi photon produces only one blob. The ``two-blob cut" uses this difference without any labels. We compute it with Invisible Cities, the NEXT collaboration's reconstruction software,\footnote{\url{https://github.com/next-exp/IC}}: the hits are voxelized, neighbouring voxels are linked into tracks, and the two ends of the longest track are found; the energy within a radius $R$ of each end is summed, and the score of an event is the smaller of the two energies. As~\cite{JHEP2021} chose $R$ on validation data, we scan voxel sizes of 5 and 10\,mm and radii from 9 to 48\,mm on the validation images and keep the setting with the highest AUROC, that is, 5\,mm voxels and $R = 42$\,mm. Voxels of 3\,mm, the resolution of the dataset, take about 100\,s per event in Invisible Cities and are therefore not scanned.

\paragraph{Sparse ResNet.} We retrain the sparse convolutional ResNet of~\cite{kekic2021next} from random initialization at each label count. We keep its architecture, its optimizer (Adam, learning rate $10^{-4}$, weight decay $10^{-4}$), its class-weighted cross-entropy and its augmentations (random shifts, flips, an exchange of the $x$ and $y$ axes, and zooms), and train with a batch size of $\min(1{,}024, K)$; the run on all labeled images is scheduled for 100K steps. The input of each voxel is its fraction of the event's total energy. We add two changes that adapt the model to our data: first, our voxels are cubes of 3\,mm instead of $10\times10\times5$\,mm, so the stem and the first downsampling stages are set to be isotropic and we average features over all voxels to get a final feature to perform classification with; second, each event is centred in a box of 90\,cm instead of the volume of NEXT-White~\cite{ferrario2019nextwhite}, the NEXT demonstrator detector whose images the model was designed. We keep the checkpoint with the highest validation AUROC.

\subsubsection{\WC}

\paragraph{fiTQun.}
\label{app:fitqun}
fiTQun~\cite{fitqunSK, fitqunT2K, fitqunUB} is the standard event reconstruction software of the Super-Kamiokande~\cite{fukuda2003superkamiokande} and T2K~\cite{T2K:2011qtm} experiments. Given a hypothesis for the particles in a recorded event, it predicts the charge and time expected at every PMT. It does this using tabulated models of Cherenkov emission, light propagation in water, and PMT response, and fits the hypothesis parameters by maximizing the likelihood of all hit and unhit PMTs. Since neither the number of particles nor their types are known in advance, fiTQun fits several competing hypotheses and classifies the event by comparing their likelihoods. For a single particle, each hypothesis ($e$, $\mu$, $\pi^\pm$) is a straight track with seven free parameters (vertex, time, direction and momentum), plus, for $\pi^\pm$, the length after which the pion stops emitting light; the parameters are found with MINUIT~\cite{James:1994vla}, seeded by a fast vertex and direction prefit. Particle identification then reduces to likelihood ratios between the best fits, \eg $\ln(L_e/L_\mu)$; notably, photons, which are nearly indistinguishable from electrons, have no hypothesis of their own and are reconstructed as electrons. An individual $\pi^0$ fit describes its two decay photons as two electron-like tracks from a common vertex, and a multi-ring fit adds tracks one at a time only if each improves the likelihood. Only light from the first trigger window is fitted. Decay electrons are found as delayed hit clusters using a sliding window that searches for peaks in the time-of-flight-corrected hit distribution. The first cluster is the primary interaction; each later one counts as a decay electron.

\paragraph{Detector tune.}
As described above, fiTQun predicts the recorded light at each PMT using detector-specific lookup tables. WAND is a new detector, so these tables must be regenerated before fiTQun can reconstruct WAND images. Detector-specific inputs to tune fiTQun for WAND are generated following the standard procedure. This includes: Cherenkov emission profiles for $e$, $\mu$ and $\pi^+$, direct-light time distributions, a scattered- and reflected-light table, PMT angular and charge responses, and the water attenuation length. The overall charge scale, per-hypothesis light scales and scattered-light timing are then fitted to single tracks fixed to their true values.

\paragraph{Particle typing.}
fiTQun's output is a set of competing fits, one per hypothesis, rather than a list of particles. To compare with Panda V2, we turn it into such a list with a decision tree whose cuts we tune on simulated events that are not used for evaluation. The number of particles is the number of rings in the best multi-ring fit. A single-ring event is first classified as electron- or muon-like from the likelihood ratio of the two single-ring fits; electron-like events are then tested against the $\pi^0$ fit, and muon-like events against the charged-pion fit. Each cut compares a likelihood ratio with a line in one fitted quantity, either the momentum of the hypothesis tested or the $\pi^0$-fit two-photon mass. In multi-ring events every ring is one particle: fiTQun only labels rings as electron- or pion-like, so we add a per-ring muon test using fiTQun's refit of each ring under each hypothesis, and we merge pairs of electron-like rings whose invariant mass is compatible with a $\pi^0$.

\subsection{Testing protocol}
\label{app:exp-test}

We evaluate each model once on the test split of its dataset (Table~\ref{tab:app-splits}), with the metrics of Table~\ref{tab:app-eval}. Where shown, intervals are 95\% percentile bootstrap intervals over 1{,}000 resamples of the test images; for NEXT, whose class counts are fixed by the simulation, the signal and background images are resampled separately. Each result comes from one training run, so the intervals reflect the sampling of the test images only, and not any variance from resampling the underlying training set.

\section{Manifold maps}\label{app:atlas}

In the manifold maps of \S\ref{sec:interp} (Fig.~\ref{fig:atlas}), we draw the encoder's point features in two dimensions with PHATE~\cite{Moon2019-lf}. PHATE was developed to visualize high-dimensional biological data, such as single-cell gene expression, in which cells form continuous trajectories and branches rather than separate clusters; unlike t-SNE~\cite{JMLR:v9:vandermaaten08a} and UMAP~\cite{McInnes2018}, which preserve mainly the nearest neighbors of each point, it preserves both local neighborhoods and the global arrangement of such trajectories. In PHATE, each point is connected to its nearest neighbors in feature space, and a random walk spreads over these connections for a number of steps; two points from which the walk spreads to the same places are similar and are placed close together in two dimensions. We run PHATE not on the points themselves but on neighborhoods of the features, small groups of nearby points found by $k$-means, and then place each point according to the positions of its two neighborhoods (\S\ref{app:atlas-construction}).

In a plot aiming to visualize the underlying manifold of a set of features, each region of the visualization should occupy an area that grows with the number of points it holds, and the map should stay the same when we refit it, such that the structure we see is a property of the encoder's features rather than of our choices in building the map. Neither requirement holds for PHATE run on neighborhoods, because each neighborhood counts as a single point: a region that $k$-means splits into more neighborhoods occupies a larger area, and the map changes with the number of neighborhoods and with the points on which they are fitted. We therefore change PHATE in two ways (\S\ref{app:atlas-changes}) and check that the resulting map is stable (\S\ref{app:atlas-stability}). We henceforth talk about how we construct these manifold maps, and assume knowledge of how PHATE~\cite{Moon2019-lf} works.

\subsection{Construction}\label{app:atlas-construction}

The map is built in six steps.

\begin{enumerate}[leftmargin=*,itemsep=2pt]
\item \textbf{Points.} A point is one voxel of the second stage of the encoder ($2\times2\times2$ input voxels). Its features are the features of all four stages at that voxel (1{,}170 features), with the features of each stage centered and scaled to the same total variance. We use three disjoint sets of points: 6.96M points to fit the neighborhoods, 25M points from other images to build the neighborhood graph, and the points of separate validation images to draw the map.
\item \textbf{Neighborhoods.} We split the points into $K = 4{,}096$ neighborhoods with $k$-means~\cite{lloyd1982kmeans,arthur2007kmeans}, about 1{,}700 points each, and fit each neighborhood with a probabilistic PCA~\cite{Tipping1999-hy} with 24 axes, so that each neighborhood has a mean, a covariance $\Sigma$ and a likelihood for every point. We partition with $k$-means rather than by fitting the mixture of probabilistic PCAs, because the map changes less when we refit the $k$-means partition with a new seed (Procrustes $R^2$ of 0.995 against 0.81 for the mixture on \LArTPC; \S\ref{app:atlas-stability}).
\item \textbf{Assignment.} We assign each point to the two neighborhoods with the highest likelihood. With the two nearest neighborhood means instead, the shape of each neighborhood would be ignored and the thin branches of the map would be lost (Table~\ref{tab:atlas-stability}).
\item \textbf{Graph.} Two neighborhoods $i$ and $j$ are connected by the number $W_{ij}$ of points assigned to both, and $W_{ii}$ is the number of points whose most likely neighborhood is $i$. The mass $\pi_i = W_{ii}/\sum_k W_{kk}$ of neighborhood $i$ is its share of the points.
\item \textbf{Layout.} In PHATE, a random walk runs on the graph for $t$ steps, with $t$ chosen at the knee of the walk's von Neumann entropy; the logarithm of the resulting transition probabilities serves as a potential, and the neighborhoods are placed in two dimensions by metric multidimensional scaling~\cite{borg2005modernMDS}, so that their distances in two dimensions match the distances between their potentials. We change this layout in two ways (\S\ref{app:atlas-changes}).
\item \textbf{Placement.} Each point is placed at the position of its most likely neighborhood plus a linear map of its whitened coordinates in that neighborhood's probabilistic PCA. The linear maps of all neighborhoods are fitted jointly~\cite{Brands_NIPS2002_8929c70f}, so that a point lands at the same place through either of its two neighborhoods.
\end{enumerate}

\subsection{Changes to PHATE}\label{app:atlas-changes}

We change four steps of PHATE: the edge weights, the random walk, the distances between potentials and the multidimensional scaling.

\paragraph{Scaling each edge by the size of its neighborhoods.} A random walk takes a time proportional to $h^2$ to leave a region of size $h$, but PHATE's walk leaves every neighborhood at the same rate. We therefore set the edge weights to $C_{ij} = W_{ij}/(h_ih_j)$, where $h_i = \sqrt{\operatorname{tr}\Sigma_i}$ is the root-mean-square radius of neighborhood $i$.

\paragraph{A random walk in continuous time.} In PHATE's walk of $t$ steps, $P^t$ with $P = D^{-1}W$, the walk jumps out of every neighborhood once per step, so the time it spends in a neighborhood cannot depend on how many points the neighborhood holds or on how large it is. We therefore replace this walk by its continuous-time form, the heat kernel $H = e^{-\tau L}$, in which each edge is a rate rather than a probability of a step: $L = \Pi^{-1}(D_C - C)$, where $C$ is the matrix of the edge weights above, $D_C$ is the diagonal matrix of the row sums of $C$, and $\Pi = \operatorname{diag}(\pi)$ is the diagonal matrix of the masses of the neighborhoods. We choose $\tau$ using PHATE's rule, at the knee of the von Neumann entropy over $\tau \le 100$, with $C$ scaled so that one unit of $\tau$ is one expected jump, which gives $\tau = 24$ for \LArTPC and 23 for \sPHENIX. The potential and its floor of $10^{-7}$, and SMACOF for multidimensional scaling, are PHATE's own choices.

\paragraph{Weighting each neighborhood by its number of points.} In PHATE, each neighborhood counts as one point, so a region of the manifold that $k$-means splits into more neighborhoods occupies a larger area of the map. We therefore weight each neighborhood by its mass $\pi$ in three parts of the layout. First, through the factor $\Pi^{-1}$ in $L$, we divide the rate at which the walk leaves each neighborhood by its mass, so that in the long run the walk spends in each neighborhood a share of its time equal to the neighborhood's share of the points, as a walk over the points themselves would. Second, when we compute the distances between potentials, we weight each neighborhood by $\sqrt{\pi}$, which is equivalent to repeating the neighborhood once per point. Third, in multidimensional scaling, we weight each pair of neighborhoods by the product of their masses, which is equivalent to applying multidimensional scaling to all points.

\subsection{Stability}\label{app:atlas-stability}

\begin{table}[t]
\centering
\footnotesize
\setlength{\tabcolsep}{4pt}
\begin{tabular}{@{}lcc@{}}
\toprule
Change & \LArTPC & \sPHENIX \\
\midrule
$K = 512$ & 0.88 / 0.89 & 0.83 / 0.80 \\
$K = 1{,}024$ & 0.97 / 0.92 & 0.92 / 0.89 \\
$K = 2{,}048$ & 0.99 / 0.99 & 0.97 / 0.94 \\
$K = 8{,}192$ & 0.99 / 0.87 & 0.90 / 0.88 \\
Class-balanced neighborhoods & 0.99 / 0.79 & 0.98 / 0.91 \\
\midrule
New $k$-means seed & 0.995 & 0.992 \\
8, 16, 48 axes per neighborhood & 0.99, 0.99, 0.99 & 0.98, 0.99, 0.99 \\
Nearest-mean assignment & 0.94 & 0.85 \\
Stage 1 only & 0.01 & 0.09 \\
Stage 2 only & 0.02 & 0.09 \\
Stage 3 only & 0.70 & 0.10 \\
Stage 4 only & 0.75 & 0.79 \\
\bottomrule
\end{tabular}
\caption{\textbf{Stability of the manifold map.} We provide Procrustes $R^2$ between the map built with one change and the final map ($K = 4{,}096$, 24 axes, all stages), on the same 200K validation points. Top: our map / PHATE's map, each compared with its own version at $K = 4{,}096$. Bottom: our map only.}
\label{tab:atlas-stability}
\end{table}

We compare two maps by their Procrustes $R^2$~\cite{Gower_Procrustes} on 200K validation points, which is 1 when the maps are identical up to rotation, reflection, scale and shift. On \LArTPC, our map stays the same ($R^2 \ge 0.99$) when we refit it, halve or double the number of neighborhoods, change the number of axes per neighborhood, or fit the neighborhoods on class-balanced points, whereas PHATE's map changes when the number of neighborhoods doubles ($R^2 = 0.87$) or the points are balanced by class ($R^2 = 0.79$; Table~\ref{tab:atlas-stability}). On \sPHENIX, both maps change more with the number of neighborhoods, ours less than PHATE's at every number of neighborhoods. Without the size of each neighborhood, the $R^2$ between the \sPHENIX map and its refit falls from 0.99 to 0.88. When we change $\tau$ by a factor of 2, the $R^2$ stays above 0.94. All four stages are needed: a map built on a single stage differs from the final map, and maps built on the first or second stage alone, or on the third stage alone for \sPHENIX, consist of disconnected islands.

\subsection{Truth probes}

\begin{figure*}[p]\centering
\includegraphics[width=\textwidth]{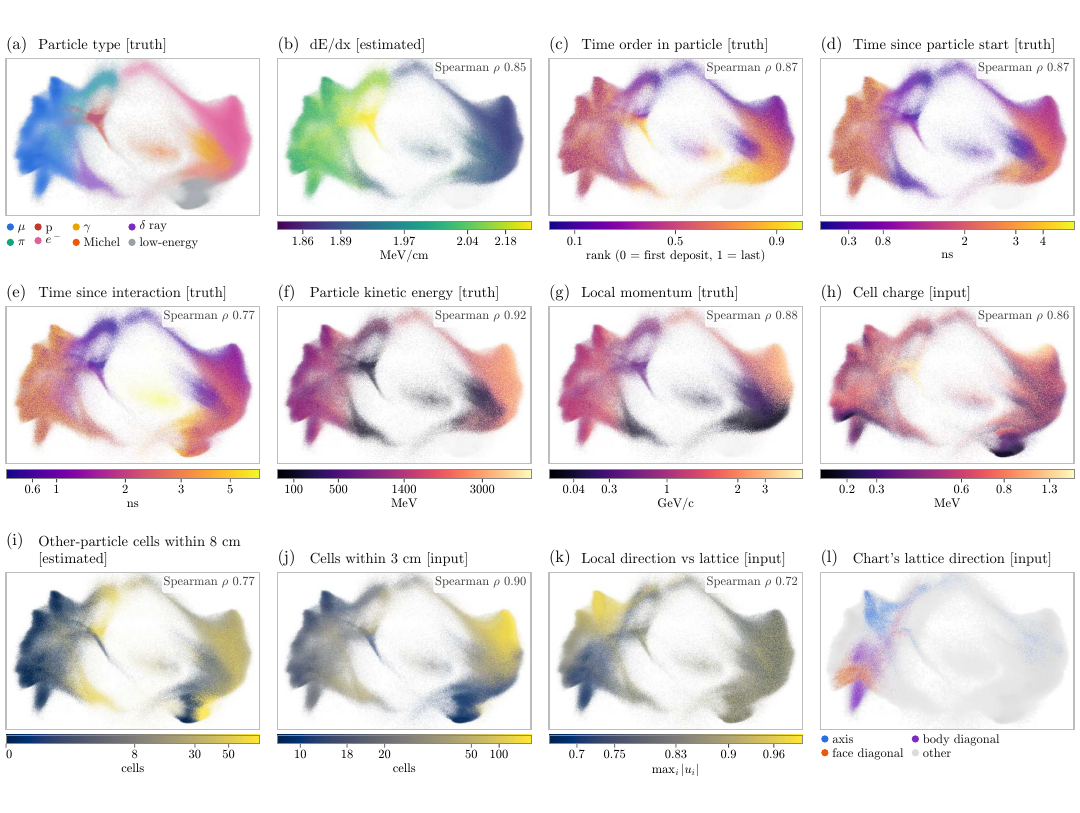}
\caption{\textbf{Extended \LArTPC manifold map.} The tag in each panel title describes the kind of quantity: [truth]  simulation truth, [estimated] computed from the truth, [input] a statistic of the encoder's input. Continuous quantities are shown on a rank color scale with ticks at quantiles, and undefined values are grey. $\rho$ is computed from a linear probe on the features, not from the map; for panels shared with Fig.~\ref{fig:atlas}, $\rho$ is the value given there, and categorical panels have no $\rho$. In (l), each point is colored by the lattice direction of its neighborhood (chart in the figure) when the neighborhood holds points within $10^\circ$ of a lattice axis, face diagonal or body diagonal at least 3 times more often than random directions would.}\label{fig:truth_lar}
\end{figure*}

\begin{figure*}[p]\centering
\includegraphics[width=\textwidth]{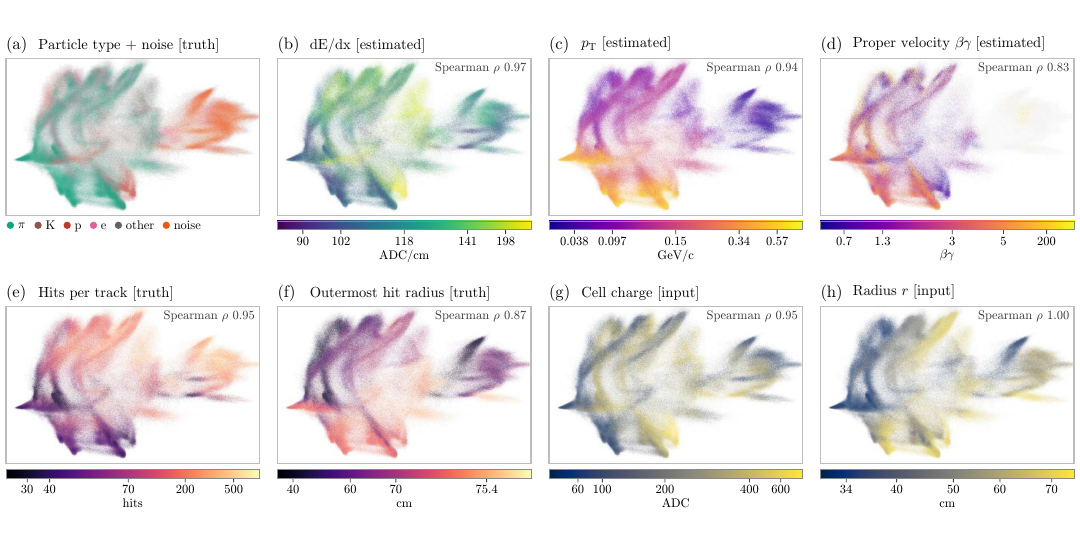}
\caption{\textbf{The \sPHENIX manifold map of Fig.~\ref{fig:atlas}, colored by further quantities}, as in Fig.~\ref{fig:truth_lar}.}\label{fig:truth_sphenix}
\end{figure*}

\begin{figure*}[p]\centering
\includegraphics[width=\textwidth]{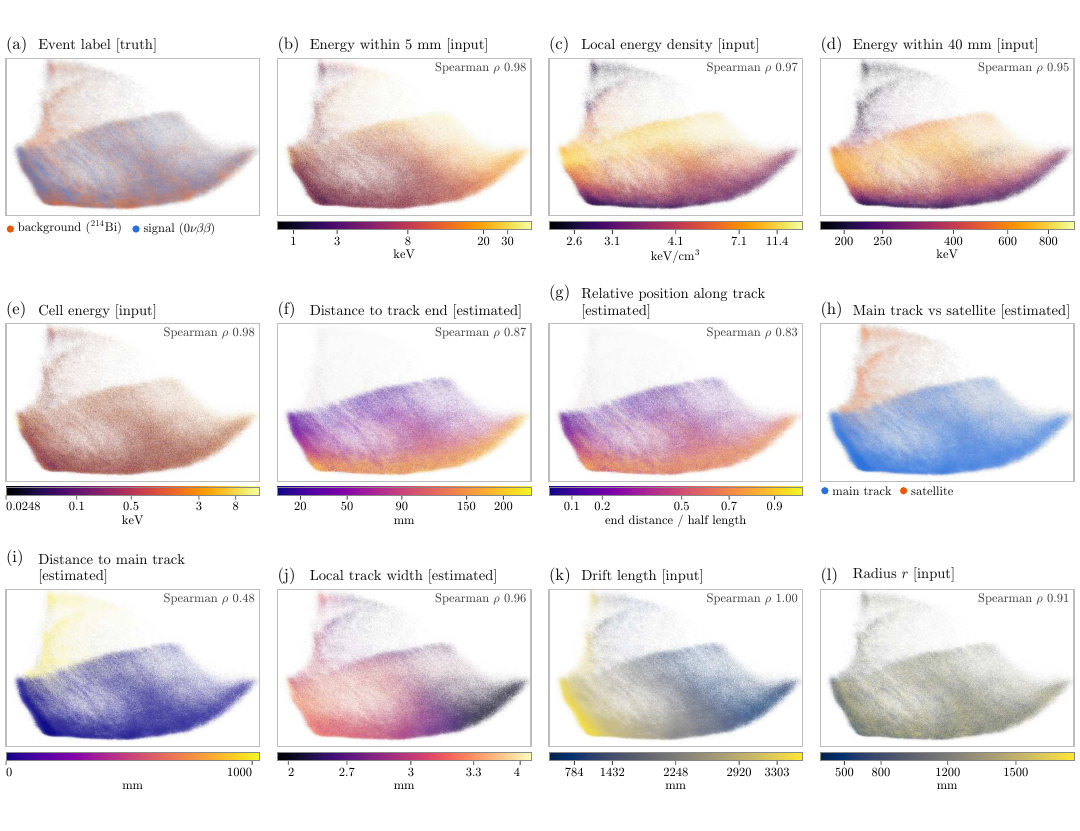}
\caption{\textbf{The \NEXT manifold map of Fig.~\ref{fig:atlas}, colored by further quantities}, as in Fig.~\ref{fig:truth_lar}.}\label{fig:truth_next}
\end{figure*}

\begin{figure*}[p]\centering
\includegraphics[width=\textwidth]{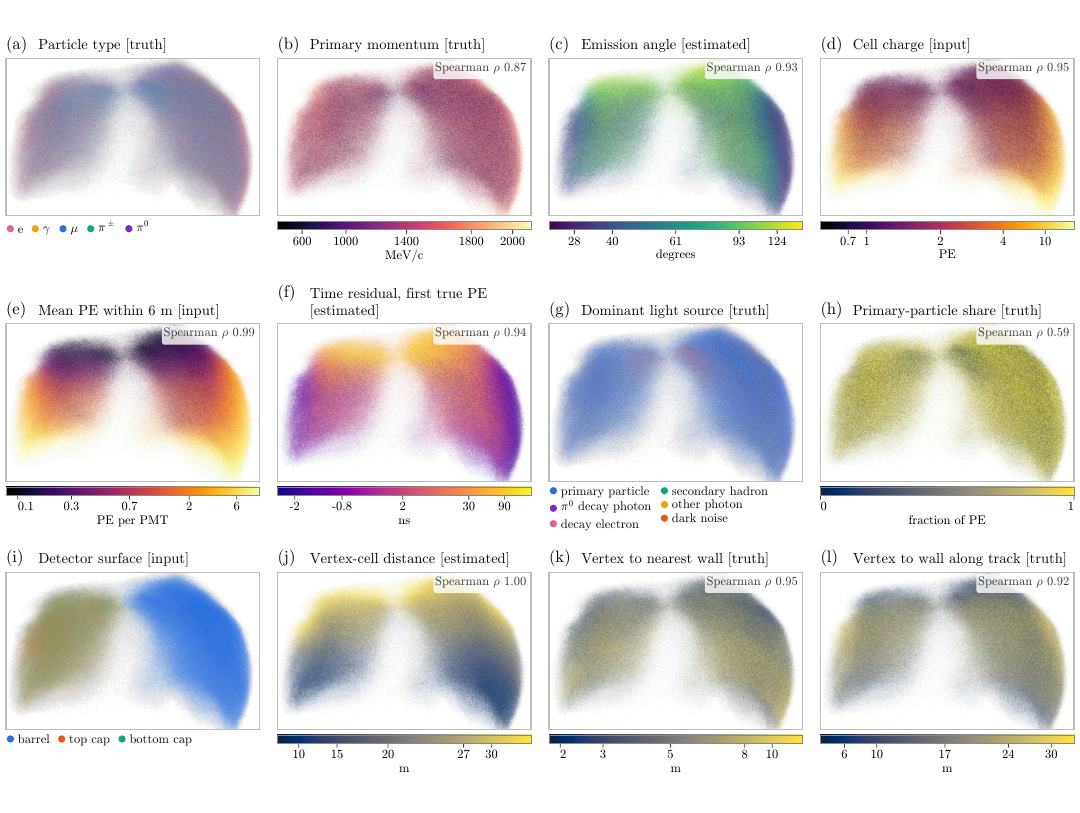}
\caption{\textbf{The WC manifold map of Fig.~\ref{fig:atlas}, colored by further quantities}, as in Fig.~\ref{fig:truth_lar}. The momentum and the two distances to the walls are the same for every point of an image, so their $\rho$ can come from the rest of the image rather than from the point itself.}\label{fig:truth_wc}
\end{figure*}

In Figs.~\ref{fig:truth_lar}--\ref{fig:truth_wc}, we show the map of each detector colored by more quantities than in Fig.~\ref{fig:atlas}: quantities of the simulation truth, quantities estimated from the truth, and statistics of the encoder's input, none of which are used to build the map.

The value $\rho$ in Fig.~\ref{fig:atlas} and Figs.~\ref{fig:truth_lar}--\ref{fig:truth_wc} is the Spearman correlation between a quantity and the prediction of a linear ridge probe on the same features of the validation points, with the ridge strength chosen on half of the images and $\rho$ measured on the other half. A high $\rho$ therefore means that the features contain the quantity linearly, whereas whether the map is arranged by the quantity is seen in the colors. We compute $\rho$ over points, except for the time order within a particle, computed within each particle of at least 10 points and averaged over particles, and for $dE/dx$, computed per particle in \LArTPC and per track in \sPHENIX, for tracks with at least 12 hits and a transverse arc of at least 10\,cm. \sPHENIX files store no momentum, so we obtain $p_\mathrm{T}$ from a circle fit, in a field of 1.4\,T, to the hits of each simulated track, or from a helix fit for tracks that close a turn. Panels tagged [input] are colored by a statistic of the encoder's input; for these, a high $\rho$ means that the quantity is available in the features, not that the map is organized by it.

\paragraph{Lattice direction in \LArTPC.} Although tracks in \LArTPC have no preferred direction, muon points are separated on the map by their direction relative to the voxel lattice. For each point, we take the local track direction $u$, the first principal axis of its 12 nearest points, and compute $\max_i|u_i|$, which is 1 along a lattice axis, 0.71 along a face diagonal and 0.58 along a body diagonal. 85\% of the variance of $\max_i|u_i|$ is explained by the neighborhood of each point, against 0.2\% when the neighborhoods are shuffled, and 121 neighborhoods, almost all of muons and mostly along face or body diagonals, hold at least half of their points within $10^\circ$ of one lattice direction, against none when shuffled. Shower points are not sorted this way. Both are shown in Fig.~\ref{fig:truth_lar}(k,\,l).

\FloatBarrier
\section{Resource usage}
\label{app:resources}

\paragraph{Pretraining.} Pretraining one PT$_\text{1M}$ encoder over the full 20M image presentations costs approximately 450, 455, 430 and 280 A100-hours for the \LArTPC, \sPHENIX, \textcolor{wc}{WC} and \NEXT detectors, respectively, or 1{,}600 A100-hours for all four. For this work, we utilize NVIDIA A100s, H200s, and a number of L40S GPUs.

\paragraph{Fine-tuning.} Training the most expressive head that we use on each frozen encoder with every labeled image, with image mixing, costs approximately 188 A100-hours for the \LArTPC (1M images), 213 for \sPHENIX (70K images), 133 for \textcolor{wc}{WC} (982K images) and 14 for \NEXT (417K images). Note that \NEXT trains on unaugmented images without image mixing, whereas the others do.

\begin{figure*}[t]
    \centering
    \includegraphics[width=\textwidth]{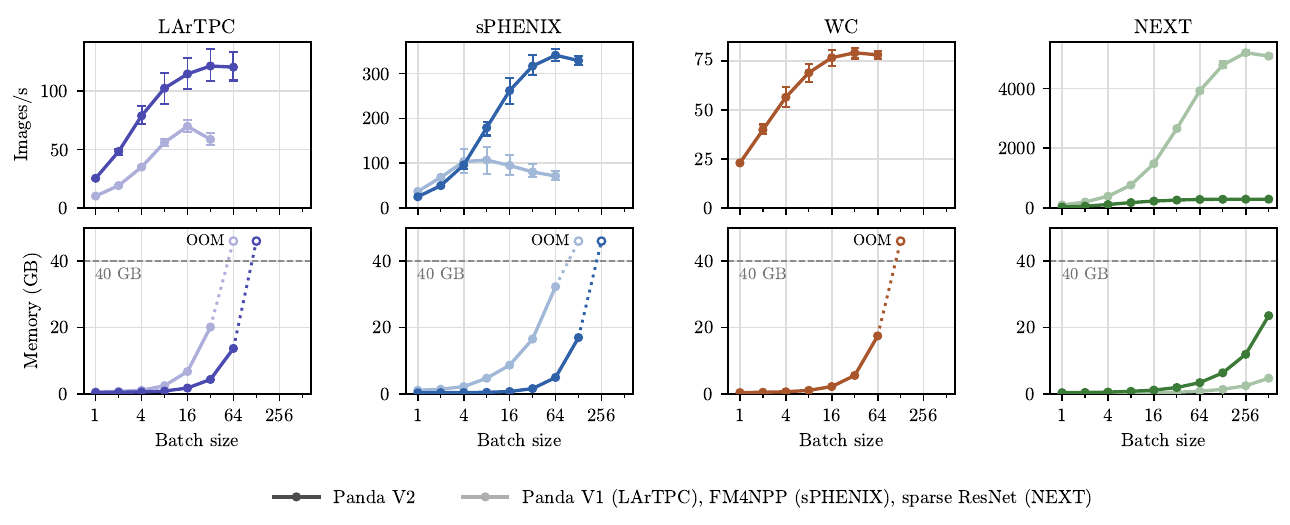}
    \caption{\textbf{Inference throughput and memory on one 40GB A100 GPU.} \textbf{Top:} images processed per second against batch size, for the forward pass of the encoder and the task head; error bars span the middle 68\% of forward passes over 100 batches. \textbf{Bottom:} peak GPU memory against batch size.}
    \label{fig:app-inference}
\end{figure*}

\paragraph{Inference.} We also measure the inference throughput and peak memory of Panda V2 against the baselines of the main text: Panda V1 on the \LArTPC, FM4NPP with Panda Detector on \sPHENIX, and the supervised sparse ResNet on \NEXT; on the \textcolor{wc}{WC} detector, we measure Panda V2 alone (Fig.~\ref{fig:app-inference}). We time the forward pass of the encoder and the task head on one 40GB A100 GPU in BF16 over 100 batches of test images, doubling the batch size from 1 until the model runs out of memory. Panda V2 processes up to 120 images per second on the \LArTPC, against 69 for Panda V1, and 340 on \sPHENIX, against 97 for FM4NPP; it also fits twice as many images in memory on both (64 vs.\ 32 and 128 vs.\ 64). On the \textcolor{wc}{WC} detector, Panda V2 processes up to 79 images per second and fits batches of up to 64 images; for reference, fiTQun on average takes greater than one minute per image. On \NEXT, however, the sparse ResNet is far faster than Panda V2, processing up to 5{,}200 images per second against 290, even in FP32, which its sparse convolutions require.

\section{Data release}

We release every pretrained encoder of this work, PT$_\text{1M}$ and PT$_K$ for all four detectors, together with snapshots of their weights taken every 500K images seen during pretraining (40 over the 20M images of PT$_\text{1M}$), such that future work may study the evolution of the representations over training. We also make available the synthetically pretrained Panda Detector V2 decoder as well as the WAND dataset. The weights and the dataset are hosted on HuggingFace and may be loaded directly into \texttt{particle-imaging-models},\footnote{\url{https://github.com/deeplearnphysics/particle-imaging-models}} the library with which we ran every experiment of this work and which supports scalable pretraining, fine-tuning and inference across heterogeneous hardware. The models themselves are able to be downloaded and loaded into an interactive Python environment with a single line: \texttt{import pimm; model = pimm.from\_pretrained(`org/repo')}.

\section{Extended result tables}
\label{app:results}

In this section, we report the results for the models of the main text in more detail. We additionally introduce parameter-efficient fine-tuning in the form of low-rank adaptation (LoRA)~\cite{hu2021loralowrankadaptationlarge} (rank 16, $\alpha$=32). We provide 95\% bootstrap intervals, and in each column of a table, the best values are shaded, darker for better, with the best made bold.

\subsection{\LArTPC}

\paragraph{Semantic segmentation.} In Table~\ref{tab:app-lar-sem}, we extend Table~\ref{tab:semseg_lar} of the main text with the 100K label count, the LoRA rows, and 95\% intervals. In Table~\ref{tab:app-lar-perclass-sem}, we break down the scores with 1K and with 1M labeled images by motif; the macro average is the mean over the five motifs. The models differ most on the rare motifs: with 1K labeled images, PTv3 trained from scratch reaches 40.0 $F_1$ on Michel electrons and 62.0 on delta rays, against 97.3 and 94.5 for Panda V2 fine-tuned with LoRA, and most of the gap that remains with 1M labeled images between the linear layer and LoRA are in delta rays (93.1 against 97.0). We investigate a test-time inference strategy to improve delta ray performance in \S\ref{app:ttt}.

\paragraph{Particle instance segmentation.} In Table~\ref{tab:app-lar-part}, we extend Table~\ref{tab:instseg_lar} of the main text in the same way.

\paragraph{Interaction instance segmentation.} In Table~\ref{tab:app-lar-int}, we report a task that is not in the main text: grouping the points of an image into interactions, \ie the groups of particles that share a vertex (\S\ref{app:data_lar}), scored with the same ARI over points grouped by interaction instead of by particle.



\subsection{\sPHENIX}

\paragraph{Semantic segmentation.} In Table~\ref{tab:app-sph-sem}, we extend Table~\ref{tab:semseg_sph} of the main text with the LoRA results and 95\% intervals. In Table~\ref{tab:app-sph-pid}, we report particle identification and noise tagging by class with 1K and with 70K labeled images. 

\paragraph{Particle instance segmentation.} Similarly, in Table~\ref{tab:app-sph-part}, we extend Table~\ref{tab:instseg_sph} with LoRA results and 95\% CI. In Table~\ref{tab:app-sph-track}, we report track finding with 1K and with 70K labeled images in the metrics of~\cite{park2025fm4npp}: track efficiency and purity as well as spacepoint efficiency and purity.


\subsection{\WC}

\paragraph{Single-particle reconstruction.} In Table~\ref{tab:app-wc}, we extend Table~\ref{tab:wc} of the main text with the LoRA results and 95\% intervals.

\paragraph{Ring reconstruction.} In Table~\ref{tab:app-wc-rings}, we compare the reconstruction of every ring of an image with fiTQun, pooled over the single-particle and two-particle samples, with 1K and with 982K labeled images, by ring finding (the $F_1$ of matched rings regardless of type), ring classification (the fraction of matched rings assigned their true type), and the particle $F_1$, in which a match counts only if its type is also correct (\S\ref{app:data_wc}).


\subsection{\NEXT}

\paragraph{Signal classification.} In Table~\ref{tab:app-next}, we extend Table~\ref{tab:next} of the main text with the LoRA results and 95\% intervals.

\subsection{Cross-detector transfer}

In Table~\ref{tab:app-transfer}, we extend Table~\ref{tab:transfer} of the main text with 95\% intervals and with the same transfer for encoders fine-tuned with LoRA. Every cell of the frozen block, including the diagonal, comes from a run of the transfer study and not from the in-domain tables. Because the input channels differ between detectors, the transfer runs train the encoder's input layer and let its batch-normalization statistics adapt, together with the decoder, whereas the frozen encoders of the in-domain tables keep both fixed. The diagonal is therefore an in-domain control under the transfer protocol, and it differs slightly from the in-domain tables: 96.0 against 95.7 ARI on \LArTPC (Table~\ref{tab:instseg_lar}) and 96.2 against 96.0 on \sPHENIX (Table~\ref{tab:instseg_sph}). On \NEXT, the transfer runs also train a classification head on mean-pooled features with one draw of 1K labeled images, whereas Table~\ref{tab:next} reports a logistic regression on the frozen features of the same draw (5.53 against 5.13 f.o.m.). On WC, the column pools particle $F_1$ over all ten of fiTQun's single- and two-particle samples, including the photon sample, so its fiTQun baseline (33.9) and diagonal (42.5) differ from Table~\ref{tab:app-wc-rings}, which pools the nine samples without photons and uses the in-domain run (35.4 and 42.0).



\FloatBarrier

\begin{table*}[p]
\centering
\scriptsize
\setlength{\tabcolsep}{2.4pt}
\begin{tabular*}{425pt}[t]{l@{\extracolsep{\fill}}r@{\extracolsep{0pt}\hspace{2\tabcolsep}}ccccc@{\hspace{0.5\tabcolsep}}c@{\hspace{0.5\tabcolsep}}ccccc}
\toprule
\multicolumn{2}{l}{\setbox0\hbox{Sem.\ Seg.\ (mF$_1$)}\rlap{\smash{\raisebox{-3.6pt}{\rule{\wd0}{\cmidrulewidth}}}}\box0} & \multicolumn{5}{c}{PT$_K$ + FT$_K$} &  & \multicolumn{5}{c}{PT$_\text{1M}$ + FT$_K$} \\
\cmidrule(lr){3-7} \cmidrule(lr){9-13}
Method / $K$ & Param. & 100 & 1K & 10K & 100K & 1M &  & 100 & 1K & 10K & 100K & 1M \\
\midrule
\multicolumn{13}{l}{\textit{Supervised}} \\
\hspace{0.5em}PTv3~\cite{wu2024pointtransformerv3simpler} & 107.1M & 63.9{\tiny\color{black!50}$\,\pm$0.05} & 77.9{\tiny\color{black!50}$\,\pm$0.10} & 94.0{\tiny\color{black!50}$\,\pm$0.06} & 97.4{\tiny\color{black!50}$\,\pm$0.05} & \cellcolor[HTML]{F4F4FF}98.1{\tiny\color{black!50}$\,\pm$0.03} & \phantom{\raisebox{0.12ex}{\tiny=}} & \llap{\raisebox{0.12ex}{\tiny=}\hspace{0.5\tabcolsep}}63.9{\tiny\color{black!50}$\,\pm$0.05} & 77.9{\tiny\color{black!50}$\,\pm$0.10} & 94.0{\tiny\color{black!50}$\,\pm$0.06} & 97.4{\tiny\color{black!50}$\,\pm$0.05} & \cellcolor[HTML]{F4F4FF}98.1{\tiny\color{black!50}$\,\pm$0.03} \\
\hspace{0.5em}Panda V2 & 53.7M & \cellcolor[HTML]{D6D6FF}\textbf{90.4}{\tiny\color{black!50}$\,\pm$0.08} & \cellcolor[HTML]{E3E3FF}95.5{\tiny\color{black!50}$\,\pm$0.05} & \cellcolor[HTML]{E3E3FF}97.4{\tiny\color{black!50}$\,\pm$0.04} & \cellcolor[HTML]{E3E3FF}98.1{\tiny\color{black!50}$\,\pm$0.03} & \cellcolor[HTML]{E3E3FF}98.2{\tiny\color{black!50}$\,\pm$0.03} & \phantom{\raisebox{0.12ex}{\tiny=}} & \llap{\raisebox{0.12ex}{\tiny=}\hspace{0.5\tabcolsep}}90.4{\tiny\color{black!50}$\,\pm$0.08} & 95.5{\tiny\color{black!50}$\,\pm$0.05} & \cellcolor[HTML]{F4F4FF}97.4{\tiny\color{black!50}$\,\pm$0.04} & \cellcolor[HTML]{E3E3FF}98.1{\tiny\color{black!50}$\,\pm$0.03} & \cellcolor[HTML]{E3E3FF}98.2{\tiny\color{black!50}$\,\pm$0.03} \\
\midrule
\multicolumn{13}{l}{\textit{Self-supervised}} \\
\hspace{0.5em}PoLAr-MAE (dec.)~\cite{polarmae} & 4.4M  & 71.2 & 80.6 & 84.6 & 83.4 & 83.4 &  & -- & -- & -- & -- & 83.4 \\
\hspace{0.5em}Panda V1 (lin.) & \llap{\raisebox{0.12ex}{\tiny$<$}\kern0.5pt}0.01M & 78.0{\tiny\color{black!50}$\,\pm$0.09} & 89.9{\tiny\color{black!50}$\,\pm$0.07} & 92.9{\tiny\color{black!50}$\,\pm$0.05} & 93.5{\tiny\color{black!50}$\,\pm$0.05} & 93.7{\tiny\color{black!50}$\,\pm$0.04} &  & 92.2{\tiny\color{black!50}$\,\pm$0.05} & 93.5{\tiny\color{black!50}$\,\pm$0.04} & 93.7{\tiny\color{black!50}$\,\pm$0.04} & 93.7{\tiny\color{black!50}$\,\pm$0.04} & 93.7{\tiny\color{black!50}$\,\pm$0.04} \\
\hspace{0.5em}Panda V2 (lin.) & \llap{\raisebox{0.12ex}{\tiny$<$}\kern0.5pt}0.01M & 82.3{\tiny\color{black!50}$\,\pm$0.09} & 94.1{\tiny\color{black!50}$\,\pm$0.06} & 96.7{\tiny\color{black!50}$\,\pm$0.04} & 97.2{\tiny\color{black!50}$\,\pm$0.03} & 97.1{\tiny\color{black!50}$\,\pm$0.04} &  & \cellcolor[HTML]{F4F4FF}95.8{\tiny\color{black!50}$\,\pm$0.04} & \cellcolor[HTML]{F4F4FF}96.9{\tiny\color{black!50}$\,\pm$0.04} & 97.1{\tiny\color{black!50}$\,\pm$0.04} & 97.1{\tiny\color{black!50}$\,\pm$0.04} & 97.1{\tiny\color{black!50}$\,\pm$0.04} \\
\hspace{0.5em}Panda V2 (dec.) & 0.3M & \cellcolor[HTML]{F4F4FF}83.0{\tiny\color{black!50}$\,\pm$0.09} & \cellcolor[HTML]{F4F4FF}94.6{\tiny\color{black!50}$\,\pm$0.06} & \cellcolor[HTML]{F4F4FF}97.2{\tiny\color{black!50}$\,\pm$0.04} & \cellcolor[HTML]{F4F4FF}97.8{\tiny\color{black!50}$\,\pm$0.03} & 97.8{\tiny\color{black!50}$\,\pm$0.03} &  & \cellcolor[HTML]{E3E3FF}95.9{\tiny\color{black!50}$\,\pm$0.05} & \cellcolor[HTML]{E3E3FF}97.2{\tiny\color{black!50}$\,\pm$0.04} & \cellcolor[HTML]{E3E3FF}97.6{\tiny\color{black!50}$\,\pm$0.03} & \cellcolor[HTML]{F4F4FF}97.8{\tiny\color{black!50}$\,\pm$0.03} & 97.8{\tiny\color{black!50}$\,\pm$0.03} \\
\hspace{0.5em}Panda V2 (LoRA) & 2.6M & \cellcolor[HTML]{E3E3FF}86.1{\tiny\color{black!50}$\,\pm$0.08} & \cellcolor[HTML]{D6D6FF}\textbf{95.9}{\tiny\color{black!50}$\,\pm$0.05} & \cellcolor[HTML]{D6D6FF}\textbf{97.8}{\tiny\color{black!50}$\,\pm$0.04} & \cellcolor[HTML]{D6D6FF}\textbf{98.4}{\tiny\color{black!50}$\,\pm$0.03} & \cellcolor[HTML]{D6D6FF}\textbf{98.5}{\tiny\color{black!50}$\,\pm$0.02} &  & \cellcolor[HTML]{D6D6FF}\textbf{96.5}{\tiny\color{black!50}$\,\pm$0.04} & \cellcolor[HTML]{D6D6FF}\textbf{97.5}{\tiny\color{black!50}$\,\pm$0.04} & \cellcolor[HTML]{D6D6FF}\textbf{98.0}{\tiny\color{black!50}$\,\pm$0.03} & \cellcolor[HTML]{D6D6FF}\textbf{98.4}{\tiny\color{black!50}$\,\pm$0.03} & \cellcolor[HTML]{D6D6FF}\textbf{98.5}{\tiny\color{black!50}$\,\pm$0.02} \\
\hspace{0.5em}\color{black!45}PoLAr-MAE (full)~\cite{polarmae} & 24.4M \color{black!45} & \color{black!45}73.5 & \color{black!45}84.0 & \color{black!45}86.5 & \color{black!45}86.5 & \color{black!45}85.8 & \color{black!45} & \color{black!45}-- & \color{black!45}-- & \color{black!45}-- & \color{black!45}-- & \color{black!45}85.8 \\
\hspace{0.5em}\color{black!45}Panda V1 (full) & \color{black!45}107.1M & \color{black!45}84.3{\tiny\color{black!50}$\,\pm$0.09} & \color{black!45}93.4{\tiny\color{black!50}$\,\pm$0.07} & \color{black!45}96.2{\tiny\color{black!50}$\,\pm$0.05} & \color{black!45}97.2{\tiny\color{black!50}$\,\pm$0.04} & \color{black!45}98.2{\tiny\color{black!50}$\,\pm$0.03} & \color{black!45} & \color{black!45}92.6{\tiny\color{black!50}$\,\pm$0.05} & \color{black!45}95.1{\tiny\color{black!50}$\,\pm$0.05} & \color{black!45}96.4{\tiny\color{black!50}$\,\pm$0.04} & \color{black!45}-- & \color{black!45}98.2{\tiny\color{black!50}$\,\pm$0.03} \\
\bottomrule
\end{tabular*}
\caption{\label{tab:app-lar-sem}\textbf{Semantic segmentation on LArTPC.} Equivalent to Table~\ref{tab:semseg_lar} of the main text, with 95\% CI, a LoRA row and the 100K label count added.}
\end{table*}

\begin{table*}[p]
\centering
\scriptsize
\setlength{\tabcolsep}{5pt}
\begin{tabular*}{330pt}{@{}l@{\extracolsep{\fill}}c@{\extracolsep{0pt}\hspace{2\tabcolsep}}ccccc}
\toprule
Sem.\ Seg.\ (F$_1$) & Macro & Shower & Track & Michel & Delta & LED \\
\midrule
\multicolumn{7}{@{}l}{\textbf{PT$_\text{1M}$ + FT$_\text{1K}$}} \\
\multicolumn{7}{@{}l}{\textit{Supervised}} \\
\hspace{0.5em}PTv3~\cite{wu2024pointtransformerv3simpler} & 77.9{\tiny\color{black!50}$\,\pm$0.10} & 96.8{\tiny\color{black!50}$\,\pm$0.03} & 97.3{\tiny\color{black!50}$\,\pm$0.02} & 40.0{\tiny\color{black!50}$\,\pm$0.48} & 62.0{\tiny\color{black!50}$\,\pm$0.17} & 93.6{\tiny\color{black!50}$\,\pm$0.02} \\
\hspace{0.5em}Panda V2 & 95.5{\tiny\color{black!50}$\,\pm$0.05} & \cellcolor[HTML]{F4F4FF}99.0{\tiny\color{black!50}$\,\pm$0.01} & \cellcolor[HTML]{F4F4FF}99.5{\tiny\color{black!50}$\,\pm$0.01} & 93.7{\tiny\color{black!50}$\,\pm$0.23} & 89.6{\tiny\color{black!50}$\,\pm$0.11} & 95.8{\tiny\color{black!50}$\,\pm$0.02} \\
\arrayrulecolor{black!20}\cmidrule[0.3pt]{1-7}\arrayrulecolor{black}
\multicolumn{7}{@{}l}{\textit{Self-supervised}} \\
\hspace{0.5em}Panda V1 (lin.) & 93.5{\tiny\color{black!50}$\,\pm$0.04} & 98.7{\tiny\color{black!50}$\,\pm$0.01} & 99.0{\tiny\color{black!50}$\,\pm$0.01} & 94.3{\tiny\color{black!50}$\,\pm$0.20} & 80.8{\tiny\color{black!50}$\,\pm$0.10} & 94.5{\tiny\color{black!50}$\,\pm$0.02} \\
\hspace{0.5em}Panda V2 (lin.) & \cellcolor[HTML]{F4F4FF}96.9{\tiny\color{black!50}$\,\pm$0.04} & \cellcolor[HTML]{E3E3FF}99.2{\tiny\color{black!50}$\,\pm$0.01} & \cellcolor[HTML]{E3E3FF}99.6{\tiny\color{black!50}$\,\pm$0.00} & \cellcolor[HTML]{F4F4FF}96.5{\tiny\color{black!50}$\,\pm$0.17} & \cellcolor[HTML]{F4F4FF}92.9{\tiny\color{black!50}$\,\pm$0.07} & \cellcolor[HTML]{F4F4FF}96.1{\tiny\color{black!50}$\,\pm$0.02} \\
\hspace{0.5em}Panda V2 (dec.) & \cellcolor[HTML]{E3E3FF}97.2{\tiny\color{black!50}$\,\pm$0.04} & \cellcolor[HTML]{E3E3FF}99.2{\tiny\color{black!50}$\,\pm$0.01} & \cellcolor[HTML]{D6D6FF}\textbf{99.7}{\tiny\color{black!50}$\,\pm$0.00} & \cellcolor[HTML]{E3E3FF}97.1{\tiny\color{black!50}$\,\pm$0.16} & \cellcolor[HTML]{E3E3FF}93.9{\tiny\color{black!50}$\,\pm$0.06} & \cellcolor[HTML]{E3E3FF}96.3{\tiny\color{black!50}$\,\pm$0.02} \\
\hspace{0.5em}Panda V2 (LoRA) & \cellcolor[HTML]{D6D6FF}\textbf{97.5}{\tiny\color{black!50}$\,\pm$0.04} & \cellcolor[HTML]{D6D6FF}\textbf{99.3}{\tiny\color{black!50}$\,\pm$0.01} & \cellcolor[HTML]{D6D6FF}\textbf{99.7}{\tiny\color{black!50}$\,\pm$0.00} & \cellcolor[HTML]{D6D6FF}\textbf{97.3}{\tiny\color{black!50}$\,\pm$0.17} & \cellcolor[HTML]{D6D6FF}\textbf{94.5}{\tiny\color{black!50}$\,\pm$0.06} & \cellcolor[HTML]{D6D6FF}\textbf{96.8}{\tiny\color{black!50}$\,\pm$0.02} \\
\color{black!45}\hspace{0.5em}Panda V1 (full) & \color{black!45}95.1{\tiny\color{black!50}$\,\pm$0.05} & \color{black!45}99.0{\tiny\color{black!50}$\,\pm$0.01} & \color{black!45}99.3{\tiny\color{black!50}$\,\pm$0.01} & \color{black!45}94.6{\tiny\color{black!50}$\,\pm$0.21} & \color{black!45}87.0{\tiny\color{black!50}$\,\pm$0.09} & \color{black!45}95.5{\tiny\color{black!50}$\,\pm$0.02} \\
\midrule
\multicolumn{7}{@{}l}{\textbf{PT$_\text{1M}$ + FT$_\text{1M}$}} \\
\multicolumn{7}{@{}l}{\textit{Supervised}} \\
\hspace{0.5em}PTv3~\cite{wu2024pointtransformerv3simpler} & \cellcolor[HTML]{F4F4FF}98.1{\tiny\color{black!50}$\,\pm$0.03} & \cellcolor[HTML]{D6D6FF}\textbf{99.4}{\tiny\color{black!50}$\,\pm$0.01} & \cellcolor[HTML]{D6D6FF}\textbf{99.8}{\tiny\color{black!50}$\,\pm$0.00} & 97.7{\tiny\color{black!50}$\,\pm$0.14} & \cellcolor[HTML]{E3E3FF}96.6{\tiny\color{black!50}$\,\pm$0.06} & \cellcolor[HTML]{F4F4FF}97.0{\tiny\color{black!50}$\,\pm$0.02} \\
\hspace{0.5em}Panda V2 & \cellcolor[HTML]{E3E3FF}98.2{\tiny\color{black!50}$\,\pm$0.03} & \cellcolor[HTML]{D6D6FF}\textbf{99.4}{\tiny\color{black!50}$\,\pm$0.00} & \cellcolor[HTML]{D6D6FF}\textbf{99.8}{\tiny\color{black!50}$\,\pm$0.00} & \cellcolor[HTML]{E3E3FF}98.4{\tiny\color{black!50}$\,\pm$0.11} & \cellcolor[HTML]{F4F4FF}95.9{\tiny\color{black!50}$\,\pm$0.05} & \cellcolor[HTML]{D6D6FF}\textbf{97.4}{\tiny\color{black!50}$\,\pm$0.02} \\
\arrayrulecolor{black!20}\cmidrule[0.3pt]{1-7}\arrayrulecolor{black}
\multicolumn{7}{@{}l}{\textit{Self-supervised}} \\
\hspace{0.5em}PoLAr-MAE (dec.)~\cite{polarmae} & 83.4 & 96.8 & 97.3 & 73.9 & 62.8 & 86.2 \\
\hspace{0.5em}Panda V1 (lin.) & 93.7{\tiny\color{black!50}$\,\pm$0.04} & 98.7{\tiny\color{black!50}$\,\pm$0.01} & \cellcolor[HTML]{F4F4FF}99.0{\tiny\color{black!50}$\,\pm$0.01} & 95.1{\tiny\color{black!50}$\,\pm$0.19} & 80.8{\tiny\color{black!50}$\,\pm$0.11} & 94.6{\tiny\color{black!50}$\,\pm$0.02} \\
\hspace{0.5em}Panda V2 (lin.) & 97.1{\tiny\color{black!50}$\,\pm$0.04} & \cellcolor[HTML]{F4F4FF}99.2{\tiny\color{black!50}$\,\pm$0.01} & \cellcolor[HTML]{E3E3FF}99.6{\tiny\color{black!50}$\,\pm$0.00} & 97.3{\tiny\color{black!50}$\,\pm$0.16} & 93.1{\tiny\color{black!50}$\,\pm$0.07} & 96.2{\tiny\color{black!50}$\,\pm$0.02} \\
\hspace{0.5em}Panda V2 (dec.) & 97.8{\tiny\color{black!50}$\,\pm$0.03} & \cellcolor[HTML]{E3E3FF}99.3{\tiny\color{black!50}$\,\pm$0.00} & \cellcolor[HTML]{D6D6FF}\textbf{99.8}{\tiny\color{black!50}$\,\pm$0.00} & \cellcolor[HTML]{F4F4FF}98.0{\tiny\color{black!50}$\,\pm$0.14} & 95.1{\tiny\color{black!50}$\,\pm$0.06} & 96.8{\tiny\color{black!50}$\,\pm$0.02} \\
\hspace{0.5em}Panda V2 (LoRA) & \cellcolor[HTML]{D6D6FF}\textbf{98.5}{\tiny\color{black!50}$\,\pm$0.02} & \cellcolor[HTML]{D6D6FF}\textbf{99.4}{\tiny\color{black!50}$\,\pm$0.00} & \cellcolor[HTML]{D6D6FF}\textbf{99.8}{\tiny\color{black!50}$\,\pm$0.00} & \cellcolor[HTML]{D6D6FF}\textbf{98.6}{\tiny\color{black!50}$\,\pm$0.11} & \cellcolor[HTML]{D6D6FF}\textbf{97.0}{\tiny\color{black!50}$\,\pm$0.04} & \cellcolor[HTML]{E3E3FF}97.3{\tiny\color{black!50}$\,\pm$0.02} \\
\color{black!45}\hspace{0.5em}PoLAr-MAE (full)~\cite{polarmae} & \color{black!45}85.7 & \color{black!45}97.3 & \color{black!45}97.4 & \color{black!45}83.2 & \color{black!45}64.0 & \color{black!45}86.7 \\
\color{black!45}\hspace{0.5em}Panda V1 (full) & \color{black!45}98.2{\tiny\color{black!50}$\,\pm$0.03} & \color{black!45}99.4{\tiny\color{black!50}$\,\pm$0.01} & \color{black!45}99.8{\tiny\color{black!50}$\,\pm$0.00} & \color{black!45}97.9{\tiny\color{black!50}$\,\pm$0.13} & \color{black!45}97.0{\tiny\color{black!50}$\,\pm$0.07} & \color{black!45}97.1{\tiny\color{black!50}$\,\pm$0.02} \\
\bottomrule
\end{tabular*}
\caption{\label{tab:app-lar-perclass-sem}\textbf{LArTPC semantic segmentation by motif}, F$_1$ in percent on the test images after fine-tuning on 1K and on 1M labeled images. Errors are 95\% CI.}
\end{table*}

\begin{table*}[p]
\centering
\scriptsize
\setlength{\tabcolsep}{2.4pt}
\begin{tabular*}{425pt}[t]{l@{\extracolsep{\fill}}r@{\extracolsep{0pt}\hspace{2\tabcolsep}}ccccc@{\hspace{0.5\tabcolsep}}c@{\hspace{0.5\tabcolsep}}ccccc}
\toprule
\multicolumn{2}{l}{\setbox0\hbox{Particle Inst.\ Seg.\ (ARI)}\rlap{\smash{\raisebox{-3.6pt}{\rule{\wd0}{\cmidrulewidth}}}}\box0} & \multicolumn{5}{c}{PT$_K$ + FT$_K$} &  & \multicolumn{5}{c}{PT$_\text{1M}$ + FT$_K$} \\
\cmidrule(lr){3-7} \cmidrule(lr){9-13}
Method / $K$ & Param. & 100 & 1K & 10K & 100K & 1M &  & 100 & 1K & 10K & 100K & 1M \\
\midrule
\multicolumn{13}{l}{\textit{Supervised}} \\
\hspace{0.5em}Panda V2 & 58.2M & \cellcolor[HTML]{D6D6FF}\textbf{93.5}{\tiny\color{black!50}$\,\pm$0.07} & \cellcolor[HTML]{D6D6FF}\textbf{96.8}{\tiny\color{black!50}$\,\pm$0.04} & \cellcolor[HTML]{E3E3FF}97.4{\tiny\color{black!50}$\,\pm$0.03} & \cellcolor[HTML]{E3E3FF}98.0{\tiny\color{black!50}$\,\pm$0.02} & \cellcolor[HTML]{F4F4FF}97.9{\tiny\color{black!50}$\,\pm$0.02} & \phantom{\raisebox{0.12ex}{\tiny=}} & \cellcolor[HTML]{D6D6FF}\llap{\raisebox{0.12ex}{\tiny=}\hspace{0.5\tabcolsep}}\textbf{93.5}{\tiny\color{black!50}$\,\pm$0.07} & \cellcolor[HTML]{E3E3FF}96.8{\tiny\color{black!50}$\,\pm$0.04} & \cellcolor[HTML]{E3E3FF}97.4{\tiny\color{black!50}$\,\pm$0.03} & \cellcolor[HTML]{E3E3FF}98.0{\tiny\color{black!50}$\,\pm$0.02} & \cellcolor[HTML]{F4F4FF}97.9{\tiny\color{black!50}$\,\pm$0.02} \\
\midrule
\multicolumn{13}{l}{\textit{Self-supervised}} \\
\hspace{0.5em}Panda V1 (dec.) & 4.5M & -- & -- & -- & -- & -- &  & 73.2{\tiny\color{black!50}$\,\pm$0.17} & 87.3{\tiny\color{black!50}$\,\pm$0.10} & \cellcolor[HTML]{F4F4FF}95.0{\tiny\color{black!50}$\,\pm$0.05} & \cellcolor[HTML]{F4F4FF}96.6{\tiny\color{black!50}$\,\pm$0.03} & 97.0{\tiny\color{black!50}$\,\pm$0.03} \\
\hspace{0.5em}Panda V2 (dec.) & 4.7M & \cellcolor[HTML]{F4F4FF}67.2{\tiny\color{black!50}$\,\pm$0.17} & \cellcolor[HTML]{F4F4FF}92.0{\tiny\color{black!50}$\,\pm$0.08} & \cellcolor[HTML]{F4F4FF}97.2{\tiny\color{black!50}$\,\pm$0.04} & \cellcolor[HTML]{E3E3FF}98.0{\tiny\color{black!50}$\,\pm$0.03} & \cellcolor[HTML]{E3E3FF}98.0{\tiny\color{black!50}$\,\pm$0.03} &  & \cellcolor[HTML]{F4F4FF}86.6{\tiny\color{black!50}$\,\pm$0.11} & \cellcolor[HTML]{F4F4FF}95.7{\tiny\color{black!50}$\,\pm$0.05} & \cellcolor[HTML]{E3E3FF}97.4{\tiny\color{black!50}$\,\pm$0.03} & \cellcolor[HTML]{E3E3FF}98.0{\tiny\color{black!50}$\,\pm$0.03} & \cellcolor[HTML]{E3E3FF}98.0{\tiny\color{black!50}$\,\pm$0.03} \\
\hspace{0.5em}Panda V2 (LoRA) & 7.0M & \cellcolor[HTML]{E3E3FF}69.8{\tiny\color{black!50}$\,\pm$0.17} & \cellcolor[HTML]{E3E3FF}95.9{\tiny\color{black!50}$\,\pm$0.05} & \cellcolor[HTML]{D6D6FF}\textbf{97.9}{\tiny\color{black!50}$\,\pm$0.03} & \cellcolor[HTML]{D6D6FF}\textbf{98.3}{\tiny\color{black!50}$\,\pm$0.03} & \cellcolor[HTML]{D6D6FF}\textbf{98.5}{\tiny\color{black!50}$\,\pm$0.02} &  & \cellcolor[HTML]{E3E3FF}92.2{\tiny\color{black!50}$\,\pm$0.08} & \cellcolor[HTML]{D6D6FF}\textbf{97.3}{\tiny\color{black!50}$\,\pm$0.03} & \cellcolor[HTML]{D6D6FF}\textbf{98.1}{\tiny\color{black!50}$\,\pm$0.03} & \cellcolor[HTML]{D6D6FF}\textbf{98.4}{\tiny\color{black!50}$\,\pm$0.02} & \cellcolor[HTML]{D6D6FF}\textbf{98.5}{\tiny\color{black!50}$\,\pm$0.02} \\
\hspace{0.5em}\color{black!45}Panda V1 (full) & \color{black!45}95.2M & \color{black!45}-- & \color{black!45}-- & \color{black!45}-- & \color{black!45}-- & \color{black!45}97.9{\tiny\color{black!50}$\,\pm$0.02} & \color{black!45} & \color{black!45}-- & \color{black!45}-- & \color{black!45}-- & \color{black!45}-- & \color{black!45}97.9{\tiny\color{black!50}$\,\pm$0.02} \\
\bottomrule
\end{tabular*}
\caption{\label{tab:app-lar-part}\textbf{Particle instance segmentation on LArTPC.} Equivalent to Table~\ref{tab:instseg_lar} of the main text, with 95\% CI, a LoRA row and the 100K label count added.}
\end{table*}

\begin{table*}[p]
\centering
\scriptsize
\setlength{\tabcolsep}{2.4pt}
\begin{tabular*}{425pt}[t]{l@{\extracolsep{\fill}}r@{\extracolsep{0pt}\hspace{2\tabcolsep}}ccccc@{\hspace{0.5\tabcolsep}}c@{\hspace{0.5\tabcolsep}}ccccc}
\toprule
\multicolumn{2}{l}{\setbox0\hbox{Interaction Inst.\ Seg.\ (ARI)}\rlap{\smash{\raisebox{-3.6pt}{\rule{\wd0}{\cmidrulewidth}}}}\box0} & \multicolumn{5}{c}{PT$_K$ + FT$_K$} &  & \multicolumn{5}{c}{PT$_\text{1M}$ + FT$_K$} \\
\cmidrule(lr){3-7} \cmidrule(lr){9-13}
Method / $K$ & Param. & 100 & 1K & 10K & 100K & 1M &  & 100 & 1K & 10K & 100K & 1M \\
\midrule
\multicolumn{13}{l}{\textit{Supervised}} \\
\hspace{0.5em}Panda V2 & 58.2M & \cellcolor[HTML]{D6D6FF}\textbf{89.3}{\tiny\color{black!50}$\,\pm$0.14} & \cellcolor[HTML]{D6D6FF}\textbf{96.8}{\tiny\color{black!50}$\,\pm$0.05} & \cellcolor[HTML]{E3E3FF}97.7{\tiny\color{black!50}$\,\pm$0.04} & \cellcolor[HTML]{E3E3FF}98.2{\tiny\color{black!50}$\,\pm$0.03} & \cellcolor[HTML]{E3E3FF}98.2{\tiny\color{black!50}$\,\pm$0.03} & \phantom{\raisebox{0.12ex}{\tiny=}} & \cellcolor[HTML]{E3E3FF}\llap{\raisebox{0.12ex}{\tiny=}\hspace{0.5\tabcolsep}}89.3{\tiny\color{black!50}$\,\pm$0.14} & \cellcolor[HTML]{E3E3FF}96.8{\tiny\color{black!50}$\,\pm$0.05} & \cellcolor[HTML]{E3E3FF}97.7{\tiny\color{black!50}$\,\pm$0.04} & \cellcolor[HTML]{E3E3FF}98.2{\tiny\color{black!50}$\,\pm$0.03} & \cellcolor[HTML]{E3E3FF}98.2{\tiny\color{black!50}$\,\pm$0.03} \\
\midrule
\multicolumn{13}{l}{\textit{Self-supervised}} \\
\hspace{0.5em}Panda V1 (dec.) & 4.3M & -- & -- & -- & -- & -- &  & 71.5{\tiny\color{black!50}$\,\pm$0.20} & 83.6{\tiny\color{black!50}$\,\pm$0.15} & 94.6{\tiny\color{black!50}$\,\pm$0.07} & 96.1{\tiny\color{black!50}$\,\pm$0.06} & 97.4{\tiny\color{black!50}$\,\pm$0.04} \\
\hspace{0.5em}Panda V2 (dec.) & 4.7M & \cellcolor[HTML]{F4F4FF}61.0{\tiny\color{black!50}$\,\pm$0.20} & \cellcolor[HTML]{F4F4FF}88.2{\tiny\color{black!50}$\,\pm$0.13} & \cellcolor[HTML]{F4F4FF}97.2{\tiny\color{black!50}$\,\pm$0.05} & \cellcolor[HTML]{F4F4FF}98.1{\tiny\color{black!50}$\,\pm$0.03} & \cellcolor[HTML]{F4F4FF}98.1{\tiny\color{black!50}$\,\pm$0.03} &  & \cellcolor[HTML]{F4F4FF}80.7{\tiny\color{black!50}$\,\pm$0.16} & \cellcolor[HTML]{F4F4FF}95.3{\tiny\color{black!50}$\,\pm$0.07} & \cellcolor[HTML]{F4F4FF}97.4{\tiny\color{black!50}$\,\pm$0.04} & \cellcolor[HTML]{F4F4FF}98.1{\tiny\color{black!50}$\,\pm$0.03} & \cellcolor[HTML]{F4F4FF}98.1{\tiny\color{black!50}$\,\pm$0.03} \\
\hspace{0.5em}Panda V2 (LoRA) & 7.0M & \cellcolor[HTML]{E3E3FF}62.3{\tiny\color{black!50}$\,\pm$0.20} & \cellcolor[HTML]{E3E3FF}95.5{\tiny\color{black!50}$\,\pm$0.07} & \cellcolor[HTML]{D6D6FF}\textbf{97.9}{\tiny\color{black!50}$\,\pm$0.04} & \cellcolor[HTML]{D6D6FF}\textbf{98.3}{\tiny\color{black!50}$\,\pm$0.03} & \cellcolor[HTML]{D6D6FF}\textbf{98.5}{\tiny\color{black!50}$\,\pm$0.02} &  & \cellcolor[HTML]{D6D6FF}\textbf{92.3}{\tiny\color{black!50}$\,\pm$0.10} & \cellcolor[HTML]{D6D6FF}\textbf{97.3}{\tiny\color{black!50}$\,\pm$0.04} & \cellcolor[HTML]{D6D6FF}\textbf{98.1}{\tiny\color{black!50}$\,\pm$0.03} & \cellcolor[HTML]{D6D6FF}\textbf{98.4}{\tiny\color{black!50}$\,\pm$0.03} & \cellcolor[HTML]{D6D6FF}\textbf{98.5}{\tiny\color{black!50}$\,\pm$0.02} \\
\hspace{0.5em}\color{black!45}Panda V1 (full) & \color{black!45}95.0M & \color{black!45}-- & \color{black!45}-- & \color{black!45}-- & \color{black!45}-- & \color{black!45}97.6{\tiny\color{black!50}$\,\pm$0.04} & \color{black!45} & \color{black!45}-- & \color{black!45}-- & \color{black!45}-- & \color{black!45}-- & \color{black!45}97.6{\tiny\color{black!50}$\,\pm$0.04} \\
\bottomrule
\end{tabular*}
\caption{\label{tab:app-lar-int}\textbf{Interaction instance segmentation on LArTPC.} Error bars are 95\% CI.}
\end{table*}

\begin{table*}[p]
\centering
\scriptsize
\setlength{\tabcolsep}{2.4pt}
\begin{tabular*}{351pt}[t]{l@{\extracolsep{\fill}}r@{\extracolsep{0pt}\hspace{2\tabcolsep}}cccc@{\hspace{0.5\tabcolsep}}c@{\hspace{0.5\tabcolsep}}cccc}
\toprule
\multicolumn{2}{l}{\setbox0\hbox{Sem.\ Seg.\ (mF$_1$)}\rlap{\smash{\raisebox{-3.6pt}{\rule{\wd0}{\cmidrulewidth}}}}\box0} & \multicolumn{4}{c}{PT$_K$ + FT$_K$} &  & \multicolumn{4}{c}{PT$_\text{1M}$ + FT$_K$} \\
\cmidrule(lr){3-6} \cmidrule(lr){8-11}
Method / $K$ & Param. & 100 & 1K & 10K & 70K &  & 100 & 1K & 10K & 70K \\
\midrule
\multicolumn{11}{l}{\textit{Supervised}} \\
\hspace{0.5em}OneFormer~\cite{kolodiazhnyi2023oneformer3dtransformerunifiedpoint,park2025fm4npp} & 45.0M & -- & -- & -- & 53.0 & \raisebox{0.12ex}{\tiny=} & -- & -- & -- & 53.0 \\
\hspace{0.5em}SAGEConv~\cite{park2025fm4npp} & 0.9M & -- & -- & -- & 53.6 & \raisebox{0.12ex}{\tiny=} & -- & -- & -- & 53.6 \\
\hspace{0.5em}Panda V2 & 54.0M & \cellcolor[HTML]{C4D7F4}\textbf{56.2}{\tiny\color{black!50}$\,\pm$0.56} & \cellcolor[HTML]{D8E5F8}72.9{\tiny\color{black!50}$\,\pm$0.53} & \cellcolor[HTML]{D8E5F8}83.3{\tiny\color{black!50}$\,\pm$0.41} & \cellcolor[HTML]{EEF3FC}83.7{\tiny\color{black!50}$\,\pm$0.33} & \phantom{\raisebox{0.12ex}{\tiny=}} & \llap{\raisebox{0.12ex}{\tiny=}\hspace{0.5\tabcolsep}}56.2{\tiny\color{black!50}$\,\pm$0.56} & 72.9{\tiny\color{black!50}$\,\pm$0.53} & 83.3{\tiny\color{black!50}$\,\pm$0.41} & \cellcolor[HTML]{EEF3FC}83.7{\tiny\color{black!50}$\,\pm$0.33} \\
\midrule
\multicolumn{11}{l}{\textit{Self-supervised}} \\
\hspace{0.5em}FM4NPP (dec.){$^\dagger$}~\cite{park2025fm4npp} & 0.7M & -- & -- & -- & -- &  & -- & -- & -- & 81.8 \\
\hspace{0.5em}Panda V2 (lin.) & \llap{\raisebox{0.12ex}{\tiny$<$}\kern0.5pt}0.01M & 51.3{\tiny\color{black!50}$\,\pm$0.40} & 69.0{\tiny\color{black!50}$\,\pm$0.39} & 78.1{\tiny\color{black!50}$\,\pm$0.32} & 83.1{\tiny\color{black!50}$\,\pm$0.26} &  & \cellcolor[HTML]{EEF3FC}79.9{\tiny\color{black!50}$\,\pm$0.35} & \cellcolor[HTML]{EEF3FC}82.9{\tiny\color{black!50}$\,\pm$0.27} & \cellcolor[HTML]{EEF3FC}84.3{\tiny\color{black!50}$\,\pm$0.23} & \cellcolor[HTML]{EEF3FC}83.7{\tiny\color{black!50}$\,\pm$0.25} \\
\hspace{0.5em}Panda V2 (mlp) & 0.3M & \cellcolor[HTML]{EEF3FC}52.8{\tiny\color{black!50}$\,\pm$0.44} & \cellcolor[HTML]{EEF3FC}71.0{\tiny\color{black!50}$\,\pm$0.46} & \cellcolor[HTML]{EEF3FC}80.7{\tiny\color{black!50}$\,\pm$0.29} & \cellcolor[HTML]{D8E5F8}85.1{\tiny\color{black!50}$\,\pm$0.23} &  & \cellcolor[HTML]{D8E5F8}82.1{\tiny\color{black!50}$\,\pm$0.36} & \cellcolor[HTML]{D8E5F8}84.6{\tiny\color{black!50}$\,\pm$0.30} & \cellcolor[HTML]{D8E5F8}85.1{\tiny\color{black!50}$\,\pm$0.22} & \cellcolor[HTML]{D8E5F8}85.3{\tiny\color{black!50}$\,\pm$0.22} \\
\hspace{0.5em}Panda V2 (LoRA) & 2.9M & \cellcolor[HTML]{D8E5F8}55.0{\tiny\color{black!50}$\,\pm$0.49} & \cellcolor[HTML]{C4D7F4}\textbf{76.9}{\tiny\color{black!50}$\,\pm$0.48} & \cellcolor[HTML]{C4D7F4}\textbf{84.2}{\tiny\color{black!50}$\,\pm$0.36} & \cellcolor[HTML]{C4D7F4}\textbf{86.2}{\tiny\color{black!50}$\,\pm$0.29} &  & \cellcolor[HTML]{C4D7F4}\textbf{85.4}{\tiny\color{black!50}$\,\pm$0.38} & \cellcolor[HTML]{C4D7F4}\textbf{86.7}{\tiny\color{black!50}$\,\pm$0.35} & \cellcolor[HTML]{C4D7F4}\textbf{87.7}{\tiny\color{black!50}$\,\pm$0.33} & \cellcolor[HTML]{C4D7F4}\textbf{86.4}{\tiny\color{black!50}$\,\pm$0.30} \\
\bottomrule
\multicolumn{11}{@{}l}{$^\dagger$Pretrained on 10M.} \\
\end{tabular*}
\caption{\label{tab:app-sph-sem}\textbf{Semantic segmentation on sPHENIX.} Equivalent to Table~\ref{tab:semseg_sph} of the main text, with 95\% CI and a LoRA row added.}
\end{table*}

\begin{table*}[p]
\centering
\scriptsize
\setlength{\tabcolsep}{3pt}
\begin{tabular}{@{}lccccccc!{\color{black!30}\vrule}cccc}
\toprule
 & \multicolumn{7}{c!{\color{black!30}\vrule}}{Particle identification} & \multicolumn{4}{c}{Noise tagging} \\
\cmidrule(lr){2-8} \cmidrule(lr){9-12}
 & & \multicolumn{6}{c!{\color{black!30}\vrule}}{F$_1$} & & \multicolumn{3}{c}{F$_1$} \\
\cmidrule(lr){3-8} \cmidrule(lr){10-12}
Method & Accuracy & Macro & Others & Pion & Kaon & Proton & Electron & Accuracy & Macro & Non-noise & Noise \\
\midrule
\multicolumn{12}{@{}l}{\textbf{PT$_\text{1M}$ + FT$_\text{1K}$}} \\
\multicolumn{12}{@{}l}{\textit{Supervised}} \\
\hspace{0.5em}Panda V2 & 86.3{\tiny\color{black!50}$\,\pm$0.28} & 72.9{\tiny\color{black!50}$\,\pm$0.53} & 55.5{\tiny\color{black!50}$\,\pm$1.98} & 90.9{\tiny\color{black!50}$\,\pm$0.21} & 37.8{\tiny\color{black!50}$\,\pm$1.71} & 86.0{\tiny\color{black!50}$\,\pm$0.50} & \cellcolor[HTML]{EEF3FC}87.4{\tiny\color{black!50}$\,\pm$0.52} & 96.8{\tiny\color{black!50}$\,\pm$0.25} & 91.7{\tiny\color{black!50}$\,\pm$0.62} & \cellcolor[HTML]{EEF3FC}98.2{\tiny\color{black!50}$\,\pm$0.14} & 85.2{\tiny\color{black!50}$\,\pm$1.11} \\
\arrayrulecolor{black!20}\cmidrule[0.3pt]{1-12}\arrayrulecolor{black}
\multicolumn{12}{@{}l}{\textit{Self-supervised}} \\
\hspace{0.5em}Panda V2 (lin.) & \cellcolor[HTML]{EEF3FC}88.7{\tiny\color{black!50}$\,\pm$0.19} & \cellcolor[HTML]{EEF3FC}82.9{\tiny\color{black!50}$\,\pm$0.27} & \cellcolor[HTML]{EEF3FC}82.8{\tiny\color{black!50}$\,\pm$0.82} & \cellcolor[HTML]{EEF3FC}91.0{\tiny\color{black!50}$\,\pm$0.16} & \cellcolor[HTML]{EEF3FC}41.6{\tiny\color{black!50}$\,\pm$0.98} & \cellcolor[HTML]{EEF3FC}89.3{\tiny\color{black!50}$\,\pm$0.42} & \cellcolor[HTML]{C4D7F4}\textbf{96.6}{\tiny\color{black!50}$\,\pm$0.20} & \cellcolor[HTML]{D8E5F8}97.6{\tiny\color{black!50}$\,\pm$0.22} & \cellcolor[HTML]{D8E5F8}93.8{\tiny\color{black!50}$\,\pm$0.53} & \cellcolor[HTML]{D8E5F8}98.6{\tiny\color{black!50}$\,\pm$0.13} & \cellcolor[HTML]{D8E5F8}89.0{\tiny\color{black!50}$\,\pm$0.95} \\
\hspace{0.5em}Panda V2 (mlp) & \cellcolor[HTML]{D8E5F8}92.7{\tiny\color{black!50}$\,\pm$0.16} & \cellcolor[HTML]{D8E5F8}84.6{\tiny\color{black!50}$\,\pm$0.30} & \cellcolor[HTML]{D8E5F8}83.7{\tiny\color{black!50}$\,\pm$0.84} & \cellcolor[HTML]{D8E5F8}94.5{\tiny\color{black!50}$\,\pm$0.12} & \cellcolor[HTML]{D8E5F8}54.7{\tiny\color{black!50}$\,\pm$1.18} & \cellcolor[HTML]{D8E5F8}92.2{\tiny\color{black!50}$\,\pm$0.32} & \cellcolor[HTML]{D8E5F8}96.5{\tiny\color{black!50}$\,\pm$0.22} & \cellcolor[HTML]{EEF3FC}97.5{\tiny\color{black!50}$\,\pm$0.24} & \cellcolor[HTML]{EEF3FC}93.5{\tiny\color{black!50}$\,\pm$0.58} & \cellcolor[HTML]{D8E5F8}98.6{\tiny\color{black!50}$\,\pm$0.14} & \cellcolor[HTML]{EEF3FC}88.4{\tiny\color{black!50}$\,\pm$1.04} \\
\hspace{0.5em}Panda V2 (LoRA) & \cellcolor[HTML]{C4D7F4}\textbf{93.9}{\tiny\color{black!50}$\,\pm$0.16} & \cellcolor[HTML]{C4D7F4}\textbf{86.7}{\tiny\color{black!50}$\,\pm$0.35} & \cellcolor[HTML]{C4D7F4}\textbf{84.3}{\tiny\color{black!50}$\,\pm$0.83} & \cellcolor[HTML]{C4D7F4}\textbf{95.5}{\tiny\color{black!50}$\,\pm$0.12} & \cellcolor[HTML]{C4D7F4}\textbf{62.3}{\tiny\color{black!50}$\,\pm$1.46} & \cellcolor[HTML]{C4D7F4}\textbf{93.3}{\tiny\color{black!50}$\,\pm$0.33} & \cellcolor[HTML]{D8E5F8}96.5{\tiny\color{black!50}$\,\pm$0.23} & \cellcolor[HTML]{C4D7F4}\textbf{97.7}{\tiny\color{black!50}$\,\pm$0.24} & \cellcolor[HTML]{C4D7F4}\textbf{94.3}{\tiny\color{black!50}$\,\pm$0.56} & \cellcolor[HTML]{C4D7F4}\textbf{98.7}{\tiny\color{black!50}$\,\pm$0.14} & \cellcolor[HTML]{C4D7F4}\textbf{89.9}{\tiny\color{black!50}$\,\pm$1.00} \\
\midrule
\multicolumn{12}{@{}l}{\textbf{PT$_\text{1M}$ + FT$_\text{70K}$}} \\
\multicolumn{12}{@{}l}{\textit{Supervised}} \\
\hspace{0.5em}GATConv~\cite{park2025fm4npp} & 69.2 & 48.9 & 16.2 & 79.2 & 1.1 & 52.3 & 59.1 & 91.0 & 73.4 & 95.1 & 47.9 \\
\hspace{0.5em}GCNConv~\cite{park2025fm4npp} & 68.9 & 48.3 & 13.8 & 79.1 & 1.4 & 51.9 & 58.2 & 91.0 & 73.2 & 95.0 & 47.8 \\
\hspace{0.5em}GraphConv~\cite{park2025fm4npp} & 70.8 & 50.6 & 21.3 & 80.2 & 1.6 & 54.2 & 62.9 & 91.9 & 77.0 & 95.5 & 56.4 \\
\hspace{0.5em}SAGEConv~\cite{park2025fm4npp} & 72.6 & 53.6 & 18.3 & 81.6 & 6.4 & 66.3 & 64.9 & 91.7 & 76.7 & 95.4 & 56.3 \\
\hspace{0.5em}OneFormer~\cite{kolodiazhnyi2023oneformer3dtransformerunifiedpoint,park2025fm4npp} & 77.0 & 53.0 & 39.7 & 83.6 & 0.0 & 57.3 & 78.7 & 96.5 & 91.7 & 98.0 & 85.2 \\
\hspace{0.5em}AdapterOnly~\cite{park2025fm4npp} & 66.3 & 43.6 & 1.9 & 77.9 & 0.0 & 50.0 & 44.0 & 91.1 & 71.3 & 95.2 & 37.9 \\
\hspace{0.5em}Panda V2 & \cellcolor[HTML]{D8E5F8}91.6{\tiny\color{black!50}$\,\pm$0.18} & \cellcolor[HTML]{EEF3FC}83.7{\tiny\color{black!50}$\,\pm$0.33} & 81.4{\tiny\color{black!50}$\,\pm$0.90} & \cellcolor[HTML]{D8E5F8}93.7{\tiny\color{black!50}$\,\pm$0.14} & \cellcolor[HTML]{D8E5F8}53.9{\tiny\color{black!50}$\,\pm$1.20} & \cellcolor[HTML]{EEF3FC}91.6{\tiny\color{black!50}$\,\pm$0.36} & 95.9{\tiny\color{black!50}$\,\pm$0.25} & \cellcolor[HTML]{D8E5F8}98.0{\tiny\color{black!50}$\,\pm$0.22} & \cellcolor[HTML]{D8E5F8}95.0{\tiny\color{black!50}$\,\pm$0.51} & \cellcolor[HTML]{D8E5F8}98.9{\tiny\color{black!50}$\,\pm$0.13} & \cellcolor[HTML]{EEF3FC}91.1{\tiny\color{black!50}$\,\pm$0.93} \\
\arrayrulecolor{black!20}\cmidrule[0.3pt]{1-12}\arrayrulecolor{black}
\multicolumn{12}{@{}l}{\textit{Self-supervised}} \\
\hspace{0.5em}FM4NPP (dec.){$^\dagger$}~\cite{park2025fm4npp} & \cellcolor[HTML]{EEF3FC}90.4 & 81.8 & -- & -- & -- & -- & -- & 97.1 & 92.8 & -- & -- \\
\hspace{0.5em}Panda V2 (lin.) & 89.2{\tiny\color{black!50}$\,\pm$0.19} & \cellcolor[HTML]{EEF3FC}83.7{\tiny\color{black!50}$\,\pm$0.25} & \cellcolor[HTML]{EEF3FC}84.1{\tiny\color{black!50}$\,\pm$0.78} & \cellcolor[HTML]{EEF3FC}91.4{\tiny\color{black!50}$\,\pm$0.16} & \cellcolor[HTML]{EEF3FC}43.0{\tiny\color{black!50}$\,\pm$0.96} & 89.9{\tiny\color{black!50}$\,\pm$0.42} & \cellcolor[HTML]{EEF3FC}96.7{\tiny\color{black!50}$\,\pm$0.20} & \cellcolor[HTML]{EEF3FC}97.9{\tiny\color{black!50}$\,\pm$0.22} & \cellcolor[HTML]{EEF3FC}94.8{\tiny\color{black!50}$\,\pm$0.51} & \cellcolor[HTML]{EEF3FC}98.8{\tiny\color{black!50}$\,\pm$0.13} & 90.7{\tiny\color{black!50}$\,\pm$0.94} \\
\hspace{0.5em}Panda V2 (mlp) & 89.2{\tiny\color{black!50}$\,\pm$0.19} & \cellcolor[HTML]{D8E5F8}85.3{\tiny\color{black!50}$\,\pm$0.22} & \cellcolor[HTML]{D8E5F8}86.2{\tiny\color{black!50}$\,\pm$0.73} & 91.2{\tiny\color{black!50}$\,\pm$0.17} & 41.8{\tiny\color{black!50}$\,\pm$0.97} & \cellcolor[HTML]{C4D7F4}\textbf{93.1}{\tiny\color{black!50}$\,\pm$0.30} & \cellcolor[HTML]{D8E5F8}97.2{\tiny\color{black!50}$\,\pm$0.19} & \cellcolor[HTML]{C4D7F4}\textbf{98.2}{\tiny\color{black!50}$\,\pm$0.20} & \cellcolor[HTML]{C4D7F4}\textbf{95.5}{\tiny\color{black!50}$\,\pm$0.47} & \cellcolor[HTML]{C4D7F4}\textbf{99.0}{\tiny\color{black!50}$\,\pm$0.11} & \cellcolor[HTML]{C4D7F4}\textbf{92.0}{\tiny\color{black!50}$\,\pm$0.84} \\
\hspace{0.5em}Panda V2 (LoRA) & \cellcolor[HTML]{C4D7F4}\textbf{93.3}{\tiny\color{black!50}$\,\pm$0.15} & \cellcolor[HTML]{C4D7F4}\textbf{86.4}{\tiny\color{black!50}$\,\pm$0.30} & \cellcolor[HTML]{C4D7F4}\textbf{86.9}{\tiny\color{black!50}$\,\pm$0.69} & \cellcolor[HTML]{C4D7F4}\textbf{94.8}{\tiny\color{black!50}$\,\pm$0.12} & \cellcolor[HTML]{C4D7F4}\textbf{57.5}{\tiny\color{black!50}$\,\pm$1.15} & \cellcolor[HTML]{D8E5F8}93.0{\tiny\color{black!50}$\,\pm$0.31} & \cellcolor[HTML]{C4D7F4}\textbf{97.3}{\tiny\color{black!50}$\,\pm$0.19} & \cellcolor[HTML]{C4D7F4}\textbf{98.2}{\tiny\color{black!50}$\,\pm$0.22} & \cellcolor[HTML]{C4D7F4}\textbf{95.5}{\tiny\color{black!50}$\,\pm$0.54} & \cellcolor[HTML]{C4D7F4}\textbf{99.0}{\tiny\color{black!50}$\,\pm$0.13} & \cellcolor[HTML]{D8E5F8}91.9{\tiny\color{black!50}$\,\pm$0.97} \\
\bottomrule
\multicolumn{11}{@{}l}{$^\dagger$Pretrained on 10M.} \\
\end{tabular}
\caption{\label{tab:app-sph-pid}\textbf{sPHENIX particle identification and noise tagging by class}, in percent on the test images after fine-tuning on 1K and on 70K labeled images. For FM4NPP, we use the accuracy and macro averages of its main results table and leave out the per-class values of its appendix, whose macro averages differ from that table. Error bars are 95\% CI.}
\end{table*}

\begin{table*}[p]
\centering
\scriptsize
\setlength{\tabcolsep}{2.4pt}
\begin{tabular*}{351pt}[t]{l@{\extracolsep{\fill}}r@{\extracolsep{0pt}\hspace{2\tabcolsep}}cccc@{\hspace{0.5\tabcolsep}}c@{\hspace{0.5\tabcolsep}}cccc}
\toprule
\multicolumn{2}{l}{\setbox0\hbox{Particle Inst.\ Seg.\ (ARI)}\rlap{\smash{\raisebox{-3.6pt}{\rule{\wd0}{\cmidrulewidth}}}}\box0} & \multicolumn{4}{c}{PT$_K$ + FT$_K$} &  & \multicolumn{4}{c}{PT$_\text{1M}$ + FT$_K$} \\
\cmidrule(lr){3-6} \cmidrule(lr){8-11}
Method / $K$ & Param. & 100 & 1K & 10K & 70K &  & 100 & 1K & 10K & 70K \\
\midrule
\multicolumn{11}{l}{\textit{Supervised}} \\
\hspace{0.5em}EggNet~\cite{calafiura2024eggnetevolvinggraphbasedgraph, park2025fm4npp} & 0.2M & -- & -- & -- & 72.6 & \raisebox{0.12ex}{\tiny=} & -- & -- & -- & 72.6 \\
\hspace{0.5em}Exa.TrkX~\cite{exatrk, park2025fm4npp} & 3.9M & -- & -- & -- & 87.7 & \raisebox{0.12ex}{\tiny=} & -- & -- & -- & 87.7 \\
\hspace{0.5em}Panda V2 & 57.6M & \cellcolor[HTML]{C4D7F4}\textbf{86.3}{\tiny\color{black!50}$\,\pm$0.40} & \cellcolor[HTML]{C4D7F4}\textbf{94.3}{\tiny\color{black!50}$\,\pm$0.23} & \cellcolor[HTML]{C4D7F4}\textbf{96.7}{\tiny\color{black!50}$\,\pm$0.19} & \cellcolor[HTML]{EEF3FC}97.6{\tiny\color{black!50}$\,\pm$0.15} & \phantom{\raisebox{0.12ex}{\tiny=}} & \cellcolor[HTML]{EEF3FC}\llap{\raisebox{0.12ex}{\tiny=}\hspace{0.5\tabcolsep}}86.3{\tiny\color{black!50}$\,\pm$0.40} & \cellcolor[HTML]{EEF3FC}94.3{\tiny\color{black!50}$\,\pm$0.23} & \cellcolor[HTML]{EEF3FC}96.7{\tiny\color{black!50}$\,\pm$0.19} & \cellcolor[HTML]{EEF3FC}97.6{\tiny\color{black!50}$\,\pm$0.15} \\
\midrule
\multicolumn{11}{l}{\textit{Self-supervised}} \\
\hspace{0.5em}FM4NPP (dec.){$^\dagger$}~\cite{park2025fm4npp} & 2.4M & -- & -- & -- & -- &  & 46.8 & 69.3 & 89.3 & 94.5 \\
\hspace{0.5em}Panda V2 (dec.) & 4.2M & \cellcolor[HTML]{EEF3FC}73.5{\tiny\color{black!50}$\,\pm$0.45} & \cellcolor[HTML]{EEF3FC}91.5{\tiny\color{black!50}$\,\pm$0.29} & \cellcolor[HTML]{EEF3FC}96.2{\tiny\color{black!50}$\,\pm$0.20} & \cellcolor[HTML]{C4D7F4}\textbf{98.0}{\tiny\color{black!50}$\,\pm$0.16} &  & \cellcolor[HTML]{D8E5F8}89.6{\tiny\color{black!50}$\,\pm$0.34} & \cellcolor[HTML]{D8E5F8}96.0{\tiny\color{black!50}$\,\pm$0.20} & \cellcolor[HTML]{D8E5F8}97.4{\tiny\color{black!50}$\,\pm$0.18} & \cellcolor[HTML]{D8E5F8}98.1{\tiny\color{black!50}$\,\pm$0.13} \\
\hspace{0.5em}Panda V2 (LoRA) & 6.4M & \cellcolor[HTML]{D8E5F8}78.8{\tiny\color{black!50}$\,\pm$0.47} & \cellcolor[HTML]{D8E5F8}92.1{\tiny\color{black!50}$\,\pm$0.31} & \cellcolor[HTML]{D8E5F8}96.6{\tiny\color{black!50}$\,\pm$0.20} & \cellcolor[HTML]{D8E5F8}97.9{\tiny\color{black!50}$\,\pm$0.22} &  & \cellcolor[HTML]{C4D7F4}\textbf{90.5}{\tiny\color{black!50}$\,\pm$0.31} & \cellcolor[HTML]{C4D7F4}\textbf{96.6}{\tiny\color{black!50}$\,\pm$0.18} & \cellcolor[HTML]{C4D7F4}\textbf{98.0}{\tiny\color{black!50}$\,\pm$0.15} & \cellcolor[HTML]{C4D7F4}\textbf{98.4}{\tiny\color{black!50}$\,\pm$0.15} \\
\bottomrule
\multicolumn{11}{@{}l}{$^\dagger$Pretrained on 10M.} \\
\end{tabular*}
\caption{\label{tab:app-sph-part}\textbf{Particle instance segmentation on sPHENIX.} Equivalent to Table~\ref{tab:instseg_sph} of the main text, with 95\% CI and a LoRA row added.}
\end{table*}

\begin{table*}[p]
\centering
\scriptsize
\setlength{\tabcolsep}{3pt}
\begin{tabular}{@{}l ccccc!{\color{black!30}\vrule}ccccc}
\toprule
 & \multicolumn{5}{c!{\color{black!30}\vrule}}{\textbf{PT$_\text{1M}$ + FT$_\text{1K}$}} & \multicolumn{5}{c}{\textbf{PT$_\text{1M}$ + FT$_\text{70K}$}} \\
\cmidrule(lr){2-6} \cmidrule(lr){7-11}
 & & \multicolumn{2}{c}{Track} & \multicolumn{2}{c!{\color{black!30}\vrule}}{Spacepoint} & & \multicolumn{2}{c}{Track} & \multicolumn{2}{c}{Spacepoint} \\
\cmidrule(lr){3-4} \cmidrule(lr){5-6} \cmidrule(lr){8-9} \cmidrule(lr){10-11}
Track finding & ARI & Efficiency & Purity & Efficiency & Purity & ARI & Efficiency & Purity & Efficiency & Purity \\
\midrule
\multicolumn{11}{@{}l}{\textit{Supervised}} \\
\hspace{0.5em}EggNet~\cite{calafiura2024eggnetevolvinggraphbasedgraph, park2025fm4npp} & -- & -- & -- & -- & -- & 72.6 & 74.2 & 75.1 & 93.0 & 92.3 \\
\hspace{0.5em}Exa.TrkX~\cite{exatrk, park2025fm4npp} & -- & -- & -- & -- & -- & 87.7 & 91.8 & 66.4 & 94.5 & \cellcolor[HTML]{EEF3FC}98.8 \\
\hspace{0.5em}AdapterOnly~\cite{park2025fm4npp} & -- & -- & -- & -- & -- & 72.4 & 78.0 & 64.5 & 89.3 & 92.1 \\
\hspace{0.5em}Panda V2 & \cellcolor[HTML]{EEF3FC}94.3{\tiny\color{black!50}$\,\pm$0.23} & \cellcolor[HTML]{EEF3FC}94.6{\tiny\color{black!50}$\,\pm$0.15} & \cellcolor[HTML]{EEF3FC}97.3{\tiny\color{black!50}$\,\pm$0.12} & \cellcolor[HTML]{EEF3FC}95.3{\tiny\color{black!50}$\,\pm$0.22} & \cellcolor[HTML]{EEF3FC}97.6{\tiny\color{black!50}$\,\pm$0.12} & \cellcolor[HTML]{EEF3FC}97.6{\tiny\color{black!50}$\,\pm$0.15} & \cellcolor[HTML]{EEF3FC}97.1{\tiny\color{black!50}$\,\pm$0.12} & \cellcolor[HTML]{EEF3FC}98.8{\tiny\color{black!50}$\,\pm$0.07} & \cellcolor[HTML]{EEF3FC}98.1{\tiny\color{black!50}$\,\pm$0.14} & \cellcolor[HTML]{EEF3FC}98.8{\tiny\color{black!50}$\,\pm$0.07} \\
\arrayrulecolor{black!20}\cmidrule[0.3pt]{1-11}\arrayrulecolor{black}
\multicolumn{11}{@{}l}{\textit{Self-supervised}} \\
\hspace{0.5em}FM4NPP (dec.){$^\dagger$}~\cite{park2025fm4npp} & -- & -- & -- & -- & -- & 94.5 & 96.1 & 93.1 & 97.6 & 98.3 \\
\hspace{0.5em}Panda V2 (dec.) & \cellcolor[HTML]{D8E5F8}96.0{\tiny\color{black!50}$\,\pm$0.20} & \cellcolor[HTML]{D8E5F8}94.9{\tiny\color{black!50}$\,\pm$0.18} & \cellcolor[HTML]{D8E5F8}97.7{\tiny\color{black!50}$\,\pm$0.11} & \cellcolor[HTML]{D8E5F8}96.4{\tiny\color{black!50}$\,\pm$0.14} & \cellcolor[HTML]{D8E5F8}98.0{\tiny\color{black!50}$\,\pm$0.13} & \cellcolor[HTML]{D8E5F8}98.1{\tiny\color{black!50}$\,\pm$0.13} & \cellcolor[HTML]{D8E5F8}97.2{\tiny\color{black!50}$\,\pm$0.14} & \cellcolor[HTML]{D8E5F8}99.0{\tiny\color{black!50}$\,\pm$0.07} & \cellcolor[HTML]{D8E5F8}98.2{\tiny\color{black!50}$\,\pm$0.12} & \cellcolor[HTML]{D8E5F8}99.0{\tiny\color{black!50}$\,\pm$0.07} \\
\hspace{0.5em}Panda V2 (LoRA) & \cellcolor[HTML]{C4D7F4}\textbf{96.6}{\tiny\color{black!50}$\,\pm$0.18} & \cellcolor[HTML]{C4D7F4}\textbf{95.5}{\tiny\color{black!50}$\,\pm$0.17} & \cellcolor[HTML]{C4D7F4}\textbf{97.9}{\tiny\color{black!50}$\,\pm$0.10} & \cellcolor[HTML]{C4D7F4}\textbf{96.6}{\tiny\color{black!50}$\,\pm$0.17} & \cellcolor[HTML]{C4D7F4}\textbf{98.4}{\tiny\color{black!50}$\,\pm$0.10} & \cellcolor[HTML]{C4D7F4}\textbf{98.4}{\tiny\color{black!50}$\,\pm$0.15} & \cellcolor[HTML]{C4D7F4}\textbf{97.9}{\tiny\color{black!50}$\,\pm$0.11} & \cellcolor[HTML]{C4D7F4}\textbf{99.4}{\tiny\color{black!50}$\,\pm$0.05} & \cellcolor[HTML]{C4D7F4}\textbf{98.7}{\tiny\color{black!50}$\,\pm$0.10} & \cellcolor[HTML]{C4D7F4}\textbf{99.3}{\tiny\color{black!50}$\,\pm$0.06} \\
\bottomrule
\end{tabular}
\caption{\label{tab:app-sph-track}\textbf{sPHENIX track finding}, in percent on the test images after fine-tuning on 1K and on 70K labeled images, in the format of the appendix of~\cite{park2025fm4npp}, from which the rows of other methods are taken. The ARI columns repeat Table~\ref{tab:app-sph-part}, which scores the grouping of hits into particles image by image; the other columns count whole tracks, as the fractions of true and of predicted tracks matched under the double-majority rule, and hits (spacepoints), as the fractions of hits in the right track, pooled over the test set. Error bars are 95\% CI.}
\end{table*}

\begin{table*}[p]
\centering
\scriptsize
\setlength{\tabcolsep}{3pt}
\begin{tabular}[t]{lrccc@{\hspace{0.8em}}ccc@{\hspace{0.8em}}ccc}
\toprule
\multicolumn{2}{l}{\setbox0\hbox{Single Particle Reco.}\rlap{\smash{\raisebox{-3.6pt}{\rule{\wd0}{\cmidrulewidth}}}}\box0} & \multicolumn{3}{c}{Classification (AUC)} & \multicolumn{3}{c}{Muon Kinematics ($\sigma_{68}$ $\downarrow$)} & \multicolumn{3}{c}{Electron Kinematics ($\sigma_{68}$ $\downarrow$)} \\
\cmidrule(lr){3-5} \cmidrule(lr){6-8} \cmidrule(lr){9-11}
Method & Param. & $\mu/e^\dagger$ & $\mu/\pi^\pm$ & $e/\pi^0$ & Vtx.\ (cm) & Dir.\ ($^\circ$) & Mom.\ (\%) & Vtx.\ (cm) & Dir.\ ($^\circ$) & Mom.\ (\%) \\
\midrule
\multicolumn{11}{l}{\textit{Classical}} \\
\hspace{0.5em}fiTQun & -- & \multicolumn{1}{c}{17{\color{black!50}$^{\scriptscriptstyle +17.77}_{\scriptscriptstyle -12.73}$}} & \multicolumn{1}{c}{82.3{\tiny\color{black!50}$\,\pm$0.86}} & \multicolumn{1}{c}{86.6{\tiny\color{black!50}$\,\pm$0.74}} & \multicolumn{1}{c}{\cellcolor[HTML]{F7D9BD}\textbf{19.0}{\tiny\color{black!50}$\,\pm$0.34}} & \multicolumn{1}{c}{\cellcolor[HTML]{F7D9BD}\textbf{1.13}{\tiny\color{black!50}$\,\pm$0.04}} & \multicolumn{1}{c}{\cellcolor[HTML]{F7D9BD}\textbf{1.9}{\tiny\color{black!50}$\,\pm$0.05}} & \multicolumn{1}{c}{\cellcolor[HTML]{F7D9BD}\textbf{20.9}{\tiny\color{black!50}$\,\pm$0.54}} & \multicolumn{1}{c}{\cellcolor[HTML]{F7D9BD}\textbf{1.79}{\tiny\color{black!50}$\,\pm$0.06}} & \multicolumn{1}{c}{\cellcolor[HTML]{F7D9BD}\textbf{2.3}{\tiny\color{black!50}$\,\pm$0.09}} \\
\midrule
\multicolumn{11}{l}{\textit{Self-supervised}} \\
\hspace{0.5em}Panda V2 (lin.) & \llap{\raisebox{0.12ex}{\tiny$<$}\kern0.5pt}0.01M & \cellcolor[HTML]{FDF4EC}0.14{\color{black!50}$^{\scriptscriptstyle +0.27}_{\scriptscriptstyle -0.14}$} & \cellcolor[HTML]{FAE6D3}99.2{\tiny\color{black!50}$\,\pm$0.22} & \cellcolor[HTML]{FDF4EC}92.4{\tiny\color{black!50}$\,\pm$0.53} & 37.0{\tiny\color{black!50}$\,\pm$0.63} & \cellcolor[HTML]{FDF4EC}2.29{\tiny\color{black!50}$\,\pm$0.05} & \cellcolor[HTML]{FDF4EC}2.2{\tiny\color{black!50}$\,\pm$0.06} & 41.3{\tiny\color{black!50}$\,\pm$0.78} & 3.22{\tiny\color{black!50}$\,\pm$0.08} & \cellcolor[HTML]{FDF4EC}3.3{\tiny\color{black!50}$\,\pm$0.10} \\
\hspace{0.5em}Panda V2 (mlp) & 0.3M & \cellcolor[HTML]{F7D9BD}\textbf{0.03}{\color{black!50}$^{\scriptscriptstyle +0.07}_{\scriptscriptstyle -0.03}$} & \cellcolor[HTML]{FDF4EC}99.0{\tiny\color{black!50}$\,\pm$0.27} & \cellcolor[HTML]{FAE6D3}93.0{\tiny\color{black!50}$\,\pm$0.48} & \cellcolor[HTML]{FDF4EC}27.0{\tiny\color{black!50}$\,\pm$0.48} & \cellcolor[HTML]{FAE6D3}1.42{\tiny\color{black!50}$\,\pm$0.03} & \cellcolor[HTML]{FAE6D3}2.0{\tiny\color{black!50}$\,\pm$0.05} & \cellcolor[HTML]{FDF4EC}30.3{\tiny\color{black!50}$\,\pm$0.55} & \cellcolor[HTML]{FDF4EC}2.51{\tiny\color{black!50}$\,\pm$0.07} & 3.4{\tiny\color{black!50}$\,\pm$0.09} \\
\hspace{0.5em}Panda V2 (LoRA) & 2.6M & \cellcolor[HTML]{FAE6D3}0.10{\color{black!50}$^{\scriptscriptstyle +0.24}_{\scriptscriptstyle -0.10}$} & \cellcolor[HTML]{F7D9BD}\textbf{99.3}{\tiny\color{black!50}$\,\pm$0.21} & \cellcolor[HTML]{F7D9BD}\textbf{95.7}{\tiny\color{black!50}$\,\pm$0.37} & \cellcolor[HTML]{FAE6D3}20.2{\tiny\color{black!50}$\,\pm$0.38} & \cellcolor[HTML]{F7D9BD}\textbf{1.13}{\tiny\color{black!50}$\,\pm$0.03} & \cellcolor[HTML]{FAE6D3}2.0{\tiny\color{black!50}$\,\pm$0.06} & \cellcolor[HTML]{FAE6D3}22.5{\tiny\color{black!50}$\,\pm$0.44} & \cellcolor[HTML]{FAE6D3}1.93{\tiny\color{black!50}$\,\pm$0.05} & \cellcolor[HTML]{FAE6D3}3.2{\tiny\color{black!50}$\,\pm$0.09} \\
\bottomrule
\multicolumn{11}{@{}l}{$^\dagger$$1-\mathrm{AUC}$, in units of $10^{-5}$ (lower is better).} \\
\end{tabular}
\caption{\label{tab:app-wc}\textbf{Single-particle reconstruction in a water Cherenkov detector.} Equivalent to Table~\ref{tab:wc} of the main text, with 95\% CI and a LoRA row added.}
\end{table*}

\begin{table*}[p]
\centering
\scriptsize
\setlength{\tabcolsep}{4pt}
\begin{tabular}{@{}l ccc!{\color{black!30}\vrule}ccc}
\toprule
 & \multicolumn{3}{c!{\color{black!30}\vrule}}{\textbf{PT$_\text{1M}$ + FT$_\text{1K}$}} & \multicolumn{3}{c}{\textbf{PT$_\text{1M}$ + FT$_\text{982K}$}} \\
\cmidrule(lr){2-4} \cmidrule(lr){5-7}
Ring reconstruction & Particle $F_1$ & Ring finding $F_1$ & Ring classification & Particle $F_1$ & Ring finding $F_1$ & Ring classification \\
\midrule
\multicolumn{7}{@{}l}{\textit{Classical}} \\
\hspace{0.5em}fiTQun & 35.4 & 47.1 & 75.2 & 35.4 & 47.1 & 75.2 \\
\arrayrulecolor{black!20}\cmidrule[0.3pt]{1-7}\arrayrulecolor{black}
\multicolumn{7}{@{}l}{\textit{Self-supervised}} \\
\hspace{0.5em}Panda V2 (dec.) & 42.0{\tiny\color{black!50}$\,\pm$0.19} & 53.7{\tiny\color{black!50}$\,\pm$0.17} & 78.2{\tiny\color{black!50}$\,\pm$0.24} & 53.8{\tiny\color{black!50}$\,\pm$0.19} & 57.9{\tiny\color{black!50}$\,\pm$0.19} & 92.9{\tiny\color{black!50}$\,\pm$0.14} \\
\hspace{0.5em}Panda V2 (LoRA) & \cellcolor[HTML]{F7D9BD}\textbf{45.1}{\tiny\color{black!50}$\,\pm$0.19} & \cellcolor[HTML]{F7D9BD}\textbf{55.5}{\tiny\color{black!50}$\,\pm$0.18} & \cellcolor[HTML]{F7D9BD}\textbf{81.3}{\tiny\color{black!50}$\,\pm$0.21} & \cellcolor[HTML]{F7D9BD}\textbf{58.4}{\tiny\color{black!50}$\,\pm$0.20} & \cellcolor[HTML]{F7D9BD}\textbf{61.5}{\tiny\color{black!50}$\,\pm$0.19} & \cellcolor[HTML]{F7D9BD}\textbf{95.0}{\tiny\color{black!50}$\,\pm$0.11} \\
\bottomrule
\end{tabular}
\caption{\label{tab:app-wc-rings}\textbf{Ring reconstruction metrics against fiTQun}, in percent, pooled over single- and two-particle test samples. Ring finding is the $F_1$ of matched rings regardless of type, ring classification the fraction of matched rings assigned their true type (see \S\ref{app:data_wc}). Error bars are 95\% CI.}
\end{table*}

\begin{table*}[p]
\centering
\scriptsize
\setlength{\tabcolsep}{2pt}
\begin{tabular}[t]{lrccccc}
\toprule
\multicolumn{2}{l}{\setbox0\hbox{Class.\ (max.\ f.o.m.)}\rlap{\smash{\raisebox{-3.6pt}{\rule{\wd0}{\cmidrulewidth}}}}\box0} & \multicolumn{5}{c}{PT$_\text{1M}$ + FT$_K$} \\
\cmidrule(lr){3-7}
Method / $K$ & Param. & 100 & 1K & 10K & 100K & 417K \\
\midrule
\multicolumn{7}{l}{\textit{Supervised / classical}} \\
\hspace{0.5em}Two-blob cut & -- & \multicolumn{5}{c}{3.21{\tiny\color{black!50}$\,\pm$0.18}} \\
\hspace{0.5em}Sparse ResNet & 33.1M & 2.39{\tiny\color{black!50}$\,\pm$0.11} & 2.48{\tiny\color{black!50}$\,\pm$0.18} & \cellcolor[HTML]{F1F8EF}4.02{\tiny\color{black!50}$\,\pm$0.22} & \cellcolor[HTML]{F1F8EF}4.63{\tiny\color{black!50}$\,\pm$0.36} & \cellcolor[HTML]{F1F8EF}5.33{\tiny\color{black!50}$\,\pm$0.39} \\
\midrule
\multicolumn{7}{l}{\textit{Self-supervised}} \\
\hspace{0.5em}Panda V2 (lin.) & \llap{\raisebox{0.12ex}{\tiny$<$}\kern0.5pt}0.01M & \cellcolor[HTML]{DFEEDB}4.37{\tiny\color{black!50}$\,\pm$0.33} & \cellcolor[HTML]{DFEEDB}5.13{\tiny\color{black!50}$\,\pm$0.38} & \cellcolor[HTML]{DFEEDB}4.91{\tiny\color{black!50}$\,\pm$0.32} & \cellcolor[HTML]{DFEEDB}5.38{\tiny\color{black!50}$\,\pm$0.40} & \cellcolor[HTML]{DFEEDB}5.41{\tiny\color{black!50}$\,\pm$0.41} \\
\hspace{0.5em}Panda V2 (LoRA) & 2.3M & \cellcolor[HTML]{CFE6C8}\textbf{4.91}{\tiny\color{black!50}$\,\pm$0.35} & \cellcolor[HTML]{CFE6C8}\textbf{5.61}{\tiny\color{black!50}$\,\pm$0.43} & \cellcolor[HTML]{CFE6C8}\textbf{5.53}{\tiny\color{black!50}$\,\pm$0.41} & \cellcolor[HTML]{CFE6C8}\textbf{6.16}{\tiny\color{black!50}$\,\pm$0.52} & \cellcolor[HTML]{CFE6C8}\textbf{5.93}{\tiny\color{black!50}$\,\pm$0.48} \\
\bottomrule
\end{tabular}
\caption{\label{tab:app-next}\textbf{Signal classification on NEXT.} Equivalent to Table~\ref{tab:next} of the main text, with 95\% CI and a LoRA row added.}
\end{table*}

\begin{table*}[p]
\centering
\scriptsize
\setlength{\tabcolsep}{2.9pt}
\begin{tabular}{l cccc}
\toprule
PT$_\text{1M}$ $\downarrow$ FT$_\text{1K}$ $\rightarrow$ & LArTPC (ARI) & sPHENIX (ARI) & WC ($F_1$) & NEXT (f.o.m.) \\
\midrule
\multicolumn{5}{@{}l}{\textit{Frozen}} \\
\hspace{0.5em}LArTPC & \cellcolor[HTML]{D6D6FF}\textbf{96.0}{\tiny\color{black!50}$\,\pm$0.05} & \cellcolor[HTML]{D8E5F8}88.0{\tiny\color{black!50}$\,\pm$0.34} & 35.5{\tiny\color{black!50}$\,\pm$0.21} & \cellcolor[HTML]{DFEEDB}2.53{\tiny\color{black!50}$\,\pm$0.12} \\
\hspace{0.5em}sPHENIX & \cellcolor[HTML]{E3E3FF}93.7{\tiny\color{black!50}$\,\pm$0.07} & \cellcolor[HTML]{C4D7F4}\textbf{96.2}{\tiny\color{black!50}$\,\pm$0.21} & \cellcolor[HTML]{FDF4EC}35.8{\tiny\color{black!50}$\,\pm$0.21} & \cellcolor[HTML]{F1F8EF}2.34{\tiny\color{black!50}$\,\pm$0.14} \\
\hspace{0.5em}WC & 89.5{\tiny\color{black!50}$\,\pm$0.09} & 82.2{\tiny\color{black!50}$\,\pm$0.45} & \cellcolor[HTML]{F7D9BD}\textbf{42.5}{\tiny\color{black!50}$\,\pm$0.21} & 2.31{\tiny\color{black!50}$\,\pm$0.14} \\
\hspace{0.5em}NEXT & \cellcolor[HTML]{F4F4FF}91.2{\tiny\color{black!50}$\,\pm$0.07} & \cellcolor[HTML]{EEF3FC}84.0{\tiny\color{black!50}$\,\pm$0.41} & \cellcolor[HTML]{FAE6D3}37.2{\tiny\color{black!50}$\,\pm$0.22} & \cellcolor[HTML]{CFE6C8}\textbf{5.53}{\tiny\color{black!50}$\,\pm$0.45} \\
\midrule
\multicolumn{5}{@{}l}{\textit{LoRA}} \\
\hspace{0.5em}LArTPC & \cellcolor[HTML]{D6D6FF}\textbf{97.2}{\tiny\color{black!50}$\,\pm$0.04} & 90.4{\tiny\color{black!50}$\,\pm$0.36} & \cellcolor[HTML]{FDF4EC}41.5{\tiny\color{black!50}$\,\pm$0.21} & \cellcolor[HTML]{F1F8EF}4.16{\tiny\color{black!50}$\,\pm$0.26} \\
\hspace{0.5em}sPHENIX & \cellcolor[HTML]{E3E3FF}96.6{\tiny\color{black!50}$\,\pm$0.04} & \cellcolor[HTML]{C4D7F4}\textbf{96.9}{\tiny\color{black!50}$\,\pm$0.18} & 40.8{\tiny\color{black!50}$\,\pm$0.21} & \cellcolor[HTML]{DFEEDB}4.28{\tiny\color{black!50}$\,\pm$0.31} \\
\hspace{0.5em}WC & 95.5{\tiny\color{black!50}$\,\pm$0.05} & \cellcolor[HTML]{D8E5F8}92.9{\tiny\color{black!50}$\,\pm$0.28} & \cellcolor[HTML]{F7D9BD}\textbf{45.6}{\tiny\color{black!50}$\,\pm$0.23} & 3.31{\tiny\color{black!50}$\,\pm$0.17} \\
\hspace{0.5em}NEXT & \cellcolor[HTML]{F4F4FF}95.9{\tiny\color{black!50}$\,\pm$0.05} & \cellcolor[HTML]{EEF3FC}91.8{\tiny\color{black!50}$\,\pm$0.32} & \cellcolor[HTML]{FAE6D3}42.4{\tiny\color{black!50}$\,\pm$0.21} & \cellcolor[HTML]{CFE6C8}\textbf{5.00}{\tiny\color{black!50}$\,\pm$0.33} \\
\midrule
Baseline & 87.3{\tiny\color{black!50}$\,\pm$0.10}\rlap{$^{a}$} & 69.3\rlap{$^{b}$} & 33.9\rlap{$^{c}$} & 3.21{\tiny\color{black!50}$\,\pm$0.18}\rlap{$^{d}$} \\
\bottomrule
\multicolumn{5}{@{}l}{$^a$Panda V1~\cite{panda}. $^b$FM4NPP~\cite{park2025fm4npp}. $^c$fiTQun~\cite{fitqunSK}. $^d$Two-blob cut~\cite{ferrario2019nextwhite,ferrario2016topological}.} \\
\end{tabular}
\caption{\label{tab:app-transfer}\textbf{Out-of-domain transfer.} Equivalent to Table~\ref{tab:transfer} of the main text, with 95\% CI and a LoRA block added. Two baselines have no interval: the FM4NPP value is read from a figure of~\cite{park2025fm4npp}, and fiTQun's ring scores are available as totals per sample only.}
\end{table*}

\end{document}